\documentclass[%
reprint,
superscriptaddress,
nofootinbib,
amsmath,amssymb,
aps,
pra,
]{revtex4-2}

\usepackage{graphicx}
\usepackage{dcolumn}
\usepackage{bm}
\usepackage{physics}
\usepackage{multirow}
\usepackage{mathtools}
\usepackage{amsmath}
\usepackage[colorlinks,linkcolor=blue,urlcolor=blue,citecolor=blue]{hyperref}
\allowdisplaybreaks[4]

\newcommand{\mcQ}{\mathcal{Q}}

\begin{document}

\title{Concatenate codes, save qubits}
\author{Satoshi Yoshida}
\email{satoshiyoshida.phys@gmail.com}
\affiliation{Department of Physics, Graduate School of Science, The University of Tokyo, 7-3-1 Hongo, Bunkyo-ku, Tokyo 113-0033, Japan}
\author{Shiro Tamiya}
\email{shiro.tamiya01@gmail.com}
\affiliation{Department of Applied Physics, Graduate School of Engineering, The University of Tokyo, 7-3-1 Hongo, Bunkyo-ku, Tokyo 113-8656, Japan}
\author{Hayata Yamasaki}
\email{hayata.yamasaki@gmail.com}
\affiliation{Department of Physics, Graduate School of Science, The University of Tokyo, 7-3-1 Hongo, Bunkyo-ku, Tokyo 113-0033, Japan}

\begin{abstract}
The essential requirement for fault-tolerant quantum computation (FTQC) is the total protocol design to achieve a fair balance of all the critical factors relevant to its practical realization, such as the space overhead, the threshold, and the modularity.
A major obstacle in realizing FTQC with conventional protocols, such as those based on the surface code and the concatenated Steane code, has been the space overhead, i.e., the required number of physical qubits per logical qubit.
Protocols based on high-rate quantum low-density parity-check (LDPC) codes gather considerable attention as a way to reduce the space overhead, but problematically, the existing fault-tolerant protocols for such quantum LDPC codes sacrifice the other factors.
Here we construct a new fault-tolerant protocol to meet these requirements simultaneously based on more recent progress on the techniques for concatenated codes rather than quantum LDPC codes, achieving a constant space overhead, a high threshold, and flexibility in modular architecture designs.
In particular, under a physical error rate of $0.1\%$, our protocol reduces the space overhead to achieve the logical CNOT error rates $10^{-10}$ and $10^{-24}$ by more than $90 \%$ and $96 \%$, respectively, compared to the protocol for the surface code.
Furthermore, our protocol achieves the threshold of $2.5 \%$ under a conventional circuit-level error model, substantially outperforming that of the surface code.
The use of concatenated codes also naturally introduces abstraction layers essential for the modularity of FTQC architectures.
These results indicate that the code-concatenation approach opens a way to significantly save qubits in realizing FTQC while fulfilling the other essential requirements for the practical protocol design.
\end{abstract}

\maketitle

The realization of fault-tolerant quantum computation (FTQC) requires the total protocol design to meet all the essential factors relevant to its practical implementation, such as the space overhead, the threshold, and the modularity.
The recent development of constant-overhead protocols~\cite{kovalev2013fault, gottesman2013fault, fawzi2018constant, yamasaki2024time, krishna2021fault, cohen2022low, krishna2021fault, tremblay2022constant} substantially reduces the space overhead, i.e., the required number of physical qubits per logical qubit, compared to the conventional protocols such as those based on the surface code \cite{bravyi1998quantum, dennis2002topological} and the concatenated Steane code \cite{steane1996simple}.  
In particular, the most recent development~\cite{yamasaki2024time} based on the concatenation of quantum Hamming codes \cite{hamming1950error, steane1996simple} is promising for the implementation of FTQC since Ref.~\cite{yamasaki2024time} explicitly clarifies the full details of the protocol for implementing logical gates and efficient decoders, making it possible to realize universal quantum computation in a fault-tolerant way.  
Toward the practical implementation, however, it is indispensable to optimize the original protocol in Ref.~\cite{yamasaki2024time} to improve its threshold, which is, by construction, at least as bad as the concatenated Steane code.
Furthermore, even a proper quantitative evaluation of the original protocol in Ref.~\cite{yamasaki2024time} was still missing due to the lack of the numerical study of the protocols based on the quantum Hamming codes.

\begin{figure}[t!]
    \centering
    \includegraphics[width=\linewidth]{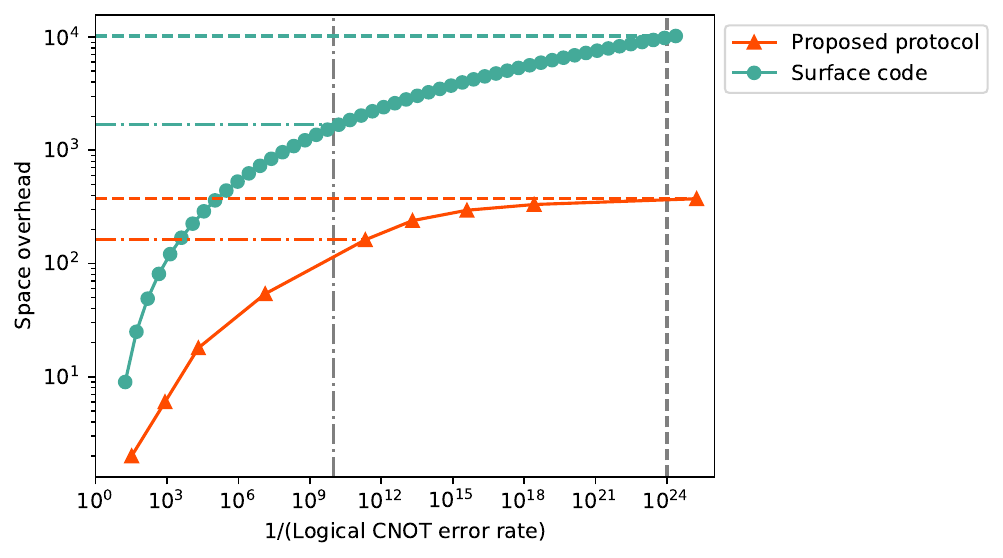}
    \caption{{\bf Comparison of space overhead of the proposed protocol with that for the surface code.} The figure plots the space overheads and logical error rates of the proposed protocol ({\color[rgb]{1,0.29411764705,0} $\blacktriangle$}) and the surface code ({\color[rgb]{0.26666666666,0.66666666666,0.6} $\bullet$}). The logical error rate is calculated under a circuit-level depolarizing error model at a physical error rate $0.1\%$. The dash-dotted lines represent the logical error rate $10^{-10}$ and the corresponding space overhead, i.e., $162$ physical qubits per logical qubit for our protocol. The dashed lines represent the logical error rate $10^{-24}$ and the corresponding space overhead, i.e., $373$ physical qubits per logical qubit for our protocol. Our protocol reduces the space overhead to achieve the logical error rates $10^{-10}$ and $10^{-24}$ by more than $90 \%$ and $96 \%$, respectively, compared to the protocol for the surface code.}
    \label{fig:overhead_total}
\end{figure}

\begin{table}
    \caption{{\bf Construction of the proposed protocol.} Our quantum code uses the level-5 $C_4/C_6$ code as an underlying quantum code ($\mcQ_0$), and on top of this, we concatenate a series of quantum Hamming codes. The second column of this table shows a quantum code to be concatenated at each level. The rightmost column of this table shows the space overhead, which is the ratio of the number of physical qubits denoted by $N$ and the number of logical qubits denoted by $K$.}
    \label{tab:concatenation}
    \begin{ruledtabular}
        \begin{tabular}{c|c|r|r|r}
             & Quantum code & $N$ & $K$ & $N/K$\\\hline
            level-1 & $C_4 (=[[4,2,2]])$ & $4$ & $2$ & $2$\\
            level-2 & $C_6 (=[[6,2,2]])$ & $12$ & $2$ & $6$\\
            level-3 & $C_6 (=[[6,2,2]])$ & $36$ & $2$ & $18$\\
            level-4 & $C_6 (=[[6,2,2]])$ & $108$ & $2$ & $54$\\
            level-5 & $C_6 (=[[6,2,2]])$ & $324$ & $2$ & $162$\\
            level-6 & $\mcQ_5 (=[[31,21,3]])$ & $1.00\times 10^4$ & $42$ & $239$\\
            level-7 & $\mcQ_6 (=[[63,51,3]])$ & $6.33\times 10^5$ & $2.14\times 10^3$ & $295$\\
            level-8 & $\mcQ_7 (=[[127,113,3]])$ & $8.04\times 10^7$ & $2.42\times 10^5$ & $332$\\
            level-9 & $\mcQ_7 (=[[127,113,3]])$ & $1.02\times 10^{10}$ & $2.74\times 10^7$ & $373$
        \end{tabular}
    \end{ruledtabular}
\end{table}

In this work, we construct an optimized fault-tolerant protocol by substantially improving the protocol in Ref.~\cite{yamasaki2024time}, achieving an extremely low space overhead and a high threshold to simultaneously outperform the surface code.
The optimization is performed based on our quantitative evaluation of the performance of the fault-tolerant protocols for various choices of quantum error-correcting codes (see Tables~\ref{tab:concatenation} and~\ref{tab:threshold}), which we carried out in a unified way under a circuit-level depolarizing error model following the convention of Ref.~\cite{PhysRevA.86.032324}.
Our numerical study makes it possible to optimize the combination of the quantum codes to be concatenated.
Our numerical results show that the threshold of the original protocol for quantum Hamming codes in Ref.~\cite{yamasaki2024time} is $\sim 10^{-8}$.
To improve the threshold, our protocol uses the $C_4/C_6$ code~\cite{knill2005quantum}, which achieves the state-of-the-art high threshold and is recently realized in experiments \cite{paetznick2024demonstration}, at the physical level; on top of the $C_4/C_6$ code, our protocol concatenates the quantum Hamming codes at the larger concatenation levels to achieve the constant space overhead.
Under a physical error rate of $0.1\%$, compared to the conventional protocol for the surface code, our protocol reduces the space overhead to achieve the logical error rate $10^{-10}$ and $10^{-24}$ by more than $90 \%$ and $96 \%$, respectively (see Fig.~\ref{fig:overhead_total}).
The threshold of our protocol is $2.5 \%$, which substantially outperforms that of the surface code (see Table~\ref{tab:threshold}).
These results establish a basis for the practical fault-tolerant protocols, especially suited for the architectures with all-to-all two-qubit gate connectivity, which is partially achieved experimentally in neutral atoms~\cite{bluvstein2023logical}, trapped ions~\cite{PhysRevX.11.041058,egan2021fault}, and theoretically proposed in optics~\cite{yamasaki2020polylogoverhead,Bourassa2021blueprintscalable,litinski2022active}.

\section*{Results}

\noindent {\bf Setting.}
We construct a space-overhead-efficient fault-tolerant protocol by optimizing the protocol presented in Ref.~\cite{yamasaki2024time}.
The original protocol in Ref.~\cite{yamasaki2024time} is based on the concatenation of a series of quantum Hamming codes with increasing code sizes.
Quantum Hamming code is a family of quantum codes $\mcQ_r$ parameterized by $r\in\{3,4,\ldots\}$, consisting of $N_r=2^r-1$ physical qubits and $K_r=N_r-2r$ logical qubits with code distance 3 \cite{hamming1950error, steane1996simple}, which is written as an $[[N_r, K_r, 3]]$ code.
By concatenating the quantum Hamming code $\mcQ_{r_{l}}$ for a sequence $(r_{l}=l+2)_{l=1,2,\ldots}$ of parameters at the concatenation level $l\in\{1, \ldots, L\}$, we obtain a quantum code consisting of $N = \prod_{l=1}^{L} N_{r_l}$ physical qubits and $K = \prod_{l=1}^{L} K_{r_l}$ logical qubits.
Its space overhead, defined by the ratio of $N$ and $K$~\cite{gottesman2013fault}, converges to a finite constant factor $\eta_{\infty}$ as
\begin{align}
    {N\over K} = \prod_{l=1}^{L} {N_{r_l} \over K_{r_l}} \to \eta_{\infty} <\infty \quad \mathrm{as}\quad L\to \infty,
\end{align}
where $\eta_\infty$ is given by $\eta_\infty\approx 36$~\cite{yamasaki2024time}.
However, the threshold of the protocol based on this quantum code is given by $\sim 10^{-8}$, as shown in Supplementary Information.
As discussed in Ref.~\cite{yamasaki2024time}, instead of $r_l=l+2$, we can also take an arbitrary sequence $(r_l)_{l=1,2,\ldots}$ satisfying $\eta_\infty = \prod_{l=1}^{\infty}{N_{r_l} \over K_{r_l}} < \infty$ to achieve the constant space overhead, and our choice of $r_l$ will be clarified below.

We optimize this original protocol by replacing the physical qubits of the original protocol with logical qubits of a finite-size quantum code $\mcQ_0$ (called an underlying quantum code).
With this replacement, we aim to improve the threshold determined at the physical level while maintaining the constant space overhead at the large concatenation levels.
Here, the logical error rate of the logical qubits of the underlying quantum code should be lower than the threshold of the original protocol so that the original protocol can further suppress the logical error rate.
If the underlying quantum code $\mcQ_0$ has $N_0$ physical qubits and $K_0$ logical qubits, the overall space overhead is given by
\begin{align}
\label{eq:space_overhead}
    {N\over K} = {N_0\over K_0} \prod_{l=1}^{L}{N_{r_l} \over K_{r_l}} \to \eta'_\infty <\infty \quad \text{as $L\to\infty$},
\end{align}
which remains a constant value given by $\eta^\prime_\infty = {N_0\over K_0} \eta_\infty$ as long as we use a fixed code as the underlying quantum code.

For our protocol, we propose the following code construction:
\begin{itemize}
    \item As an underlying quantum code, we use the $C_4/C_6$ code~\cite{knill2005quantum} as first $L'$ levels of the concatenated code, where the 4-qubit code denoted by $C_4 (= [[4,2,2]])$ is concatenated with the 6-qubit code denoted by $C_6 (=[[6,2,2]])$ for $L'-1$ times.
    \item On top of the underlying quantum code, i.e., at the concatenation levels $L'+1, L'+2, \ldots, L$, we concatenate quantum Hamming codes $\mcQ_{r_l}$ for an optimized choice of the sequence $(r_l)_{l=1,2,\ldots}$ of parameters, where $\mcQ_{r_l}$ is used at the concatenation level $L^\prime+l$.
\end{itemize}
The $C_4/C_6$ code is adopted as the underlying quantum code since it achieves the state-of-the-art high threshold.
To avoid the increase of overhead, we use a non-post-selected protocol of the $C_4/C_6$ code in Ref.~\cite{knill2005quantum} rather than a post-selected one that excludes the error-detected events.

\begin{table}
    \caption{{\bf Comparison of the error threshold of the underlying codes and the required space overhead to achieve the logical error rate $10^{-24}$ of the overall quantum codes obtained by concatenating the underlying codes with the optimized series of the quantum Hamming codes.}
    The table shows the error threshold of the underlying codes (the $C_4/C_6$ code, the surface code, the concatenated Steane code, and the $C_4$/Steane code) and the required space overhead to achieve the logical error rate $10^{-24}$ under the physical error rates $p=0.01\%, 0.1\%, 1\%$ for the overall quantum codes obtained by concatenating the underlying codes with the optimized series of the quantum Hamming codes.  Bold values represent the minimum space overheads among the four quantum codes under the same physical error rates. Note that for $p=1\%$, the $C_4/C_6$ code is the only one among the four codes that can suppress the logical error rate; similarly, for $p=0.1\%$, the concatenated Steane code cannot suppress the logical error rate. We remark that the threshold values for the $C_4/C_6$ code, the concatenated Steane code and the surface codes are slightly lower than those shown in Refs.~\cite{goto2013fault, goto2016minimizing, vuillot2019code} due to the difference of the error model.}
    \label{tab:threshold}
    \begin{ruledtabular}
        \begin{tabular}{c|r|rrr}
            \multirow{2}{*}{Underlying code} & \multirow{2}{*}{Threshold} & \multicolumn{3}{c}{Overall space overhead} \\
             & & $p=0.01\%$ & $p=0.1\%$ & $p=1\%$\\\hline
            $C_4/C_6$ code & $2.5 \%$ & ${\bf 1.0\times 10^2}$ & ${\bf 3.7\times 10^2}$ & ${\bf 6.2\times 10^3}$\\
            Surface code & $0.31 \%$ & $4.3\times 10^2$ & $4.5\times 10^3$ & -\\
            Steane code & $0.030 \%$ & $6.1\times 10^3$ & - & -\\
            $C_4$/Steane code & $0.14 \%$ & $3.3\times 10^2$ & $9.3\times 10^4$ & -
        \end{tabular}
    \end{ruledtabular}
\end{table}

To estimate the space overhead and the threshold, we evaluate the logical CNOT error rate of the fault-tolerant protocols based on the $C_4/C_6$ code and the quantum Hamming codes.
The logical CNOT error rate is evaluated at each concatenation level using the Monte Carlo sampling method in Refs.~\cite{goto2014step, goto2016minimizing}, which is based on the reference entanglement method~\cite{schumacher1996sending, knill2005quantum}.
By convention, we describe the noise on physical qubits by a circuit-level depolarizing error model (see Methods for the details of the simulation method and the error model).
In the simulation, we assume no geometrical constraints on manipulating quantum gates, which is applicable to neutral atoms~\cite{bluvstein2023logical}, trapped ions~\cite{PhysRevX.11.041058,egan2021fault}, and optics~\cite{yamasaki2020polylogoverhead,Bourassa2021blueprintscalable,litinski2022active}.
Our numerical results show that by using the $C_4/C_6$ code as the underlying quantum code, our protocol achieves a high threshold $2.5 \%$ (see Table~\ref{tab:threshold}), where we use the non-post-selected protocol of the $C_4/C_6$ code rather than the post-selected one in Ref.~\cite{knill2005quantum}.
We optimize the combination of the quantum codes, i.e., the choice of parameters $L'$, $L$, and $r_l$, based on our simulation results so as to reduce the space overhead.
In particular, the optimized parameters that we found are $L'=5$, $L=9$, and $r_1=5, r_2=6, r_3=r_4=7$ (see Table \ref{tab:concatenation}).
Note that the quantum codes $\mcQ_3, \mcQ_4$ in the original protocol of Ref.~\cite{yamasaki2024time} are skipped to improve the space overhead of our protocol.
The quantum code $\mcQ_7$ is used twice since the quantum code $\mcQ_8$ in level-9 is not expected to reduce the logical error to $10^{-24}$.
To avoid the combinatorial explosion arising from the combinations of these parameters, we performed a level-by-level numerical simulation at each concatenation level (see Methods for the details). 
With this technique, our simulation makes it possible to flexibly optimize the combination of the quantum codes to be concatenated for designing our protocol.

\noindent {\bf Large-scale resource estimation.}
Under a physical error rate of $0.1\%$,
we compare the space overhead of our proposed protocol to achieve the logical CNOT error rates $10^{-10}$ and $10^{-24}$ with a conventional protocol for the surface code.
Note that another conventional protocol using the concatenated Steane code cannot suppress the logical error rate under the physical error rate $0.1\%$ since the threshold is larger than $0.1\%$ (see Table \ref{tab:threshold}).
Factoring of a $2048$-bit integer using Shor's algorithm \cite{shor1994algorithms} requires the logical error rate $10^{-10}$~\cite{gidney2021factor}, which is relevant to the currently used cryptosystem RSA-2048 \cite{rivest1978method, barker2016nist}.  
The logical error rate $\sim 10^{-24}$ is a rough estimate of the logical error rate of classical computation (see Methods for the details of these estimations).

As shown in Fig.~\ref{fig:overhead_total}, the surface code requires the space overhead $\sim 1.7 \times 10^3$ and $\sim 10.2 \times 10^3$ to achieve the logical error rates $\sim 10^{-10}$ and $\sim 10^{-24}$, respectively.
On the other hand, our protocol only requires the space overheads $\sim 162$ and $\sim 373$ to achieve the same logical error rates, saving the space overheads by more than $90 \%$ and $96 \%$, respectively, compared to the surface code.
Note that our protocol achieves constant space overhead while the protocol for the surface code (as well as that for the concatenated Steane code) has growing space overhead; thus, in principle, the advantage of our protocol can be arbitrarily large as the target logical error rate becomes small.
However, our contribution here is to clarify that our protocol indeed offers a space-overhead advantage by orders of magnitude in the practical regimes.

\begin{figure}
    \centering
    \includegraphics[width=\linewidth]{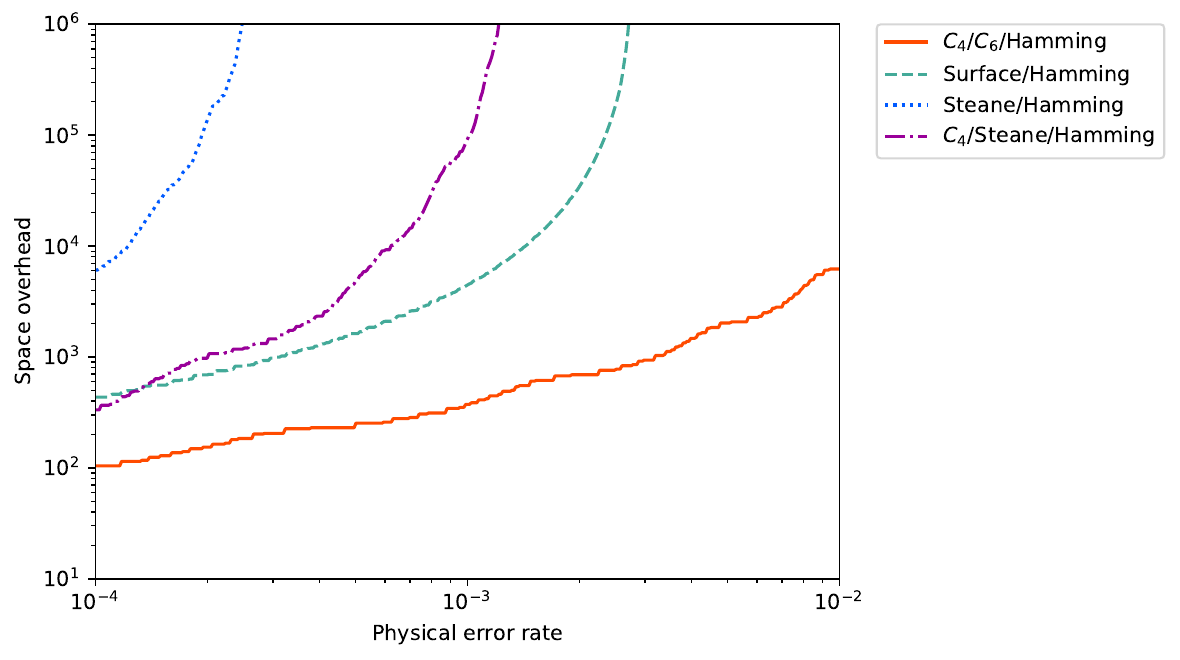}
    \caption{{\bf Comparison on space overheads of the underlying code (the $C_4/C_6$ code, the surface code, the concatenated Steane code, and the $C_4$/Steane code) concatenated with the optimal series of the quantum Hamming codes.} The horizontal axis shows the physical error rate, and the vertical axis shows the space overhead to achieve the logical error rate $10^{-24}$. The simulation is performed under the circuit-level depolarizing error model.}
    \label{fig:overhead_underlying}
\end{figure}

\noindent {\bf Comparison on underlying quantum codes.}
The quantum code for our protocol shown in Table \ref{tab:concatenation} is obtained by optimizing the underlying quantum code, and under the physical error rate $0.1\%$, our optimized choice of the underlying quantum code turns out to be the level-$5$ $C_4/C_6$ code.
Here, we show this optimization procedure in more detail.
For this optimization, we compare four candidate quantum codes: the $C_4/C_6$ code \cite{knill2005quantum}, the surface code \cite{bravyi1998quantum, dennis2002topological}, the concatenated Steane code \cite{steane2003overhead}, and the $C_4$/Steane code.
The $C_4$/Steane code is newly constructed in this work by concatenating the $[[4,2,2]]$ code (i.e., the $C_4$ code)  with the Steane code (see Supplementary Information for details).
For each underlying code, we optimize the concatenation level or the distance of the underlying code and the series of the quantum Hamming codes, and compare the required overall space overhead to achieve the logical error rate $10^{-24}$.

In Fig.~\ref{fig:overhead_underlying} and Table \ref{tab:threshold}, we compare the thresholds of these four underlying quantum codes and the space overheads of the overall quantum codes obtained by concatenating the underlying codes with the series of quantum Hamming codes to achieve the logical error rate $10^{-24}$ at the physical error rates $p\in[0.01\%, 1\%]$.
In Fig.~\ref{fig:overhead_underlying}, we call the concatenated code of $X$ and the series of quantum Hamming codes as $X/\mathrm{Hamming}$ for $X\in\{C_4/C_6, \mathrm{surface}, \mathrm{Steane}, C_4/\mathrm{Steane}\}$.
For a fair comparison, we performed the numerical simulation of implementing logical CNOT gates for all these four codes under the aforementioned circuit-level depolarizing error model.
For the decoding of the surface code, we use the minimum-weight perfect matching decoder~\cite{higgott2021pymatching,higgott2023sparse}, and for the other concatenated codes, we use a hard-decision decoder to cover practical situations where the efficiency of implementing the
decoder matters (see Supplementary Information for more details).
Conventionally, the threshold for the surface code is evaluated by implementing a quantum memory (i.e., the logical identity gate)~\cite{PhysRevA.86.032324}, but for a fair comparison, we here evaluate that by the logical CNOT gate, which is implemented by lattice surgery~\cite{horsman2012surface, vuillot2019code} and is simulated using the method in Ref.~\cite{Gidney2022stability} (see Supplementary Information for details).
Similarly, Ref.~\cite{steane2003overhead} evaluates the threshold for the concatenated Steane code by implementing the logical identity gate, but we evaluate that by the transversal implementation of the logical CNOT gate.
Note that the thresholds evaluated by the logical CNOT gate may be worse than those by the logical identity gate~\cite{vuillot2019code}, but our setting of the numerical simulation is motivated by the fact that the realization of quantum memory by just implementing the logical identity gate is insufficient for universal quantum computation.
We also remark that various numerical simulations have been performed in the literature under different error models from ours, e.g., for the surface code in Refs.~\cite{10.5555/2011814.2011815,chamberland2018deep}, for the concatenated Steane code in Refs.~\cite{10.5555/2011814.2011815,goto2016minimizing}, and for the $C_4/C_6$ code in Refs.~\cite{knill2005quantum,goto2013fault}, but our contribution here is to perform the numerical simulation of all the codes under the same circuit-level error model in a unified way to make a direct, systematic comparison.

As shown in Fig.~\ref{fig:overhead_underlying} and Table \ref{tab:threshold}, the $C_4/C_6/\mathrm{Hamming}$ code has the minimum space overhead for physical error rates $p\in [0.01\%, 1\%]$.
At the same time, the space overheads of the $\mathrm{Surface}/\mathrm{Hamming}$ and $C_4/\mathrm{Steane}/\mathrm{Hamming}$ codes are of the same magnitude as that of the $C_4/C_6/\mathrm{Hamming}$ code for a low physical error rate $p\sim 0.01\%$.

\noindent {\bf Comparison with the quantum low-density parity-check (LDPC) code.}
We have so far offered a quantitative analysis of our protocol based on the code-concatenation approach.
We here compare this approach with another existing approach toward low-overhead FTQC based on the high-rate quantum LDPC codes originally proposed in Refs.~\cite{kovalev2013fault, gottesman2013fault, fawzi2018constant}.

The crucial difference between our approach based on concatenated codes and the approach based on quantum LDPC codes is modularity.
In the approach of quantum LDPC code, one needs to realize a single large-size code block.
To suppress the logical error rate more and more, each code block may become arbitrarily large, yet an essential assumption for the fault tolerance of the quantum LDPC codes is to keep the physical error rates constant~\cite{gottesman2013fault, fawzi2018constant}.
In experiments, problematically, it is in principle challenging to arbitrarily increase the number of qubits in a single quantum device without increasing physical error rates~\cite{xu2023constantoverhead,PRXQuantum.4.040319}.
By contrast, in the code-concatenation approach, we can realize a fixed-size code at each level of the code concatenation by putting finite efforts into improving a quantum device; that is, each fixed-size code serves as a fixed-size abstraction layer in the implementation that is stored in a single module.
As shown in Ref~\cite{yamasaki2024time}, as we increase the concatenation levels, the logical error rates are suppressed doubly exponentially, whereas the required number of gates for implementing each gadget grows much more slowly.
Once the error rates are suppressed by a concatenated code at some concatenation level, the low error rate of each logical gate provides a margin for using more logical gates (i.e., tolerating more architectural overhead) to implement FTQC at the higher concatenation levels, which provides flexibility for scalable architecture design.
For example, once we develop finite-size devices implementing the fixed-size code, we can further scale up FTQC by combining these error-suppressed devices by using quantum channels to connect these devices and implement another fixed-size code to be concatenated at the next concatenation level.
These quantum channels can be lossier than the physical gates in each device since the quantum states that will go through the channels are already encoded.
In this way, our code-concatenation approach offers modularity, an essential requirement for the FTQC architectures.

Apart from the modularity, another advantage is that our protocol based on concatenated codes can implement logical gates faster than the existing protocols for quantum LDPC codes.
In the protocol for quantum LDPC codes in Refs.~\cite{gottesman2013fault, fawzi2018constant}, almost all gates, including most of the Clifford gates, are implemented by gate teleportation using auxiliary code blocks; to maintain constant space overhead, gates must be applied sequentially, which incurs the polynomial time overhead.
Other Clifford gate schemes are proposed based on code deformation \cite{krishna2021fault} and lattice surgery \cite{cohen2022low}, but they also introduce additional overheads.
In particular, the code deformation scheme may introduce an additional time overhead that may be worse than the gate teleportation method~ \cite{krishna2021fault}.
The lattice surgery scheme requires a large patch of the surface code, which makes the space overhead of the overall protocol non-constant if we want to attain low time overhead~ \cite{cohen2022low}.
Apart from these schemes for logical gate implementations, a stabilizer measurement scheme for a constant-space-overhead quantum LDPC code in thin planar connectivity is presented in Ref.~\cite{tremblay2022constant}.
This protocol implements a quantum memory (i.e., the logical identity gate), but to implement universal quantum computation in a fault-tolerant way, we need to add the components to implement state preparation and logical gates, which incur the overhead issues in the same way as the above.
More recent protocols in Refs.~\cite{pattison2023hierarchical,bravyi2023highthreshold} aim to improve the implementability of quantum LDPC codes, but in the same way, these protocols can only be used as the quantum memory; problematically, it is currently unknown how to realize logical gates with these protocols, and it is also unknown how to achieve constant-space-overhead FTQC based on these protocols without sacrificing their implementability.
In contrast with these protocols, our protocol can implement universal quantum computation within constant space overhead and quasi-polylogarithmic time overhead, by using the concatenated code rather than quantum LDPC codes, as shown in Ref.~\cite{yamasaki2024time}.

We remark that, due to this difference, it is not straightforward to obtain numerical results on the existing protocols for the high-rate quantum LDPC codes in the same setting as our protocol; however, if one develops more efficient protocols achieving universal quantum computation using the high-rate quantum LDPC codes, the current numerical results on comparing our protocol with those of the surface code and the concatenated Steane code also serve as a useful baseline for further comparison, which we leave for future work.
We also point out that in the current status, even if one wants to implement constant-space-overhead FTQC using quantum LDPC codes, one eventually needs to use concatenated codes in combination.
In particular, as shown in Refs.~\cite{gottesman2013fault, fawzi2018constant}, the existing constant-space-overhead fault-tolerant protocols for such quantum LDPC codes rely on concatenated codes for preparation of logical $\ket{0}$ states, e.g., by using the encoding procedure implemented by the concatenated Steane code~\cite{christandl2022fault}.
Thus, even though a part of the protocol using the high-rate quantum LDPC codes may be efficient, the part relying on the concatenated codes may become a bottleneck in practice, which should be taken into account in future work for a fair comparison of the overall protocols.

\section*{Discussion}

In this work, we have constructed a low-overhead, high-threshold, modular protocol for FTQC based on the recent progress on the code-concatenation approach in Ref.~\cite{yamasaki2024time}.
To design our protocol, we have performed thorough numerical simulations of the performance of fault-tolerant protocols for various quantum codes, under the same circuit-level error model in a unified way, as shown in Figs.~\ref{fig:overhead_total} and~\ref{fig:overhead_underlying} and Tables~\ref{tab:concatenation} and~\ref{tab:threshold}.
Based on these numerical results, we have proposed an optimized protocol, which we have designed by seeking an optimized combination of the underlying quantum code at the physical level and the series of quantum Hamming codes at higher concatenation levels.
The proposed protocol (Table~\ref{tab:concatenation}) uses a fixed-size $C_4/C_6$ code at the physical level to attain a high threshold and, on top of this underlying quantum code, concatenate the quantum Hamming codes to achieve the constant space overhead.
This proposed protocol achieves a substantial saving of the space overhead compared to that of the surface code (Fig.~\ref{fig:overhead_total}), has a higher threshold $2.5\%$ than those of the surface code and the concatenated Steane code (Table~\ref{tab:threshold}), and offers modularity owing to the code-concatenation approach.

At the same time, as shown in Fig.~\ref{fig:overhead_underlying}, our results show that other choices of the underlying quantum code can achieve a similar space overhead to the $C_4/C_6$ code depending on the physical error rate; in particular, we find that the surface code and the $C_4$/Steane code that we have developed in this work can achieve a similar space overhead compared to the $C_4/C_6$ code at the physical error rate $0.01\%$.
This result implies that the underlying code can be further optimized by considering other factors of the fault-tolerant quantum computation such as the connectivity.
Since our protocol is based on concatenated codes, the proposed protocol has flexibility in the choice of the underlying quantum code and the sequence of quantum Hamming codes to be concatenated, which will also be useful for further optimization of fault-tolerant protocols depending on the advances of experimental technologies in the future.

We have constructed our fault-tolerant protocol without assuming geometrical constraints on quantum gates.
Non-local interactions are indispensable to avoid the growing space overhead of FTQC on large scales, which has been a major obstacle to implementing FTQC.
By combining the swapping technique in Ref.~\cite{gottesman2000fault} with the protocol shown in Ref.~\cite{yamasaki2024time}, we can implement the constant space overhead protocol with two-dimensional nearest-neighbor gates or one-dimensional next-nearest-neighbor gates\footnote{Reference~\cite{baspin2023lower} shows that polylogarithmic space overhead is required for $d$-dimensional implementation of protocols with quantum LDPC codes for any $d$, but this limitation does not apply to protocols with concatenated codes~\cite{pattison2023hierarchical}.}.
However, it is unclear whether the constant space overhead protocol can be implemented using one-dimensional nearest-neighbor gates without sacrificing the distance of codes at each concatenation level (see also footnote d in Ref.~\cite{svore2007noise}).
By contrast, all-to-all connectivity of physical gates is indeed becoming possible in various experimental platforms, such as neutral atoms~\cite{bluvstein2023logical}, trapped ions~\cite{PhysRevX.11.041058,egan2021fault}, and optics~\cite{yamasaki2020polylogoverhead,Bourassa2021blueprintscalable,litinski2022active}; in such cases,  the proposed protocol substantially reduces the space overhead compared to the surface code, as shown in Fig.~\ref{fig:overhead_total}.
Consequently, our protocol lends increased importance to such physical platforms with all-to-all connectivity; at the same time, the technological progress on the experimental side may also lead to extra factors to be considered for practical FTQC protocols, and our results and techniques constitute a basis for further optimization of the fault-tolerant protocols in these platforms.

\begin{figure}
    \centering
    \includegraphics[width=\linewidth]{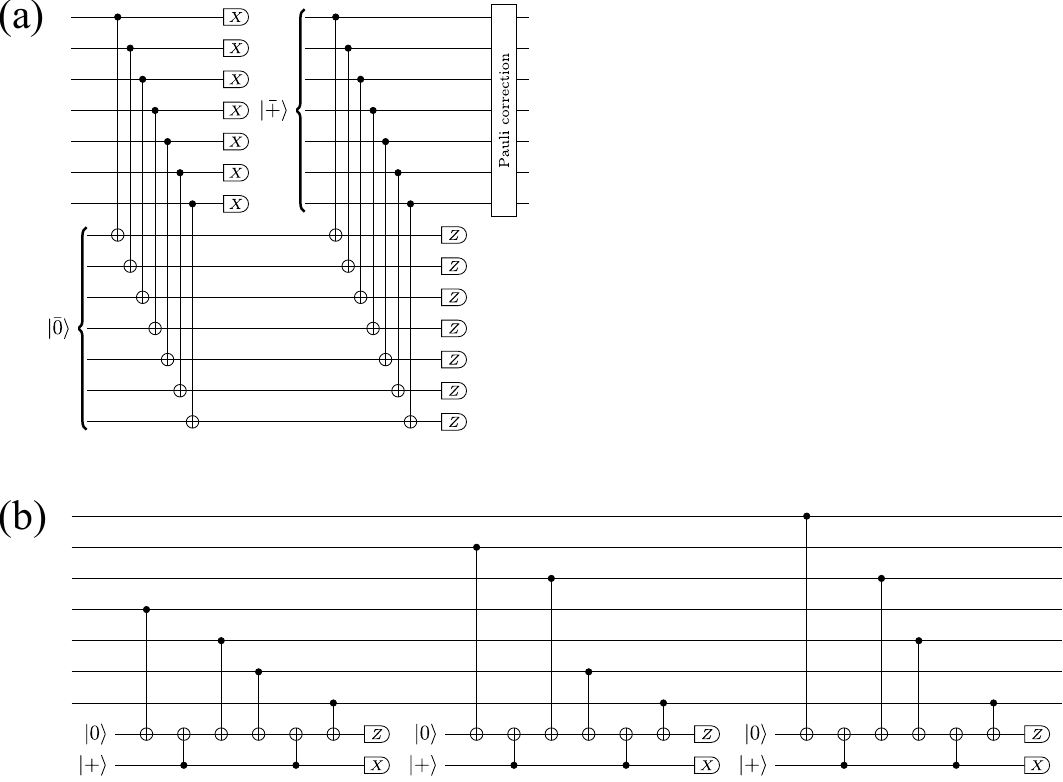}
    \caption{{\bf Comparison of  Knill's error correction gadget with the flag-qubit error correction gadget.} (a) A variant of Knill's error correction gadget for the Steane code \cite{knill2005quantum, yamasaki2024time}.
    A copy of the code block is used for the fault-tolerant syndrome extraction. The symbols $\ket{\bar{0}}$ and $\ket{\bar{+}}$ in the circuit represent the logical zero and plus states, respectively.
    (b) The flag-qubit fault-tolerant $Z$ syndrome extraction of the Steane code using two auxiliary qubits \cite{chao2018fault, reichardt2020fault}.
    This circuit followed by the fault-tolerant $X$ syndrome extraction, which is similarly constructed, constitutes the error correction gadget.
    }
    \label{fig:7qubit_comparison}
\end{figure}

Lastly, we remark that the fault-tolerant protocol requires auxiliary qubits to extract the syndrome of the quantum code and prepare the magic states.
Since Ref.~\cite{yamasaki2024time} shows the full fault-tolerant protocol which leaves the space overhead constant even including the auxiliary qubits, our proposed protocol also achieves the constant space overhead including the auxiliary qubits in the asymptotic regime.
However, the original protocol in Ref.~\cite{yamasaki2024time} is not optimized to reduce the number of auxiliary qubits, so it still remains unclear if the space overhead including auxiliary qubits may be significantly better than the conventional protocols based on the surface code at a finite regime.
One way to circumvent this problem is to reduce the number of auxiliary qubits by using the flag qubits \cite{yoder2017surface,chao2018quantum,chamberland2018flag,chao2020flag,chao2018fault,chamberland2019fault,goto2014step}.
The fault-tolerant protocol constructed in Ref.~\cite{yamasaki2024time} uses a variant of Knill's error correction gadget, which uses a copy of the code block for the fault-tolerant syndrome extraction (see Fig.~\ref{fig:7qubit_comparison}~(a)).
On the other hand, the flag-qubit protocol shown in Ref.~\cite{chao2018quantum} requires two auxiliary qubits for the fault-tolerant syndrome extraction of the quantum Hamming codes (see Fig.~\ref{fig:7qubit_comparison}~(b)).
The idea of flag qubits is extended to the error correction of arbitrary stabilizer codes \cite{chamberland2018flag, chao2020flag} and the fault-tolerant gate operations \cite{goto2014step, chao2018fault,chamberland2019fault}.
Based on these ideas, it is essential to optimize the entire fault-tolerant protocol, as has been recently done for many-hypercube codes in Ref.~\cite{goto2024high}.
We leave it as future work to optimize the full fault-tolerant protocol using the idea of the flag qubits to reduce the number of auxiliary qubits and evaluate its performance.

\section*{Methods}

In Methods, after summarizing the notations, we first describe the error model used in the numerical simulation and the Monte Carlo simulation method to evaluate the logical CNOT error rate.
Then, we provide the details of our estimation of the required logical error rate of quantum computation, based on the evaluation of the CNOT gate counts of the quantum circuit implementing Shor's algorithm for 2048-bit RSA integer factoring and the required error rates for the classical computation.
Finally, we present our method for estimating the logical CNOT error rate of the large-scale concatenated codes using the small-scale level-by-level simulation results at each concatenation level.

\noindent {\bf Notation.}
The computational basis (also called the $Z$ basis) of a qubit $\mathbb{C}^2$ is denoted by $\{\ket{0}, \ket{1}\}$, and the complementary basis (also called the $X$ basis) $\{\ket{+}, \ket{-}\}$ is defined by $\ket{\pm}\coloneqq {1\over \sqrt{2}}(\ket{0}\pm \ket{1})$.
By the convention of Ref.~\cite{nielsen2010quantum}, we use the following notation on $1$-qubit and $2$-qubit unitaries:
\begin{align}
    I &= \left(\begin{matrix}
        1 & 0\\
        0 & 1
    \end{matrix}\right),\\
    X &= \left(\begin{matrix}
        0 & 1\\
        1 & 0
    \end{matrix}\right),\\
    Y &= \left(\begin{matrix}
        0 & -i\\
        i & 0
    \end{matrix}\right),\\
    Z &= \left(\begin{matrix}
        1 & 0\\
        0 & -1
    \end{matrix}\right),\\
    H &= {\frac{1}{\sqrt{2}}} \left(\begin{matrix}
        1 & 1\\
        1 & -1
    \end{matrix}\right),\\
    S &= \left(\begin{matrix}
        1 & 0\\
        0 & i
    \end{matrix}\right),\\
    \mathrm{CNOT} &= \left(\begin{matrix}
        1 & 0 & 0 & 0\\
        0 & 1 & 0 & 0\\
        0 & 0 & 0 & 1\\
        0 & 0 & 1 & 0
    \end{matrix}\right),
\end{align}
where the $1$-qubit and $2$-qubit unitaries are shown in the matrix representations in the computational bases $\{\ket{0}, \ket{1}\} \subset \mathbb{C}^2$ and $\{\ket{0}\otimes\ket{0}, \ket{0}\otimes\ket{1}, \ket{1}\otimes\ket{0}, \ket{1}\otimes\ket{1}\} \subset \mathbb{C}^2\otimes \mathbb{C}^2$, respectively.
See also Ref.~\cite{gottesman2010introduction} for terminology on FTQC\@.

\noindent {\bf Error model.}
In this work, the stabilizer circuits for describing the fault-tolerant protocols are composed of state preparations of $\ket{0}$ and $\ket{+}$, measurements in the $Z$ and $X$ bases, single-qubit gates $I, X, Y, Z, H, S$, and a two-qubit CNOT gate.
Each of these preparation, measurement, and gate operations in a circuit is called a location in the circuit.
By the convention of Ref.~\cite{PhysRevA.86.032324}, we use a circuit-level depolarizing error model.
In this model, independent and ideally distributed (IID) Pauli errors randomly occur at each location, i.e., after state preparations and gates, and before measurements.
By convention, we ignore the error and the runtime of polynomial-time classical computation used for decoding in the fault-tolerant protocols.

\begin{figure}
    \centering
    \includegraphics[width=\linewidth]{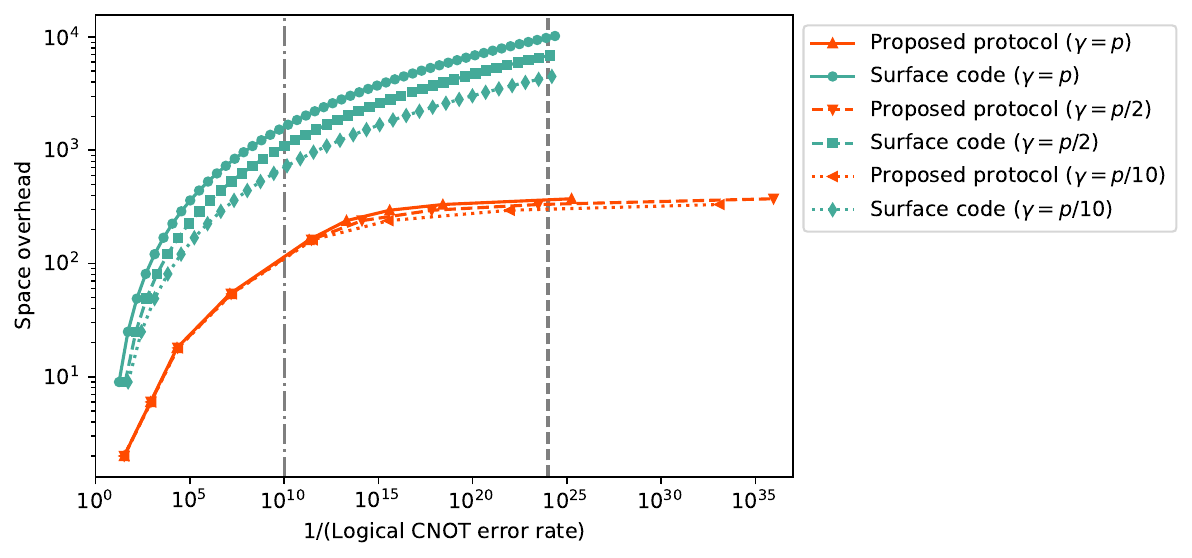}
    \caption{{\bf Comparison of space overhead of the proposed protocol with that for the surface code under three error models ($\gamma=p$, $\gamma=p/2$ and $\gamma=p/10$).}
    The figure plots the space overheads and logical error rates of the proposed protocol shown in Table~\ref{tab:concatenation} ({\color[rgb]{1,0.29411764705,0} $\blacktriangle$} for $\gamma=p$, {\color[rgb]{1,0.29411764705,0} $\blacktriangledown$} for $\gamma=p/2$ and {\color[rgb]{1,0.29411764705,0} $\blacktriangleleft$} for $\gamma=p/10$) and the surface code ({\color[rgb]{0.26666666666,0.66666666666,0.6} $\bullet$} for $\gamma=p$, {\color[rgb]{0.26666666666,0.66666666666,0.6} $\blacksquare$} for $\gamma=p/2$ and {\color[rgb]{0.26666666666,0.66666666666,0.6} $\blacklozenge$} for $\gamma=p/10$).
    The plots for the error model $\gamma=p$ are the same as shown in Fig.~\ref{fig:overhead_total}.
    The logical error rate is calculated under a circuit-level depolarizing error model at a physical error rate $0.1\%$. The dash-dotted and lines represent the logical error rate $10^{-10}$ and $10^{-24}$, respectively.
    Our protocol reduces the space overhead compared to the protocol for the surface code in a similar magnitude as in the error model used in Fig.~\ref{fig:overhead_total}.}
    \label{fig:overhead_total_all_error_models}
\end{figure}

The probabilities of the errors are given using a single parameter $p$ (called the physical error rate) as follows.
State preparations of $\ket{0}$ and $\ket{+}$ are followed by $X$ and $Z$ gates, respectively, with probability $p$.
Measurements in $Z$ and $X$ bases follow $X$ and $Z$ gates, respectively, with probability $p$.
One-qubit gates $I, X, Y, Z, H,S$ are followed by one of the $3$ possible non-identity Pauli operators $\{X, Y, Z\}$, each with probability $p/3$.
A two-qubit gate $\mathrm{CNOT}$ is followed by one of the $15$ possible non-identity Pauli products acting on $2$ qubits $\{\sigma_1\otimes \sigma_2\}_{(\sigma_1, \sigma_2)\in\{I,X,Y,Z\}^2}\setminus \{I\otimes I\}$, each with probability $p/15$.
We also investigate different error models, where the Pauli error rate on the identity gate $I$ is changed from $p/3$ to $\gamma/3$, where $\gamma$ is taken to be $\gamma=p/2$ and $\gamma=p/10$.
As shown in Fig.~\ref{fig:overhead_total_all_error_models}, our protocol shown in Table~\ref{tab:concatenation} significantly reduces the space overhead compared to that for the surface code in these error models (see Supplementary Information for more details).

\begin{figure}
    \centering
    \includegraphics[width=\linewidth]{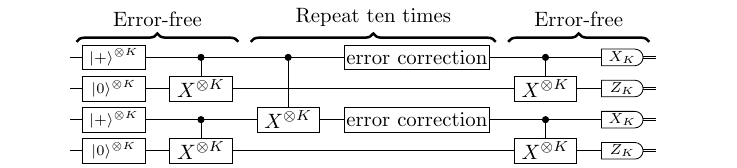}
    \caption{{\bf A quantum circuit for the reference entanglement method~\cite{goto2014step, goto2016minimizing,schumacher1996sending, knill2005quantum} to estimate a logical CNOT error rate.} In this simulation,  starting from two error-free logical Bell states, we apply a gate gadget of the logical $\mathrm{CNOT}^{\otimes K}$ gate followed by the error correction gadget ten times, using a noisy circuit.
For the surface code, we use the lattice surgery \cite{horsman2012surface, vuillot2019code} to implement logical CNOT gates, which includes the error correction.  For the other codes, we implement logical CNOT gates transversally and use Knill's error correction gadget \cite{knill2005quantum} for error correction.
Finally, we apply the error-free logical Bell measurement on the output quantum state to estimate the logical error rate. 
The symbol with $X_K$ ($Z_K$) denotes the measurements in $X$ ($Z$) basis for all the $K$ logical qubits in a code block.
}
    \label{fig:simulation_method}
\end{figure}

\noindent {\bf Simulation to evaluate logical CNOT error rates.}
In our numerical simulation, we evaluate the logical CNOT error rate using the Monte Carlo sampling method presented in Refs.~\cite{goto2014step, goto2016minimizing}, which is based on the reference entanglement method~\cite{schumacher1996sending, knill2005quantum}.
For a quantum code consisting of $N$ physical qubits and $K$ logical qubits,
the circuit that we use for the Monte Carlo sampling method is illustrated in Fig.~\ref{fig:simulation_method}, where we assume that random Pauli errors occur at each location of the circuit according to the error model described above.
In particular, starting from two error-free logical Bell states, we repeatedly apply a gate gadget of the logical $\mathrm{CNOT}^{\otimes K}$ gate followed by an error correction gadget, which is repeated ten times.
For all the quantum codes (which are Calderbank-Shor-Steane (CSS) codes in this work) except for the surface code, we implement the logical CNOT gates transversally and use Knill's error correction gadget \cite{knill2005quantum} for error correction. 
Note that Steane's error correction gadget~\cite{PhysRevLett.78.2252,gottesman2010introduction} may also have a similar performance to Knill's, but in practice, Knill's error correction gadget is more robust to leakage errors than Steane's~\cite{PhysRevA.71.042322} and thus is preferable to use if possible.
We also note that the addressable CNOT gate can be performed via gate teleportation using a certain stabilizer state \cite{glancy2006entanglement}, and the threshold of the distillation of a logical stabilizer state is much larger than that of the logical CNOT gate.
Therefore, we expect the threshold of the addressable CNOT gate is comparable with that of the transversal CNOT gate.
For the surface code, by convention, we use the lattice surgery~\cite{horsman2012surface, vuillot2019code} to implement the logical CNOT gates, which includes the error correction.
Note that the transversal implementation of the logical CNOT gate is also possible for the surface code \cite{sahay2024error}, but we performed our numerical simulation based on the lattice surgery since the lattice surgery is more widely used in the literature on resource estimation for FTQC, such as Refs.~\cite{PRXQuantum.2.030305,yoshioka2023hunting}, and the threshold of the lattice surgery CNOT is better than the transversal CNOT as shown in Ref.~\cite{sahay2024error}.
Then, we apply the error-free logical Bell measurement on the output quantum state.
Any measurement outcomes that do not result in all zeros for the $k$th logical qubits in four code blocks are counted as logical errors on the $k$th logical qubit for $k\in\{1, \ldots, K\}$.
We evaluate the logical CNOT error rate by dividing the empirical logical error probability in the simulation by ten and averaging over the $K$ logical qubits.
Since the quantum circuit in Fig.~\ref{fig:simulation_method}, including Pauli errors, is composed of Clifford gates, the sampling of measurement outcomes is efficiently simulated by a stabilizer circuit simulator; in particular, our simulation is conducted with \textsc{Stim}~\cite{gidney2021stim}.

\noindent {\bf Logical error rate required for 2048-bit RSA integer factoring.}
The security of the RSA cryptosystem is ensured by the classical hardness of integer factoring, and factoring 2048-bit integers given as the product of two similar-size prime numbers, which is called RSA integers in Ref.~\cite{gidney2021factor} leads to breaking RSA-2048.
Previous works have investigated efficient algorithms for RSA integer factoring based on Shor's algorithm~\cite{shor1994algorithms}.
In particular, Ref.~\cite{gidney2021factor} proposes an $n$-bit RSA integer factoring algorithm using $0.3 n^3+0.0005 n^3 \lg n$ Toffoli gates.
Since a Toffoli gate can be decomposed into $6$ CNOT gates and single-qubit gates~\cite{nielsen2010quantum}, this algorithm can be implemented by $1.8 n^3+0.003 n^3 \lg n$ CNOT gates.
For $n=2048$, it requires $\sim 10^{10}$ CNOT gates.
Thus, we require a logical error rate $\sim 10^{-10}$ to run this algorithm.

\noindent {\bf Required error rate for classical computation.}
The required error rate for classical computation is estimated by taking an inverse of the number of elementary gates in a large-scale classical computation that is currently available.
In particular, we consider a situation where the supercomputer Fugaku~\cite{fugaku} is run for a month.  The peak performance at double precision of Fugaku in the normal mode is given $488\;\mathrm{petaflops} \sim 5\times 10^{17} \;\mathrm{s}^{-1}$~\cite{fugaku}.
If we run it for $1\;\mathrm{month} \sim 2.6\times 10^6\;\mathrm{s}$, then the number of elementary gates is roughly estimated as $\sim 10^{24}$.
Thus, an upper bound of the logical error rate of classical computation is roughly estimated as $\sim 10^{-24}$.

\noindent {\bf Estimation of logical error rates of large-scale quantum codes from small-scale level-by-level simulations.}
In this work, we use an underlying quantum code $\mcQ_0$ concatenated with a series of quantum Hamming codes $\mcQ_{r_1},\mcQ_{r_2},\ldots,\mcQ_{r_L}$.  The logical error rate of the overall quantum code under the physical error rate $p$ is evaluated from the level-by-level numerical simulation as
\begin{align}
    P(p) = P_{r_{L}}^{(r_L+1)} \circ \cdots \circ P_{r_2}^{(r_3)} \circ P_{r_1}^{(r_2)} \circ P_0 (p),\label{eq:overall_logical_error}
\end{align}
where $P_0(p)$ is the logical error rate of $\mcQ_0$ under the physical error rate $p$, and $P_{r_l}^{(r_{l+1})}(p)$ is that of the quantum Hamming code $\mcQ_{r_l}$.
Note that the protocol depends on the quantum code concatenated above (i.e., $\mcQ_{r_{l+1}}$), thus the logical error rate of the level-$l$ protocol depends on $r_l$ and $r_{l+1}$ (see Ref.~\cite{yamasaki2024time} for the detail).
For the top code $\mcQ_{r_L}$, we implement the protocol such that we can concatenate the quantum code $\mcQ_{r_L+1}$ if needed.
Thus, the logical error rate on the top code $\mcQ_{r_L}$ is evaluated as $P_{r_{L}}^{(r_L+1)}$.
This estimation gives the upper bound of the logical error rate in the cases where the logical CNOT gates (rather than initial-state preparation of $\ket{0}$ and $\ket{+}$, single-qubit Pauli and Clifford gates, and measurements in $Z$ and $X$ bases) have the largest error rate in the set of elementary operations for the stabilizer circuits, which usually holds true since the gadget for the CNOT gate is the largest.
The logical error rates $P_0(p)$ and $P_{r_l}(p)$ for each $l\in\{1,\ldots,L\}$ are estimated by the numerical simulation using the circuit described in Fig.~\ref{fig:simulation_method}.
See Supplementary Information for more details.

With our numerical simulation, we obtain the parameters of the following fitting curves of the logical error rates (see Supplementary Information for more details).
For the quantum Hamming code $\mcQ_{r_l}$ with parameter $r_l$, due to distance $3$, $P_{r}(p)$ is approximated for $r_l\in\{3,4,5,6,7\}, r_{l+1}\in \{r_{l}+1, \ldots, \max(r_{l}+1, 7)\}$ by the following fitting curve
\begin{align}
    P_{r_l}^{(r_{l+1})}(p) = a_{r_l}^{(r_{l+1})} p^2.\label{eq:hamming_code_logical_error}
\end{align}
The logical error rate of the level-$l$ $C_4/C_6$ code, denoted by $P_{C_4/C_6}^{(l)}(p)$, is approximated by a fitting curve
\begin{align}
    P_{C_4/C_6}^{(l)}(p) = A_{C_4/C_6} (B_{C_4/C_6}p)^{F_l},\label{eq:c4c6_logical_error}
\end{align}
where $F_l$ is the Fibonacci number defined by $F_1=1$, $F_2=2$, and $F_{l}=F_{l-1}+F_{l-2}$ for $l>2$ \cite{knill2005quantum}.
The threshold $p^{(\mathrm{th})}_{C_4/C_6}$ for the $C_4/C_6$ code is estimated by 
\begin{align}
    p^{(\mathrm{th})}_{C_4/C_6} = (B_{C_4/C_6})^{-1}.
\end{align}
The logical error rate of the surface code with code distance $d$, denoted by $P_\mathrm{surface}^{(d)}(p)$, is approximated by a fitting curve
\begin{align}
    P_\mathrm{surface}^{(d)}(p) = A_\mathrm{surface}(B_\mathrm{surface} p)^{d+1\over 2}.\label{eq:surface_code_logical_error}
\end{align}
Based on the critical exponent method in Ref.~\cite{Wang_2003}, the threshold $p^{(\mathrm{th})}_\mathrm{surface}$ of the surface code is estimated as a fitting parameter of another fitting curve given by
\begin{align}
    P_\mathrm{surface}^{(d)\prime}(p) &= C_\mathrm{surface} + D_\mathrm{surface} x +E_\mathrm{surface} x^2,\\
    x &= (p-p^{(\mathrm{th})}_\mathrm{surface})d^{1/\mu}.
\end{align}
The logical error rate of the level-$l$ concatenated Steane code, denoted by $P_{\mathrm{Steane}}^{(l)}(p)$, is approximated for $l\in\{1,2\}$ by a fitting curve
\begin{align}
\label{eq:steane_code_logical_error_1_2}
    P_{\mathrm{Steane}}^{(l)}(p) &= a_\mathrm{Steane}^{(l)} p^{2^l}.
\end{align}
For $l\geq 3$, due to the limitation of computational resources, we did not directly perform the numerical simulation to determine $a_\mathrm{Steane}^{(l)}$ in~\eqref{eq:steane_code_logical_error_1_2}, but using the results for $l\in\{1,2\}$ in~\eqref{eq:steane_code_logical_error_1_2}, we recursively evaluate the logical error rates $P_{\mathrm{Steane}}^{(l)}(p)$ of level-$l$ concatenated Steane code as
\begin{align}
    P_{\mathrm{Steane}}^{(l)}(p)
    =\begin{cases}
        P_{\mathrm{Steane}}^{(1)}\circ P_{\mathrm{Steane}}^{(l-1)} (p) & (l \text{ is odd})\\
        P_{\mathrm{Steane}}^{(2)}\circ P_{\mathrm{Steane}}^{(l-2)} (p) & (l \text{ is even})
    \end{cases}.\label{eq:steane_code_logical_error}
\end{align}
The threshold $p^{(\mathrm{th})}_\mathrm{Steane}$ of the concatenated Steane code is estimated by that satisfying $P_\mathrm{Steane}^{(2)}(p^{(\mathrm{th})}_\mathrm{Steane}) = p^{(\mathrm{th})}_\mathrm{Steane}$, i.e.,
\begin{align}
    p^{(\mathrm{th})}_\mathrm{Steane} = [a_\mathrm{Steane}^{(2)}]^{-1/3}.
\end{align}
The logical error rates of the level-$l$ $C_4$/Steane codes for $l\in\{1, 2\}$, denoted by $P_{C_4/\mathrm{Steane}}^{(l)}(p)$, are approximated by fitting curves
\begin{align}
    P_{C_4/\mathrm{Steane}}^{(1)}(p) &= a_{C_4/\mathrm{Steane}}^{(1)} p,\\
    P_{C_4/\mathrm{Steane}}^{(2)}(p) &= a_{C_4/\mathrm{Steane}}^{(2)} p^3,\label{eq:c4steane_l2_logical_error}
\end{align}
where $a_{C_4/\mathrm{Steane}}^{(1)}$ is given by $a_{C_4/\mathrm{Steane}}^{(1)} = A_{C_4/C_6} B_{C_4/C_6}$ from the logical error rate of the level-1 $C_4/C_6$ since the level-1 $C_4$/Steane code coincides with the level-1 $C_4/C_6$ code.
For $l\geq 3$, similar to the concatenated Steane code, logical error rates $P_{C_4/\mathrm{Steane}}^{(l)}(p)$ of the level-$l$ $C_4$/Steane code are evaluated by
\begin{align}
    P_{C_4/\mathrm{Steane}}^{(l)}(p) = P_\mathrm{Steane}^{(l-2)}\circ P_{C_4/\mathrm{Steane}}^{(2)}(p).
    \label{eq:c4steane_logical_error}
\end{align}
Since the $C_4$/Steane code at concatenation levels $2$ and higher becomes the same as the concatenated Steane code, the threshold $p^{(\mathrm{th})}_{C_4/\mathrm{Steane}}$ of the $C_4$/Steane code is determined by the physical error rate that can be suppressed below $p^{(\mathrm{th})}_\mathrm{Steane}$ at level $2$, estimated as that satisfying $P_{C_4/\mathrm{Steane}}^{(2)}(p^{(\mathrm{th})}_{C_4/\mathrm{Steane}}) = p^{(\mathrm{th})}_\mathrm{Steane}$, i.e.,
\begin{align}
    p^{(\mathrm{th})}_\mathrm{Steane} = [a_\mathrm{Steane}^{(2)}]^{-1/9} [a_{C_4/\mathrm{Steane}}^{(2)}]^{-1/3}.
\end{align}
Using the fitting parameters of these fitting curves obtained from the level-by-level numerical simulations, we evaluate the overall logical error rate according to~\eqref{eq:overall_logical_error}.

\section*{Data Availability}

The datasets generated and/or analysed during the current study are available in the GitHub repository, \url{https://github.com/sy3104/concatenated_code_threshold}.

\section*{Code Availability}

The source codes for the simulation of the underlying codes (except for the surface code) and the quantum Hamming codes are available in GitHub and can be accessed via this link \url{https://github.com/sy3104/concatenated_code_threshold}.

\begin{acknowledgements}

S.Y.\ was supported by Japan Society for the Promotion of Science (JSPS) KAKENHI Grant Number 23KJ0734, FoPM, WINGS Program, the University of Tokyo, and DAIKIN Fellowship Program, the University of Tokyo.
S.T.\ was supported by JST [Moonshot R\&D][Grant Number JPMJMS2061], JSPS KAKENHI Grant Number 23KJ0521, and FoPM, WINGS Program, the University of Tokyo.
H.Y.\ was supported by JST PRESTO Grant Number JPMJPR201A, JPMJPR23FC, JSPS KAKENHI Grant Number JP23K19970, and MEXT Quantum Leap Flagship Program (MEXT QLEAP) JPMXS0118069605, JPMXS0120351339\@.
The quantum circuits shown in this paper are drawn using \textsc{qpic} \cite{qpic}.

\end{acknowledgements}

\section*{Author Contributions}

S.Y., S.T.\ and H.Y.\ contributed to the conception of the work, the analysis and interpretation in the work, and the preparation and revision of the manuscript.

\section*{Competing Interests}

All authors declare no financial or non-financial competing interests.

\clearpage

\onecolumngrid
\appendix

\renewcommand{\thetable}{S\arabic{table}}
\renewcommand{\thefigure}{S\arabic{figure}}
\setcounter{table}{0}
\setcounter{figure}{0}

\section*{Supplementary Information}

Supplementary Information of ``Concatenate codes, save qubits'' is organized as follows.
In Sec.~\ref{appendix_sec:protocol}, we present the details of the fault-tolerant protocols for the concatenated quantum Hamming code, the $C_4/C_6$ code, the surface code, the concatenated Steane code, and the $C_4$/Steane code.
In Sec.~\ref{appendix_sec:threshold}, we show the numerical results of the logical CNOT error rates for these quantum codes.
In Sec.~\ref{appendix_sec:other_error_models}, we compare the error threshold and the space overhead of the proposed protocol with respect to the surface code on different error models.

\section{Implementation of fault-tolerant protocols}
\label{appendix_sec:protocol}

\begin{figure}[t]
    \begin{minipage}[l]{0.05\linewidth}
    \begin{flushleft}
        (a)
    \end{flushleft}
    \end{minipage}
    \begin{minipage}[l]{0.9\linewidth}
    \begin{flushleft}
        \includegraphics[width=\linewidth]{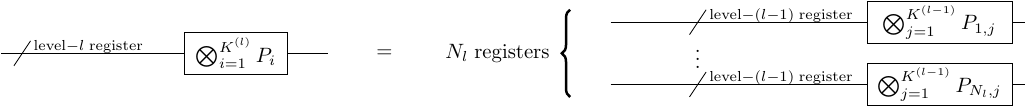}
    \end{flushleft}
    \end{minipage}\\
    \vspace{30pt}
    \begin{minipage}[l]{0.05\linewidth}
    \begin{flushleft}
        (b)
    \end{flushleft}
    \end{minipage}
    \begin{minipage}[l]{0.9\linewidth}
    \begin{flushleft}
        \includegraphics[width=\linewidth]{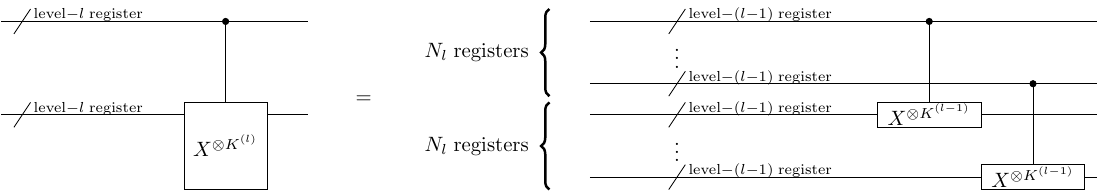}
    \end{flushleft}
    \end{minipage}\\
    \vspace{30pt}
    \begin{minipage}[l]{0.05\linewidth}
    \begin{flushleft}
        (c)
    \end{flushleft}
    \end{minipage}
    \begin{minipage}[l]{0.9\linewidth}
    \begin{flushleft}
        \includegraphics[width=\linewidth]{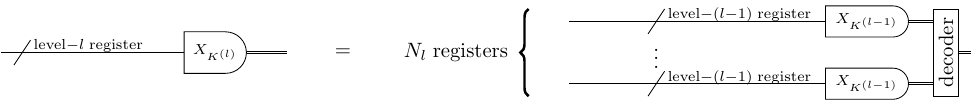}
        \\\;\\
        \includegraphics[width=\linewidth]{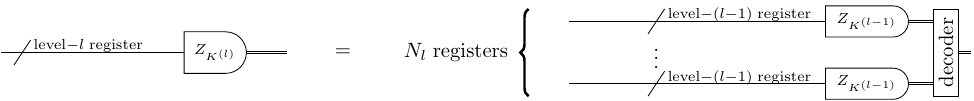}
    \end{flushleft}
    \end{minipage}
    
    \caption{Level-$l$ Pauli gate, CNOT gate, and measurement gadgets for the CSS codes using level-$(l-1)$ operations, which are used for the concatenated quantum Hamming codes, the $C_4/C_6$ code, the concatenated Steane code, and the $C_4$/Steane code in our analysis.  (a) The level-$l$ Pauli gate gadget implements the logical Pauli operator $\bigotimes_{i=1}^{K^{(l)}} P_{i}$ for $P_{i} \in \{I, X, Y, Z\}$.  It is implemented by the level-$(l-1)$ Pauli gadget as $\bigotimes_{n=1}^{N_l} \bigotimes_{j=1}^{K^{(l-1)}} P_{n,j}$, where $P_{n,j}$ is chosen from the logical Pauli operators $\{I, X, Y, Z\}$ of the level-$(l-1)$ code, which will be explained for each code in~\eqref{eq:logical_operator_hamming}, \eqref{eq:logical_operator_c4}, \eqref{eq:logical_operator_c4c6}, \eqref{eq:logical_operator_steane} and \eqref{eq:logical_operator_c4steane}.  (b) The level-$l$ CNOT gate gadget implements the logical $\mathrm{CNOT}^{\otimes K^{(l)}}$ gate. It is implemented by the $N_l$ transversal level-$(l-1)$ CNOT gate gadgets as shown on the right-hand side.  (c) The $X$ ($Z$) measurement gadget implements the measurement of the logical $X$ ($Z$) operator on the $i$th logical qubit for $i\in\{1, \ldots, K^{(l)}\}$. It is implemented by the transversal level-$(l-1)$ $X$ ($Z$) measurement gadgets, followed by the classical computation decoding the measurement outcomes of level-$(l-1)$ logical operators. See Sec.~\ref{subsec:decoder} for the details of the decoder.}
    \label{fig:css_cnot}
\end{figure}

In this section, we summarize the details of the implementation of fault-tolerant protocols for the quantum codes relevant to our analysis.
For a concatenated code, the set of logical qubits of the concatenated code at the concatenation level $l$ is called a level-$l$ register, where a level-$0$ register refers to a physical qubit~\cite{yamasaki2024time}.
For the concatenated quantum Hamming codes, the $C_4/C_6$ code, the concatenated Steane code, and the $C_4$/Steane code (which are the Calberback-Shor-Steane (CSS) codes),
the Pauli gate gadgets, the CNOT gate gadget, and the measurement gadget are implemented transversally as shown in Fig.~\ref{fig:css_cnot}.
To run the circuits for the simulation, the error correction gadget and the initial-state preparation gadget are also required, and we will describe these gadgets in this section.
In Sec.~\ref{subsec:hamming}, we describe the protocol for the concatenated quantum Hamming code.
In Sec.~\ref{subsec:underlying}, we describe the protocols for the underlying quantum codes, i.e., the $C_4/C_6$ code, the surface code, the concatenated Steane code, and the $C_4$/Steane code.
In addition, the measurement gadgets include the classical processing of decoding using the measurement outcomes, and in Sec.~\ref{subsec:decoder}, we describe the decoders.

\subsection{Concatenated quantum Hamming code}
\label{subsec:hamming}
\begin{figure}
    \centering
    \includegraphics[width=\linewidth]{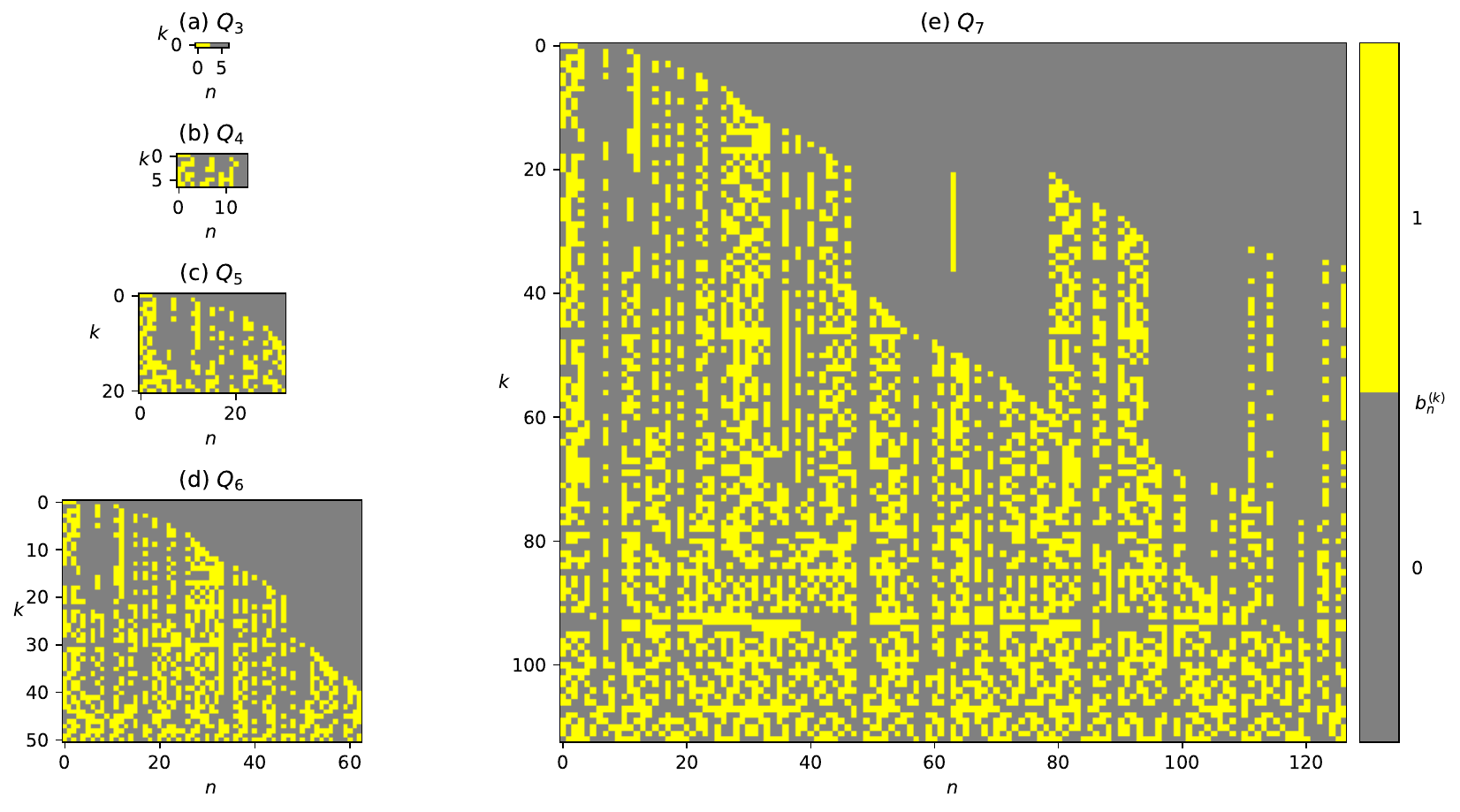}
    \caption{The explicit values of $b_n^{(k)}$ that define the logical Pauli operators in Eq.~\eqref{eq:logical_operator_hamming} for $\mcQ_3, \mcQ_4, \mcQ_5, \mcQ_6, \mcQ_7$.}
    \label{fig:hamming_logical_operator}
\end{figure}

\begin{figure}
    \centering
    \includegraphics[width=\linewidth]{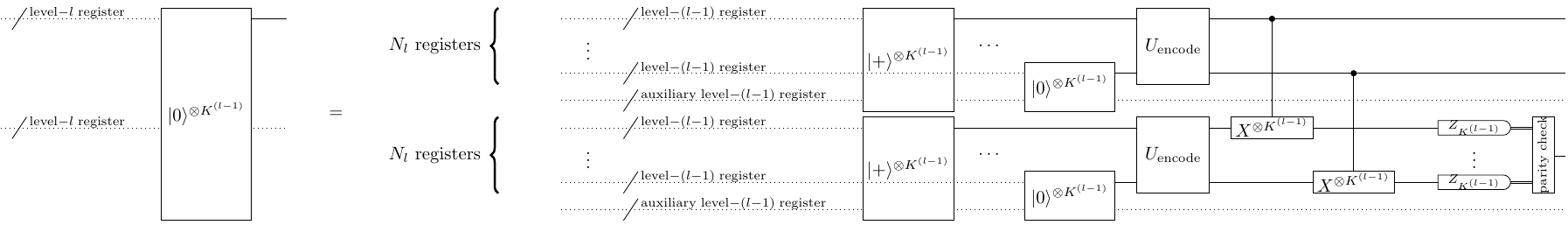}
    \\\;\\
    \includegraphics[width=\linewidth]{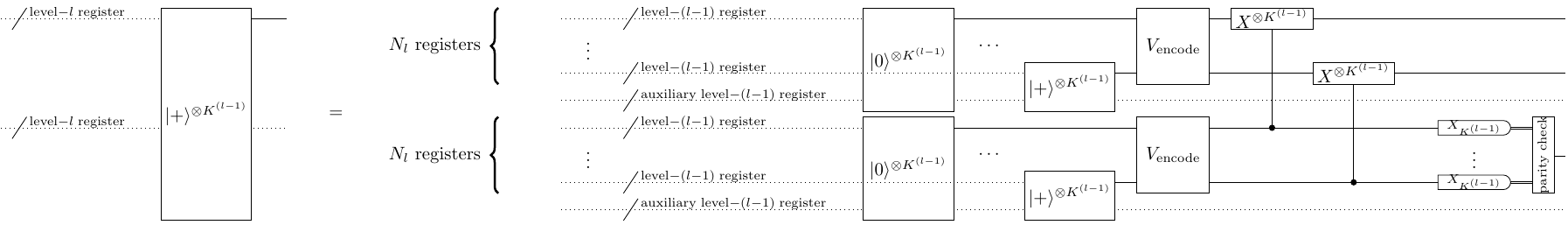}
    \caption{Level-$l$ initial-state preparation gadgets for the logical $\ket{0}$ ($\ket{+}$) state of the concatenated quantum Hamming code are implemented by using the level-$(l-1)$ gadgets.
    The  $Z$ ($X$) stabilizer generators and the logical $Z$ ($X$) operator are measured for verification from the measurement outcomes.
    If the verification fails, the output quantum state is discarded, and the initial-state preparation is rerun without additional verification.}
    \label{fig:hamming_zero_preparation}
\end{figure}

We summarize the details of the protocol for the concatenated quantum Hamming code.
A level-$l$ register refers to $K^{(l)} = \prod_{l'=1}^{l} K_{r_{l'}}$ logical qubits of the concatenated quantum Hamming code at the concatenation level $l\in\{1,2,\ldots\}$, as shown in Ref.~\cite{yamasaki2024time}.
To form a level-$l$ register, we use $N_{r_l}$ level-$(l-1)$ registers; in particular, from each of the $N_{r_l}$ level-$(l-1)$ registers, we pick up the $k$th qubit ($k\in\{1,\ldots,K^{(l-1)}\}$) and encode $K_{r_l}$ out of $K^{(l)}$ qubits of the level-$l$ register into these picked $N_{r_l}$ qubits as the $K_{r_l}$ logical qubits of the quantum Hamming code $\mcQ_{r_l}$.
The logical Pauli operators acting on the $i$th logical qubit of the level-$l$ register for $l\geq 2$, denoted by $P_i^{(1)}$ for $P\in\{I, X, Y, Z\}$, are written in terms of the level-$(l-1)$ logical Pauli operators acting on the $j$th logical qubit of the $n$th level-$(l-1)$ register, denoted by $P_{n,j}^{(l-1)}$ for $P\in\{I, X, Y, Z\}$, as
\begin{align}
\begin{split}
    X_{i}^{(l)} &= \bigotimes_{n=1}^{N_{r_l}}X_{n,j}^{(l-1) b_n^{(k)}},\\
    Z_{i}^{(l)} &= \bigotimes_{n=1}^{N_{r_l}}Z_{n,j}^{(l-1) b_n^{(k)}},
\end{split}\label{eq:logical_operator_hamming}
\end{align}
where $i=K^{(l-1)} (k-1) + j$ for $k\in\{1, \ldots, K_{r_l}\}$ and $j\in \{1, \ldots, K^{(l-1)}\}$, and $b_n^{(k)}$ represent the logical operators of the quantum Hamming code $\mcQ_{r_l}$.
The explicit forms of the logical operators, i.e., $b_n^{(k)}$ in~\eqref{eq:logical_operator_hamming}, can be determined by the method shown in Refs.~\cite{gottesman1997stabilizer,wilde2009logical}
(see Fig.~\ref{fig:hamming_logical_operator} for the values of $b_n^{(k)}$ for $\mcQ_3, \mcQ_4, \mcQ_5, \mcQ_6, \mcQ_7$).
The level-$l$ initial-state preparation gadget for the logical $\ket{0}$ ($\ket{+})$ of the concatenated quantum Hamming code is recursively defined using the level-$(l-1)$ gadgets as shown in Fig.~\ref{fig:hamming_zero_preparation}.
The  $Z$ ($X$) stabilizer generators and the logical $Z$ ($X$) operator are measured for verification from the measurement outcomes.
If the verification fails, the output quantum state is discarded, and the initial-state preparation is rerun without additional verification.
In our simulation, the leading-order effect of the verification failure is included in the estimation of the logical CNOT error rate as
\begin{align}
    P_L = P_\mathrm{CNOT}^{(0)} + P_{\mathrm{verification}} \sum_{i} P_{\mathrm{CNOT}}^{(i)},\label{eq:estimation_logical_error_rate}
\end{align}
where $P_\mathrm{CNOT}^{(0)}$ is the logical CNOT error rate evaluated in the post-selected simulation runs that all the verification succeed, $P_{\mathrm{verification}}$ is the failure probability of the verification, and $P_{\mathrm{CNOT}}^{(i)}$ is the logical CNOT error rate evaluated in the post-selected simulation runs that all the verifications but the $i$th one succeed.
We use the error correction gadget shown in Ref.~\cite{yamasaki2024time}.
See also Ref.~\cite{yamasaki2024time} for details of the full fault-tolerant protocol for implementing universal quantum computation using the quantum Hamming code while we have described here a part of the protocol relevant to our analysis.

Initial-state preparation unitaries $U_\mathrm{encode}$ for the quantum Hamming code $\mcQ_r = [[N_r, K_r, 3]]$ for $r\in\{3,4,5,6,7\}$ are constructed using Steane's Latin rectangle encoding method~\cite{steane2002fast}.
In the initial state preparation of the logical $\ket{0}$ state, the $2^{i-1}$th qubits for $i\in\{1, \ldots, r\}$ are initialized to be $\ket{+}$ states, and the $j$th qubits for $j\in\{1, \ldots, K_r\}\setminus \{2^0, \ldots, 2^{r-1}\}$ are initialized to be $\ket{0}$ states.
Steane's Latin rectangle $L$ for the quantum Hamming codes $[[N_r, K_r, 3]]$ is given by a $r\times N_r$ matrix whose elements $L_{i,j}$ for $i\in\{1, \ldots, r\}$ and $j\in\{1, \ldots, N_r\}$ specify the ordering of the CNOT gates to be applied.
If $L_{i,j} = l$ for $l\in\{1, \ldots\}$, a CNOT gate is applied between the $2^{i-1}$th qubit (control) and the $j$th qubit (target) on the depth $l$.
If $L_{i,j}=0$, no CNOT gate is applied.
The initial state preparation of the logical $\ket{+}$ state is done by replacing $\ket{0}$ ($\ket{+}$) with $\ket{+}$ ($\ket{0})$, swapping the control qubit and target qubit of the CNOT gates, and replacing the $Z$ measurements with the $X$ measurements in the initial state preparation of the logical $\ket{0}$ state.
In particular, we use the Latin rectangles $L_r$ for the $\mcQ_r$ codes for $r\in \{3,4,5,6,7\}$ given by
\begin{align}
    L_3 &= \left( \begin{matrix}
        0 & 0 & 2 & 0 & 1 & 0 & 0\\
        0 & 0 & 1 & 0 & 0 & 3 & 2\\
        0 & 0 & 0 & 0 & 2 & 1 & 0
    \end{matrix} \right),\\
    L_4 &= \left( \begin{matrix}
        0 & 0 & 3 & 0 & 7 & 0 & 1 & 0 & 4 & 0 & 5 & 0 & 2 & 0 & 6\\
        0 & 0 & 5 & 0 & 0 & 1 & 3 & 0 & 0 & 4 & 6 & 0 & 0 & 7 & 2\\
        0 & 0 & 0 & 0 & 2 & 7 & 6 & 0 & 0 & 0 & 0 & 1 & 4 & 3 & 5\\
        0 & 0 & 0 & 0 & 0 & 0 & 0 & 0 & 2 & 3 & 7 & 6 & 1 & 5 & 4
    \end{matrix} \right),\\
    L_5 &= \left(\begin{matrix}
        0 & 0 & 7 & 0 & 2 & 0 & 9 & 0 & 14 & 0 & 3 & 0 & 11 & 0 & 13 & 0 & 12 & 0 & 1 & 0 & 8 & 0 & 6 & 0 & 15 & 0 & 4 & 0 & 10 & 0 & 5\\ 
        0 & 0 & 8 & 0 & 0 & 9 & 15 & 0 & 0 & 13 & 11 & 0 & 0 & 5 & 14 & 0 & 0 & 2 & 12 & 0 & 0 & 1 & 10 & 0 & 0 & 6 & 7 & 0 & 0 & 4 & 3\\ 
        0 & 0 & 0 & 0 & 14 & 12 & 4 & 0 & 0 & 0 & 0 & 8 & 3 & 1 & 10 & 0 & 0 & 0 & 0 & 2 & 13 & 6 & 9 & 0 & 0 & 0 & 0 & 11 & 5 & 7 & 15\\ 
        0 & 0 & 0 & 0 & 0 & 0 & 0 & 0 & 1 & 10 & 7 & 5 & 4 & 6 & 15 & 0 & 0 & 0 & 0 & 0 & 0 & 0 & 0 & 3 & 12 & 14 & 8 & 9 & 13 & 2 & 11\\ 
        0 & 0 & 0 & 0 & 0 & 0 & 0 & 0 & 0 & 0 & 0 & 0 & 0 & 0 & 0 & 0 & 15 & 1 & 13 & 7 & 11 & 4 & 12 & 8 & 5 & 3 & 9 & 6 & 2 & 14 & 10
    \end{matrix}\right),\\
    L_6 &= \left(\begin{matrix}L_{6,1} & L_{6,2} & L_{6,3}\end{matrix}\right),\\
    L_7 &= \left(\begin{matrix}L_{7,1} & L_{7,2} & L_{7,3} & L_{7,4} & L_{7,5} \end{matrix}\right),
\end{align}
where $L_6$ and $L_7$ are given by horizontally concatenating the matrices defined as
\begin{align}
    L_{6,1} &=\left(\begin{matrix}
0 & 0 & 13 & 0 & 24 & 0 & 5 & 0 & 6 & 0 & 3 & 0 & 7 & 0 & 10 & 0 & 27 & 0 & 28 & 0 & 20\\ 
0 & 0 & 15 & 0 & 0 & 28 & 13 & 0 & 0 & 19 & 25 & 0 & 0 & 26 & 16 & 0 & 0 & 12 & 30 & 0 & 0\\ 
0 & 0 & 0 & 0 & 15 & 9 & 26 & 0 & 0 & 0 & 0 & 24 & 3 & 17 & 25 & 0 & 0 & 0 & 0 & 6 & 8\\ 
0 & 0 & 0 & 0 & 0 & 0 & 0 & 0 & 16 & 14 & 12 & 27 & 19 & 5 & 20 & 0 & 0 & 0 & 0 & 0 & 0\\ 
0 & 0 & 0 & 0 & 0 & 0 & 0 & 0 & 0 & 0 & 0 & 0 & 0 & 0 & 0 & 0 & 4 & 8 & 18 & 23 & 11\\ 
0 & 0 & 0 & 0 & 0 & 0 & 0 & 0 & 0 & 0 & 0 & 0 & 0 & 0 & 0 & 0 & 0 & 0 & 0 & 0 & 0\\ 
    \end{matrix}\right),\\
    L_{6,2} &=\left(\begin{matrix}
0 & 17 & 0 & 16 & 0 & 9 & 0 & 26 & 0 & 22 & 0 & 18 & 0 & 23 & 0 & 1 & 0 & 2 & 0 & 4 & 0\\ 
8 & 3 & 0 & 0 & 9 & 14 & 0 & 0 & 23 & 2 & 0 & 0 & 24 & 20 & 0 & 0 & 29 & 7 & 0 & 0 & 6\\ 
27 & 4 & 0 & 0 & 0 & 0 & 30 & 19 & 10 & 29 & 0 & 0 & 0 & 0 & 2 & 5 & 14 & 31 & 0 & 0 & 0\\ 
0 & 0 & 11 & 18 & 21 & 10 & 1 & 28 & 15 & 30 & 0 & 0 & 0 & 0 & 0 & 0 & 0 & 0 & 8 & 6 & 3\\ 
19 & 27 & 2 & 21 & 13 & 5 & 15 & 31 & 14 & 17 & 0 & 0 & 0 & 0 & 0 & 0 & 0 & 0 & 0 & 0 & 0\\ 
0 & 0 & 0 & 0 & 0 & 0 & 0 & 0 & 0 & 0 & 0 & 23 & 6 & 3 & 22 & 4 & 20 & 10 & 19 & 9 & 13\\ 
    \end{matrix}\right),\\
    L_{6,3}&=\left(\begin{matrix}
29 & 0 & 19 & 0 & 31 & 0 & 11 & 0 & 8 & 0 & 25 & 0 & 12 & 0 & 14 & 0 & 30 & 0 & 21 & 0 & 15\\ 
22 & 0 & 0 & 17 & 21 & 0 & 0 & 4 & 31 & 0 & 0 & 5 & 11 & 0 & 0 & 27 & 10 & 0 & 0 & 18 & 1\\ 
0 & 20 & 7 & 23 & 11 & 0 & 0 & 0 & 0 & 13 & 21 & 28 & 22 & 0 & 0 & 0 & 0 & 18 & 16 & 1 & 12\\ 
2 & 29 & 13 & 7 & 9 & 0 & 0 & 0 & 0 & 0 & 0 & 0 & 0 & 26 & 22 & 23 & 24 & 31 & 17 & 4 & 25\\ 
0 & 0 & 0 & 0 & 0 & 24 & 1 & 30 & 6 & 16 & 28 & 29 & 20 & 22 & 10 & 9 & 3 & 25 & 7 & 12 & 26\\ 
18 & 17 & 14 & 25 & 7 & 15 & 24 & 27 & 16 & 12 & 29 & 8 & 1 & 2 & 5 & 30 & 21 & 28 & 26 & 11 & 31\\ 
    \end{matrix}\right),\\
    L_{7,1}&=\left(\begin{matrix}
0 & 0 & 19 & 0 & 1 & 0 & 21 & 0 & 54 & 0 & 10 & 0 & 55 & 0 & 45 & 0 & 61 & 0 & 47 & 0 & 33 & 0 & 51 & 0 & 2 & 0\\ 
0 & 0 & 17 & 0 & 0 & 50 & 59 & 0 & 0 & 41 & 38 & 0 & 0 & 28 & 51 & 0 & 0 & 39 & 23 & 0 & 0 & 54 & 4 & 0 & 0 & 42\\ 
0 & 0 & 0 & 0 & 59 & 18 & 33 & 0 & 0 & 0 & 0 & 61 & 40 & 41 & 37 & 0 & 0 & 0 & 0 & 24 & 63 & 29 & 31 & 0 & 0 & 0\\ 
0 & 0 & 0 & 0 & 0 & 0 & 0 & 0 & 35 & 23 & 24 & 15 & 26 & 14 & 33 & 0 & 0 & 0 & 0 & 0 & 0 & 0 & 0 & 12 & 4 & 57\\ 
0 & 0 & 0 & 0 & 0 & 0 & 0 & 0 & 0 & 0 & 0 & 0 & 0 & 0 & 0 & 0 & 15 & 32 & 21 & 47 & 13 & 33 & 41 & 27 & 60 & 39\\ 
0 & 0 & 0 & 0 & 0 & 0 & 0 & 0 & 0 & 0 & 0 & 0 & 0 & 0 & 0 & 0 & 0 & 0 & 0 & 0 & 0 & 0 & 0 & 0 & 0 & 0\\ 
0 & 0 & 0 & 0 & 0 & 0 & 0 & 0 & 0 & 0 & 0 & 0 & 0 & 0 & 0 & 0 & 0 & 0 & 0 & 0 & 0 & 0 & 0 & 0 & 0 & 0\\ 
    \end{matrix}\right),\\
    L_{7,2}&=\left(\begin{matrix}
59 & 0 & 31 & 0 & 17 & 0 & 24 & 0 & 63 & 0 & 13 & 0 & 29 & 0 & 53 & 0 & 30 & 0 & 16 & 0 & 27 & 0 & 22 & 0 & 34 & 0\\ 
26 & 0 & 0 & 21 & 19 & 0 & 0 & 36 & 62 & 0 & 0 & 46 & 34 & 0 & 0 & 16 & 29 & 0 & 0 & 45 & 57 & 0 & 0 & 27 & 61 & 0\\ 
0 & 21 & 46 & 17 & 6 & 0 & 0 & 0 & 0 & 47 & 28 & 50 & 3 & 0 & 0 & 0 & 0 & 4 & 1 & 11 & 43 & 0 & 0 & 0 & 0 & 56\\ 
5 & 47 & 40 & 16 & 18 & 0 & 0 & 0 & 0 & 0 & 0 & 0 & 0 & 20 & 10 & 17 & 28 & 54 & 2 & 32 & 49 & 0 & 0 & 0 & 0 & 0\\ 
45 & 44 & 50 & 51 & 26 & 0 & 0 & 0 & 0 & 0 & 0 & 0 & 0 & 0 & 0 & 0 & 0 & 0 & 0 & 0 & 0 & 38 & 9 & 55 & 62 & 40\\ 
0 & 0 & 0 & 0 & 0 & 0 & 23 & 62 & 9 & 26 & 53 & 60 & 14 & 46 & 19 & 51 & 58 & 56 & 47 & 35 & 3 & 11 & 34 & 13 & 32 & 54\\ 
0 & 0 & 0 & 0 & 0 & 0 & 0 & 0 & 0 & 0 & 0 & 0 & 0 & 0 & 0 & 0 & 0 & 0 & 0 & 0 & 0 & 0 & 0 & 0 & 0 & 0\\ 
    \end{matrix}\right),\\
    L_{7,3}&=\left(\begin{matrix}
37 & 0 & 35 & 0 & 26 & 0 & 42 & 0 & 23 & 0 & 3 & 0 & 41 & 0 & 40 & 0 & 32 & 0 & 11 & 0 & 20 & 0 & 44 & 0 & 50 & 0\\ 
0 & 52 & 6 & 0 & 0 & 58 & 12 & 0 & 0 & 7 & 55 & 0 & 0 & 18 & 56 & 0 & 0 & 49 & 47 & 0 & 0 & 31 & 14 & 0 & 0 & 8\\ 
8 & 19 & 36 & 0 & 0 & 0 & 0 & 13 & 32 & 27 & 51 & 0 & 0 & 0 & 0 & 34 & 30 & 53 & 14 & 0 & 0 & 0 & 0 & 57 & 42 & 10\\ 
0 & 0 & 0 & 22 & 21 & 63 & 13 & 58 & 60 & 42 & 43 & 0 & 0 & 0 & 0 & 0 & 0 & 0 & 0 & 8 & 29 & 53 & 7 & 56 & 3 & 1\\ 
4 & 11 & 7 & 48 & 37 & 31 & 58 & 34 & 17 & 22 & 56 & 0 & 0 & 0 & 0 & 0 & 0 & 0 & 0 & 0 & 0 & 0 & 0 & 0 & 0 & 0\\ 
16 & 41 & 59 & 31 & 27 & 5 & 15 & 2 & 4 & 50 & 8 & 0 & 0 & 0 & 0 & 0 & 0 & 0 & 0 & 0 & 0 & 0 & 0 & 0 & 0 & 0\\ 
0 & 0 & 0 & 0 & 0 & 0 & 0 & 0 & 0 & 0 & 0 & 0 & 56 & 47 & 4 & 1 & 59 & 16 & 6 & 46 & 49 & 2 & 30 & 23 & 21 & 17\\ 
    \end{matrix}\right),\\
    L_{7,4}&=\left(\begin{matrix}
7 & 0 & 57 & 0 & 25 & 0 & 39 & 0 & 43 & 0 & 6 & 0 & 56 & 0 & 46 & 0 & 15 & 0 & 36 & 0 & 9 & 0 & 60 & 0 & 52 & 0\\ 
9 & 0 & 0 & 2 & 15 & 0 & 0 & 40 & 22 & 0 & 0 & 43 & 25 & 0 & 0 & 24 & 5 & 0 & 0 & 11 & 48 & 0 & 0 & 20 & 13 & 0\\ 
58 & 0 & 0 & 0 & 0 & 39 & 60 & 15 & 26 & 0 & 0 & 0 & 0 & 52 & 38 & 7 & 54 & 0 & 0 & 0 & 0 & 12 & 62 & 9 & 45 & 0\\ 
34 & 0 & 0 & 0 & 0 & 0 & 0 & 0 & 0 & 61 & 55 & 31 & 51 & 59 & 6 & 52 & 41 & 0 & 0 & 0 & 0 & 0 & 0 & 0 & 0 & 48\\ 
0 & 52 & 25 & 42 & 6 & 36 & 8 & 2 & 28 & 20 & 24 & 1 & 35 & 5 & 16 & 49 & 30 & 0 & 0 & 0 & 0 & 0 & 0 & 0 & 0 & 0\\ 
0 & 0 & 0 & 0 & 0 & 0 & 0 & 0 & 0 & 0 & 0 & 0 & 0 & 0 & 0 & 0 & 0 & 44 & 6 & 7 & 29 & 17 & 55 & 52 & 21 & 1\\ 
51 & 28 & 55 & 38 & 24 & 54 & 53 & 14 & 20 & 48 & 36 & 42 & 18 & 27 & 50 & 61 & 29 & 22 & 19 & 9 & 10 & 5 & 15 & 13 & 7 & 31\\ 
    \end{matrix}\right),\\
    L_{7,5}&=\left(\begin{matrix}
62 & 0 & 4 & 0 & 58 & 0 & 49 & 0 & 18 & 0 & 12 & 0 & 38 & 0 & 14 & 0 & 5 & 0 & 28 & 0 & 48 & 0 & 8\\ 
0 & 37 & 63 & 0 & 0 & 30 & 10 & 0 & 0 & 32 & 53 & 0 & 0 & 1 & 3 & 0 & 0 & 44 & 33 & 0 & 0 & 60 & 35\\ 
0 & 0 & 0 & 55 & 44 & 5 & 23 & 0 & 0 & 0 & 0 & 49 & 16 & 20 & 22 & 0 & 0 & 0 & 0 & 48 & 35 & 2 & 25\\ 
9 & 45 & 27 & 19 & 37 & 36 & 11 & 0 & 0 & 0 & 0 & 0 & 0 & 0 & 0 & 30 & 38 & 25 & 62 & 50 & 46 & 39 & 44\\ 
0 & 0 & 0 & 0 & 0 & 0 & 0 & 54 & 57 & 43 & 29 & 3 & 63 & 12 & 46 & 19 & 14 & 23 & 10 & 53 & 59 & 61 & 18\\ 
20 & 28 & 30 & 39 & 24 & 18 & 42 & 25 & 40 & 57 & 37 & 33 & 22 & 61 & 38 & 36 & 49 & 10 & 63 & 12 & 43 & 45 & 48\\ 
63 & 52 & 26 & 12 & 45 & 35 & 37 & 41 & 8 & 39 & 33 & 32 & 62 & 25 & 60 & 34 & 44 & 3 & 57 & 11 & 40 & 58 & 43\\ 
    \end{matrix}\right).
\end{align}
The Latin rectangle for the $[[7,1,3]]$ code is taken from Ref.~\cite{paetznick2011fault}, and the others are heuristically chosen to minimize the circuit depth as much as possible.

\subsection{Underlying quantum codes}
\label{subsec:underlying}
In this section, we describe the protocols for the underlying quantum codes, i.e., the $C_4/C_6$ code, the surface code, the concatenated Steane code, and the $C_4$/Steane code.
In Sec.~\ref{appendix:c4c6}, we describe the $C_4/C_6$ code.
In Sec.~\ref{appendix:surface}, we describe the surface code.
In Sec.~\ref{appendix:steane}, we describe the concatenated Steane code.
In Sec.~\ref{appendix:c4steane}, we describe the $C_4$/Steane code.

\subsubsection{$C_4/C_6$ code}
\label{appendix:c4c6}
We summarize the details of the protocol for the $C_4/C_6$ code.
We call the two logical qubits of the $C_4$ code (i.e., the $[[4,2,2]]$ code) a level-$1$ register.
Similarly, the level-$l$ register for $l\in \{2, 3, \ldots \}$ refers to the two logical qubits of the $C_4/C_6$ code at the concatenation level $l$.
To form the level-$l$ register, the $C_4/C_6$ code uses three level-$(l-1)$ registers (i.e., six qubits) of the level-$(l-1)$ code to encode the level-$l$ register as the logical qubits of the $C_6$ code, as shown in Ref.~\cite{knill2005quantum}.
The logical Pauli operators acting on the $i$th logical qubit of the level-$1$ register, denoted by $P_i^{(1)}$ for $P\in\{I, X, Y, Z\}$, are given by the physical Pauli operators as \cite{knill2005quantum}
\begin{align}
\begin{split}
    X_1^{(l)} \otimes I_2^{(l)} &=
        X\otimes X\otimes I\otimes I,\\
    Z_1^{(l)} \otimes I_2^{(l)}
    &=
        Z\otimes I\otimes Z\otimes I,\\
    I_1^{(l)} \otimes X_2^{(l)}
    &=
        I\otimes X\otimes I\otimes X,\\
    I_1^{(l)} \otimes Z_2^{(l)}
    &=
        I\otimes I\otimes Z\otimes Z.
\end{split}\label{eq:logical_operator_c4}
\end{align}
The logical Pauli operators acting on the $i$th logical qubit of the level-$l$ register for $l\geq 2$, denoted by $P_i^{(1)}$ for $P\in\{I, X, Y, Z\}$, are given by the level-$(l-1)$ logical Pauli operators acting on the $j$th logical qubit of the $n$th level-$(l-1)$ register, denoted by $P_{n,j}^{(l-1)}$ for $P\in\{I, X, Y, Z\}$, as \cite{knill2005quantum}
\begin{align}
\begin{split}
    X_1^{(l)} \otimes I_2^{(l)}
    &=
        I_{1,1}^{(l-1)}\otimes X_{1,2}^{(l-1)} \otimes X_{2,1}^{(l-1)} \otimes I_{2,2}^{(l-1)}\otimes I_{3,1}^{(l-1)}\otimes I_{3,2}^{(l-1)},\\
    Z_1^{(l)} \otimes I_2^{(l)}
    &=
        I_{1,1}^{(l-1)}\otimes I_{1,2}^{(l-1)} \otimes Z_{2,1}^{(l-1)} \otimes Z_{2,2}^{(l-1)}\otimes I_{3,1}^{(l-1)}\otimes Z_{3,2}^{(l-1)},\\
    I_1^{(l)} \otimes X_2^{(l)}
    &=
        X_{1,1}^{(l-1)}\otimes I_{1,2}^{(l-1)} \otimes X_{2,1}^{(l-1)} \otimes X_{2,2}^{(l-1)}\otimes I_{3,1}^{(l-1)}\otimes I_{3,2}^{(l-1)},\\
    I_1^{(l)} \otimes Z_2^{(l)}
    &=
        I_{1,1}^{(l-1)}\otimes I_{1,2}^{(l-1)} \otimes I_{2,1}^{(l-1)} \otimes Z_{2,2}^{(l-1)}\otimes Z_{3,1}^{(l-1)}\otimes I_{3,2}^{(l-1)}.
\end{split}\label{eq:logical_operator_c4c6}
\end{align}

\begin{figure}
    \centering
    \includegraphics[width=\linewidth]{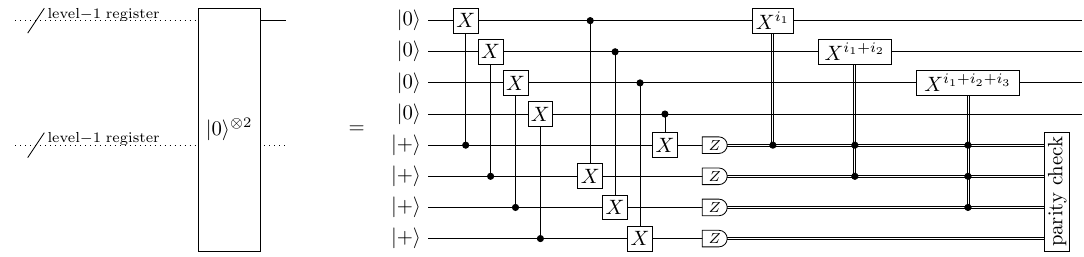}
    \\\;\\
    \includegraphics[width=\linewidth]{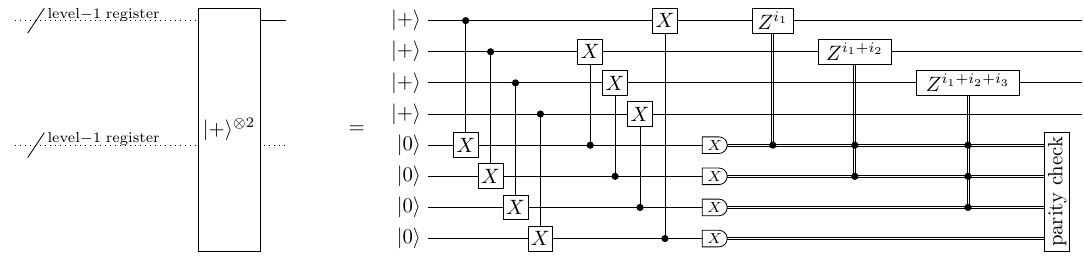}
    \caption{Level-1 initial-state preparation gadgets for the $C_4/C_6$ code and the $C_4$/Steane code implements preparations of the logical $\ket{0}^{\otimes 2}$ and $\ket{+}^{\otimes 2}$ states. Pauli gates are applied depending on the measurement outcomes $i_j$ of the $(j+4)$th qubits for $j\in\{1, \ldots, 4\}$.  The parity of the measurement outcomes is checked for verification. If $i_1+i_2+i_3+i_4 \neq 0\;(\mathrm{mod}\;2)$ holds, the output quantum state is discarded and the initial-state preparation is rerun.}
    \label{fig:c4_zero_preparation}
\end{figure}

\begin{figure}
    \centering
    \includegraphics[width=\linewidth]{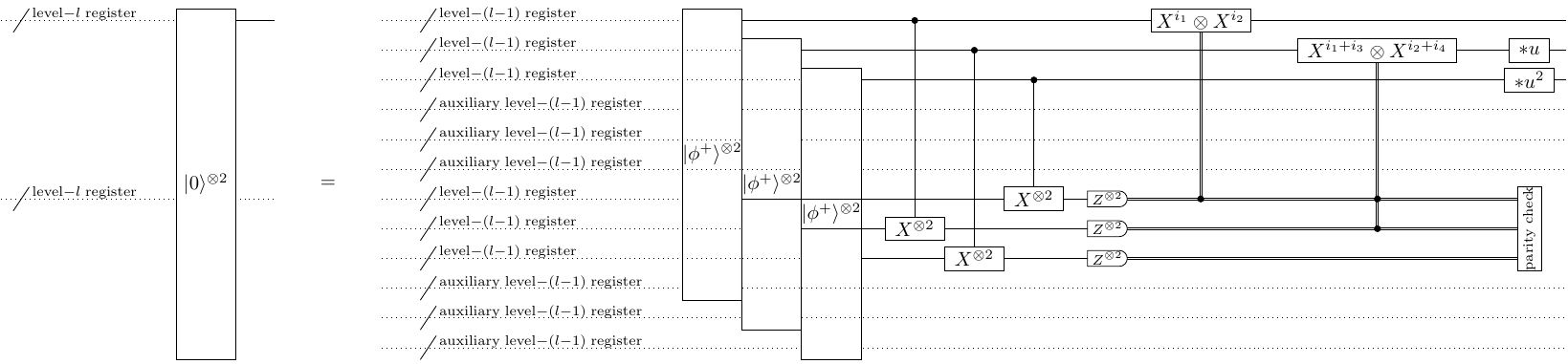}
    \\\;\\
    \includegraphics[width=\linewidth]{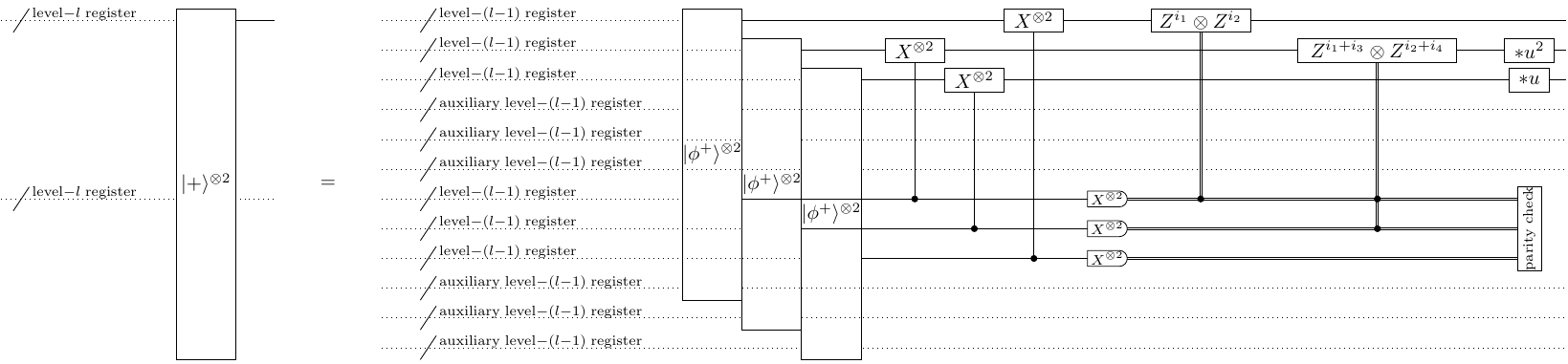}
    \caption{Level-$l$ initial-state preparation gadgets ($l\geq 2$) for the $C_4/C_6$ code implements preparations of the logical $\ket{0}^{\otimes 2}$ and $\ket{+}^{\otimes 2}$ states, implemented by using level-$(l-1)$ gadgets.  Pauli gates are applied depending on the measurement outcomes $(i_{2j-1}, i_{2j})$ of the $(j+6)$th code block for $j\in\{1, \ldots, 3\}$.  The parity of the measurement outcomes is checked for verification. If $i_1+i_3+i_5 \neq 0\;(\mathrm{mod}\;2)$ or $i_2+i_4+i_6 \neq 0\;(\mathrm{mod}\;2)$ hold, the output quantum state is discarded and the initial-state preparation is rerun.}
    \label{fig:c6_zero_preparation}
\end{figure}

\begin{figure}
    \centering
    \begin{minipage}[l]{0.05\linewidth}
    \begin{flushleft}
        (a)
    \end{flushleft}
    \end{minipage}
    \begin{minipage}[l]{0.9\linewidth}
    \begin{flushleft}
        \includegraphics{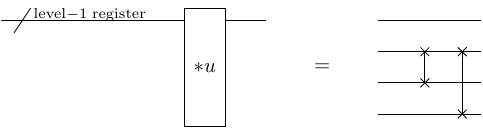}
        \\\;\\
        \includegraphics{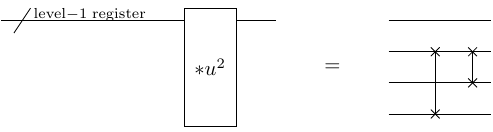}
    \end{flushleft}
    \end{minipage}\\
    \vspace{30pt}
    \begin{minipage}[l]{0.05\linewidth}
    \begin{flushleft}
        (b)
    \end{flushleft}
    \end{minipage}
    \begin{minipage}[l]{0.9\linewidth}
    \begin{flushleft}
        \includegraphics{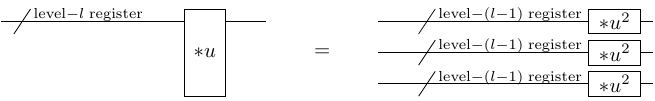}
        \\\;\\
        \includegraphics{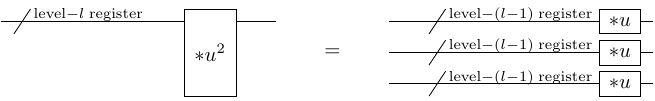}
    \end{flushleft}
    \end{minipage}
    \caption{(a) Level-$1$ $\ast u$ and $\ast u^2$ gate gadgets for the $C_4/C_6$ code implement the logical $\ast u$ and $\ast u^2$ operations given in~\eqref{eq:uu2}. (b) Level-$l$ ($\l\geq 2$) $\ast u$ and $\ast u^2$ gate gadgets for the $C_4/C_6$ code are implemented by using the level-$(l-1)$ gadgets.}
    \label{fig:c6_u}
\end{figure}

\begin{figure}
    \centering
    \begin{minipage}[l]{0.05\linewidth}
    \begin{flushleft}
        (a)
    \end{flushleft}
    \end{minipage}
    \begin{minipage}[l]{0.9\linewidth}
    \begin{flushleft}
        \includegraphics[width=\linewidth]{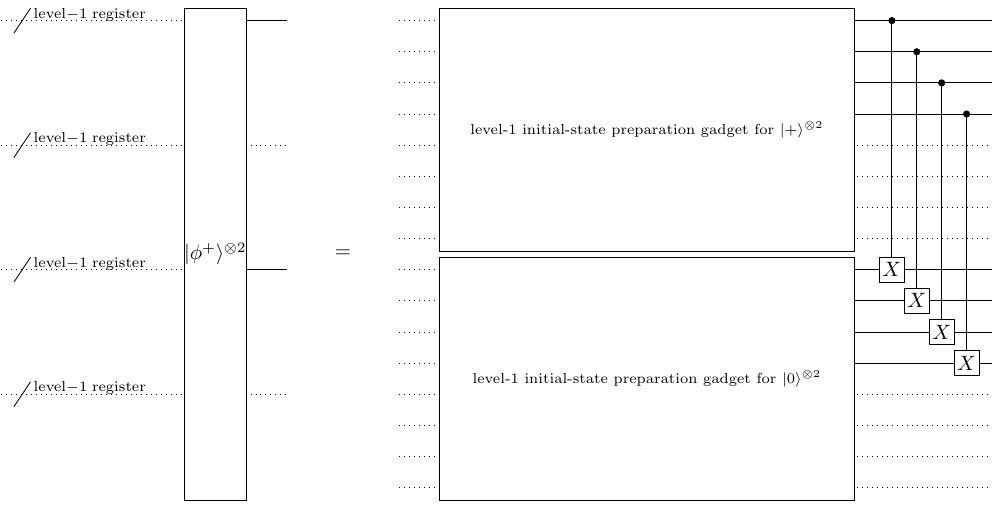}
    \end{flushleft}
    \end{minipage}\\
    \vspace{30pt}
    \begin{minipage}[l]{0.05\linewidth}
    \begin{flushleft}
        (b)
    \end{flushleft}
    \end{minipage}
    \begin{minipage}[l]{0.9\linewidth}
    \begin{flushleft}
        \includegraphics[width=\linewidth]{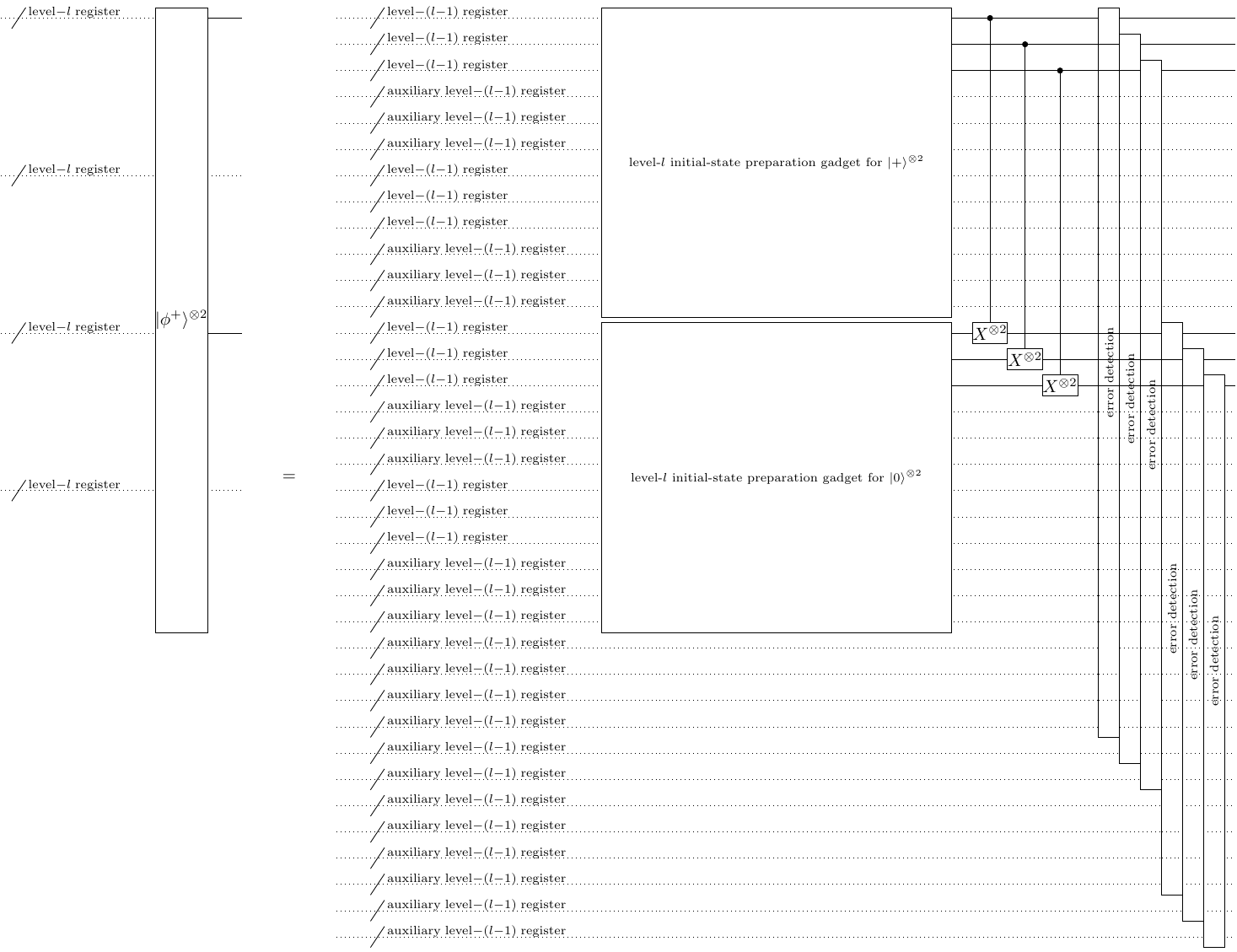}
    \end{flushleft}
    \end{minipage}
    \caption{(a) A level-$1$ Bell-state preparation gadget for the $C_4/C_6$ code implements preparation of the logical Bell state $\ket{\phi^+}^{\otimes 2} = \left({\ket{00}+\ket{11} \over \sqrt{2}}\right)^{\otimes 2}$. (b) A level-$l$ ($l\geq 2$) Bell-state preparation gadget for the $C_4/C_6$ code is implemented by using the level-$(l-1)$ gadgets.}
    \label{fig:c4c6_bell_preparation}
\end{figure}

\begin{figure}
    \centering
    \includegraphics[width=\linewidth]{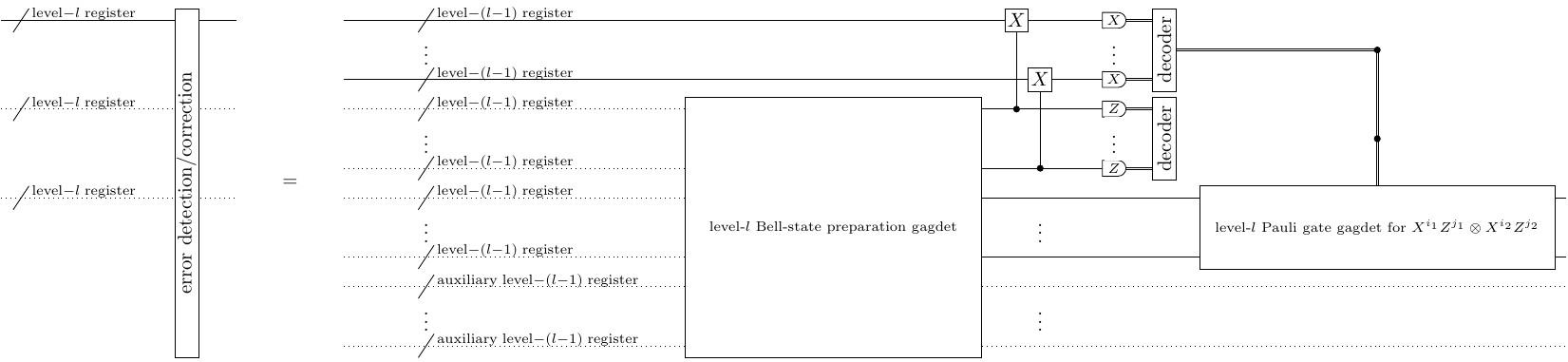}
    \caption{Level-$l$ error detection and error correction gadgets for the $C_4/C_6$ code using the level-$(l-1)$ gadgets.  }
    \label{fig:c4c6_ec}
\end{figure}

Level-$l$ initial-state preparation gadgets of the $C_4/C_6$ code are recursively defined using level-$(l-1)$ gadgets as shown in Figs.~\ref{fig:c4_zero_preparation} and \ref{fig:c6_zero_preparation}.
The initial-state preparation gadget uses $\ast u$ and $\ast u^2$ gate gadgets~\cite{knill2005quantum}, which are shown in Fig.~\ref{fig:c6_u}, implementing the logical $2$-qubit unitary operations given by
\begin{align}
\begin{split}
    \ast u &= \mathrm{CNOT} \cdot \mathrm{SWAP},\\
    \ast u^2 &= \mathrm{SWAP} \cdot \mathrm{CNOT},
\end{split}
\label{eq:uu2}
\end{align}
where $\mathrm{SWAP}$ is defined by
\begin{align}
    \mathrm{SWAP}
    =
    \left(\begin{matrix}
        1 & 0 & 0 & 0\\
        0 & 0 & 1 & 0\\
        0 & 1 & 0 & 0\\
        0 & 0 & 0 & 1
    \end{matrix}\right).
\end{align}
The parity of the measurement outcomes is checked for verification. If it fails, the output quantum state is discarded, and the initial-state preparation gadget is rerun.
Using the Bell-state preparation gadget shown in Fig.~\ref{fig:c4c6_bell_preparation} \cite{goto2013fault}, we implement Knill's error correction gadget as shown in Fig.~\ref{fig:c4c6_ec}.
In the error correction and detection gadgets, measurement outcomes of $X$ and $Z$ measurements are decoded to apply logical Pauli gates for correcting byproducts.
In the error correction gadget, if an uncorrectable error is detected in the decoding process, random numbers are assigned to the logical measurement outcomes.
In the error detection gadget, if an uncorrectable error is detected in the decoding process, the output quantum state is discarded, which incurs an erasure error.
In the Bell-state preparation gadget, an error detection gadget is applied after preparing the logical Bell state.
If an uncorrectable error is detected in the error detection, the output quantum state is discarded and the Bell-preparation gadget is rerun by replacing the error detection gadget with an error correction gadget.
Since the effect of the verification failure on the logical CNOT error rate is in a sub-leading order, we omit to include this effect in the numerical simulation.
See also Ref.~\cite{knill2005quantum} for details of the full fault-tolerant protocol for implementing universal quantum computation using the $C_4/C_6$ code while we have described here a part of the protocol relevant to our analysis.
Note that the protocol described here is the non-post-selected protocol while Ref.~\cite{knill2005quantum} also proposes a post-selected protocol, which we do not use to avoid the increase of overhead.
See also Refs.~\cite{4031377,DBLP:journals/qic/AliferisGP08,reichardt2009error} regarding the theoretical analysis of fault tolerance of this type of protocol.



\subsubsection{Surface code}
\label{appendix:surface}
\begin{figure}
    \centering
    \includegraphics[width=0.9\linewidth]{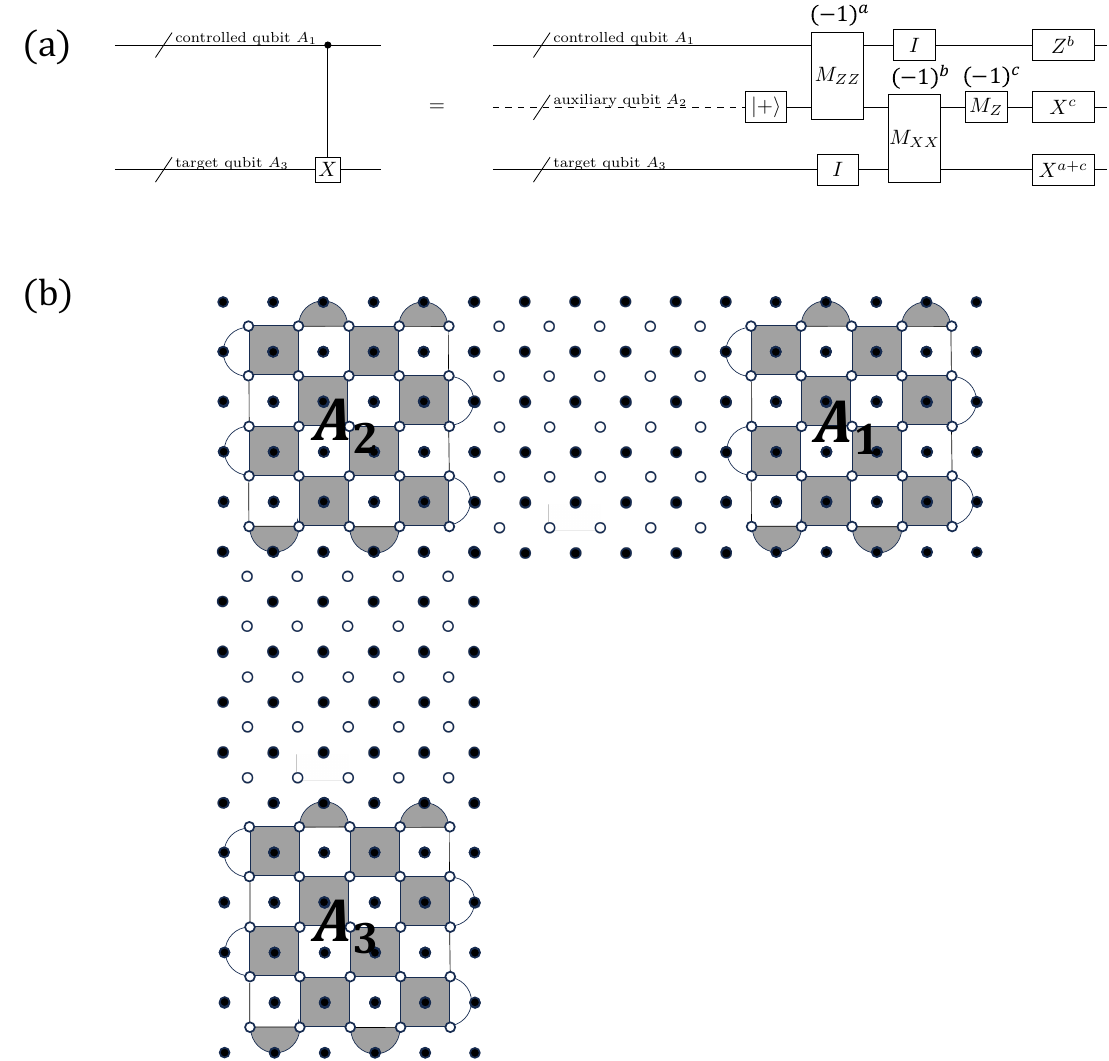}
    \caption{(a) A quantum circuit for performing a CNOT gate. A measurement operation of a Pauli operator $P$, denoted by $M_P$, is represented by a box, with each measurement outcome displayed above it. (b) The layout of physical qubits for performing a logical CNOT gate between logical qubits encoded in a rotated surface code with the distance $5$. Each circle represents a physical qubit. When defining surface codes on these physical qubits, the white circles serve as data physical qubits, and the black circles as auxiliary physical qubits for syndrome measurement. Each black (and white) region represents an $X(Z)$-type stabilizer generator of the surface code, acting on the data qubits within its region as Pauli $X(Z)$ operators, respectively.}
    \label{fig:LS-circuit}
\end{figure}

We summarize the details of the protocol for the surface code and its numerical simulation.
The surface code is a planar version~\cite{freedman2001projective,bravyi1998quantum} of the toric code~\cite{Kitaev1997QuantumEC,kitaev2003fault}, and we here consider a rotated version~\cite{Bombin_2007} of the planar surface code that requires fewer auxiliary qubits for the syndrome measurement.
The distance-$d$ rotated surface code is a $[[d^2, 1, d]]$ code, defined on a square lattice consisting of $d\times d$ data physical qubits.
In the rotated surface codes, as shown in Fig.~\ref{fig:LS-circuit}, data qubits are located at the vertices of the square plaquettes, while auxiliary qubits are placed at the centers of the squares; the $X$-type and $Z$-type stabilizer generators are arranged in an alternating checkerboard pattern.

We employ the lattice surgery~\cite{horsman2012surface,vuillot2019code} to implement a logical CNOT gate on logical qubits encoded in the surface codes.
The lattice surgery is a widely used technique for measuring logical Pauli operators acting on logical qubits encoded in the specially separated code blocks of the surface code only using the nearest-neighbor interaction of physical qubits aligned in a two-dimensional plane.
The lattice surgery also provides a way to perform a logical CNOT gate between logical qubits of the surface code blocks, which is given by a quantum circuit shown in Fig.~\ref{fig:LS-circuit}~(a).
The circuit is described by logical $I$, $X$, $Z$ gates, and the measurements of logical $X\otimes X$, $Z\otimes Z$, and $Z$ operators, denoted by $M_{X X}, M_{Z Z},$ and $M_Z$, respectively, and implemented by the lattice surgery.
The layout of physical qubits for performing a logical CNOT gate through the lattice surgery is shown in Fig.~\ref{fig:LS-circuit}~(b).
The space between code blocks of surface codes in our layout is called the routing space, which should be at least as large as the size of each code block to allow for the lattice surgery between distant code blocks~\cite{Chamberland_2022, fowler2019low}.
Also note that the lattice surgery, in combination with magic state injection and magic state distillation, leads to a protocol for implementing universal quantum computation while we here present a lattice-surgery part relevant to our analysis; see Ref.~\cite{horsman2012surface} for further details of the protocol.

Given a physical error rate $p$ of the circuit-level depolarizing error model and the distance $d$, we evaluate the logical error rates of logical $I$, $X$, $Z$, $M_{XX}$, $M_{ZZ}$, and $M_{Z}$ operations, in the circuit to perform the CNOT gate in Fig.~\ref{fig:LS-circuit}~(a).
Note that due to the limitation of the computational resources, it was hard to directly simulate the physical circuit to implement the circuit in Fig.~\ref{fig:LS-circuit}~(a) at the logical level, but the strategy here is to model the logical errors for each operation in the circuit of Fig.~\ref{fig:LS-circuit}~(a) and simulate a noisy version of the circuit in Fig.~\ref{fig:LS-circuit}~(a) with this error model.
In particular, we estimate the logical error rate of $I$, $M_{XX}$, and $M_{ZZ}$ is estimated through the memory experiment and stability experiment based on the method in Ref.~\cite{Gidney_2022}.
On the one hand, the memory experiment evaluates the probability of logical $X$ or $Z$ errors occurring on the logical qubit encoded in the surface code after $t$ rounds of syndrome measurement; on the other hand, the stability experiment evaluates the logical error probability of the product of the measurement outcomes of multiple stabilizer generators after $t$ rounds of syndrome measurement~\cite{Gidney2022stability}.
We use the minimum-weight perfect matching algorithm for decoding implemented by PyMatching package~\cite{higgott2021pymatching,higgott2023sparse}.

For the $\ket{+}$-state preparation operation in Fig.~\ref{fig:LS-circuit}~(a), we initialize the logical qubit of surface codes in the logical state $\ket{+}$ by initializing all data physical qubits of the surface code in the physical state $\ket{+}$, measuring all stabilizer generators, and running the decoder to correct errors.
However, since these operations can be performed simultaneously with the subsequent $M_{ZZ}$ operation using lattice surgery, we assume that we can subsume the logical error rate of the $\ket{+}$ state preparation operation into the logical error rate of the subsequent lattice surgery and thus can ignore it here.

For the $M_{ZZ}$ operation in Fig.~\ref{fig:LS-circuit}~(a) (and the $M_{XX}$ operation as well), the measurement outcome of the logical $Z\otimes Z$ operator is determined by the product of measurement outcomes of $Z$-type stabilizer generators in the routing space.
We estimate the probability $p_{\mathrm{stab}}$ of incorrectly reading the product of the logical measurement outcome by the stability experiment of the code block with size $d\times d$ with $d$ rounds of syndrome measurements.
The measurement outcome of the $M_{ZZ}$ operation is flipped with the logical error rate $p^{\mathrm{stab}}_{M_{ZZ}}$.
Along with measuring $Z\otimes Z$, the error correction is performed on the merged code block with size $d\times 3d$, where $d$ is the code distance.
We estimate the probability that the logical $X$ error and $Z$ error occur, denoted by $p_X$ and $p_Z$, respectively, through the memory experiment of the merged code block with size $d\times 3d$ with $d$ rounds of syndrome measurements.
A logical $X$ error during the error correction in implementing the logical $M_{ZZ}$ operation leads to a logical $X\otimes X$ error acting on the control and auxiliary logical qubits at the logical error rate $p^X_{M_{ZZ}}$.
In addition, a logical $Z$ error during error correction in the $M_{ZZ}$ operation leads to a $Z$ error acting on the controlled logical qubit at the logical error rate is $p^Z_{M_{ZZ}}$.
We numerically simulate these logical $X\otimes X$ and $Z$ errors in addition to the errors in reading the logical measurement outcomes.

For the identity operation $I$ in Fig.~\ref{fig:LS-circuit}~(a), we estimate the probability of the logical $X$ error and $Z$ error, denoted by $p^X_{I}$ and $p^Z_{I}$, respectively, through the memory experiment of the code block with size $d\times d$ with $d$ rounds of syndrome measurements.
Note that the logical identity operation is performed with $d$ rounds of syndrome measurements here because the number of the time steps for performing the $M_{ZZ}$ and $M_{XX}$ operations is also $d$ rounds.
With the logical error rate $p^X_{I}(p^Z_{I})$, the logical identity operation $I$ suffers from the logical Pauli $X(Z)$ errors.

To estimate the logical error rate of the $M_Z$ operation, we use the memory experiment by starting with a (noiseless) logical qubit in the logical state $\ket{0}$ and performing $Z$-basis measurements on data physical qubits.
Subsequently, we calculate the $Z$-type stabilizer generators by multiplying the measurement outcomes of the data qubits, correcting errors, and deducing the logical measurement outcomes of the logical $Z$ operator.
The logical measurement outcome of the logical $Z$ operator is flipped with the logical error rate $p_{M_Z}$ in the $M_Z$ operation.

As for the Pauli operations for correction operations, we can execute the logical Pauli operations classically by changing the Pauli frame~\cite{PhysRevA.86.032324}.
We assume that they can be performed without noise, depending on the measurement outcomes of $M_{ZZ}$, $M_{XX}$, and $M_Z$ operations.

In this way, for distance $d=5,7,9,11$ and various physical error rates $p$, we evaluate the logical CNOT error rate of the surface code by simulating the circuit in Fig.~\ref{fig:LS-circuit}.

\subsubsection{Concatenated Steane code}
\label{appendix:steane}
\begin{figure}
    \centering
    \includegraphics[width=\linewidth]{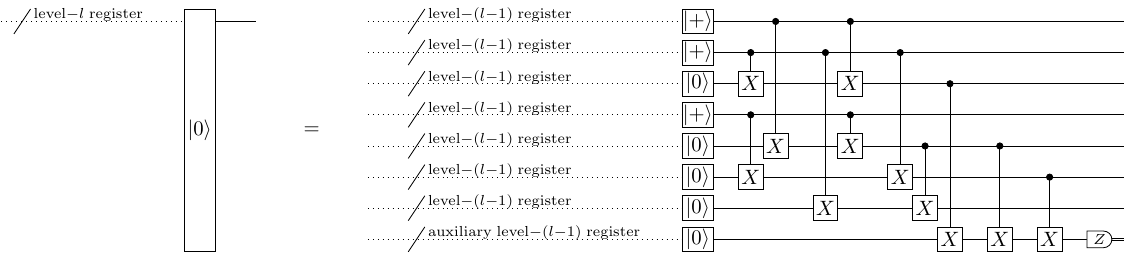}
    \\\;\\
    \includegraphics[width=\linewidth]{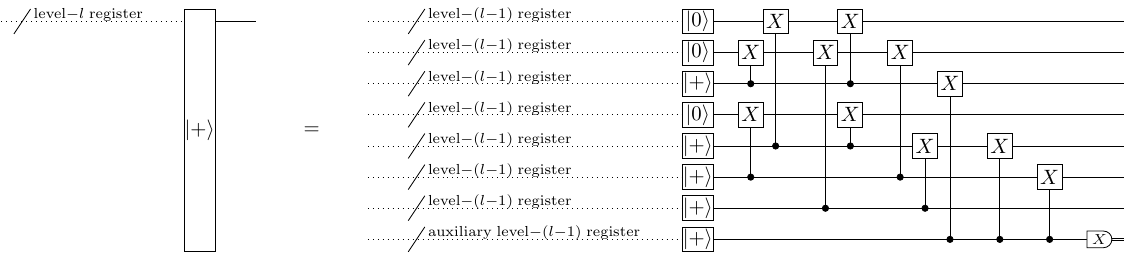}
    \caption{A level-$l$ initial-state preparation gadget for the concatenated Steane code proposed in Ref.~\cite{goto2016minimizing}. It implements preparations of the logical $\ket{0}$ ($\ket{+}$) state. The auxiliary qubit is measured for verification.  If the measurement outcome is not zero, the output quantum state is discarded and the initial-state preparation is rerun.}
    \label{fig:steane_zero_preparation_goto}
\end{figure}

We summarize the details of the protocol for the concatenated Steane code.
The protocol for the concatenated Steane code can considered to be a special case of that for the concatenated quantum Hamming code in Ref.~\cite{yamasaki2024time}, which has been presented in Sec.~\ref{subsec:hamming}, but for completeness, we here present the details relevant to our analysis.
A level-$l$ register for $l\in \{1, 2, \ldots \}$ refers to the logical qubit of the concatenated Steane code (i.e., the $[[7^l,1,3^l]]$ code).
To form a level-$l$ register, we use seven level-$(l-1)$ registers (seven qubits) of the level-$(l-1)$ code to encode the level-$l$ register as the logical qubit of the Steane code.
The logical Pauli operators, denoted by $P^{(l)}$ for $P\in\{I,X,Y,Z\}$, are given by the level-$(l-1)$ logical Pauli operators acting on the $n$th code block, denoted by $P_n$ for $P\in\{I,X,Y,Z\}$, as
\begin{align}
\begin{split}
    X^{(l)} &= X_1^{(l-1)} \otimes X_2^{(l-1)} \otimes X_3^{(l-1)} \otimes I_4^{(l-1)} \otimes I_5^{(l-1)} \otimes I_6^{(l-1)} \otimes I_7^{(l-1)},\\
    Z^{(l)} &= Z_1^{(l-1)} \otimes Z_2^{(l-1)} \otimes Z_3^{(l-1)} \otimes I_4^{(l-1)} \otimes I_5^{(l-1)} \otimes I_6^{(l-1)} \otimes I_7^{(l-1)}.
\end{split}\label{eq:logical_operator_steane}
\end{align}
For each concatenation level $l$, the level-$l$ initial-state preparation gadget for the logical $\ket{0}$ ($\ket{+})$ state of the concatenated Steane code is recursively defined using the level-$(l-1)$ gadgets as shown in Fig.~\ref{fig:steane_zero_preparation_goto}, as introduced in Ref.~\cite{goto2016minimizing}.
The measurement outcome of the auxiliary qubit in Fig.~\ref{fig:steane_zero_preparation_goto} is used for the verification; if it is non-zero, then the outcome state is discarded, and the initial-state preparation is rerun.
Since the effect of the verification failure on the logical CNOT error rate is in a sub-leading order, we omit to include this effect in the numerical simulation.

\begin{figure}
    \centering
    \includegraphics[width=\linewidth]{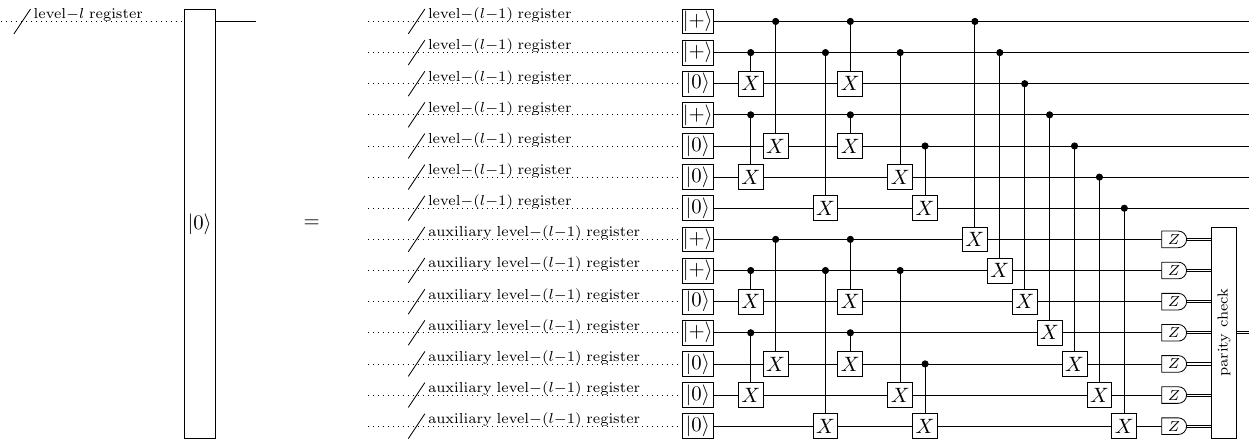}
    \\\;\\
    \includegraphics[width=\linewidth]{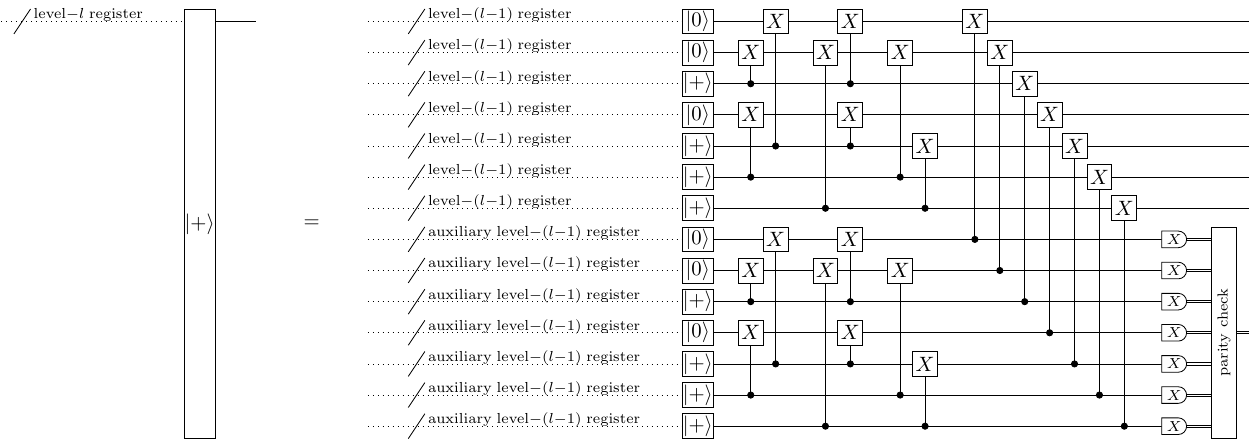}
    \caption{A conventional method for a level-$l$ initial-state preparation gadget for the concatenated Steane code using the level-$(l-1)$ gadgets. It implements preparation of the logical $\ket{0}$ ($\ket{+}$) state.  The  $Z$ ($X$) stabilizer generators and the logical $Z$ ($X$) operator are measured for verification from the measurement outcomes $i_j$ of the $(j+7)$th qubits for $j\in\{1, \ldots, 7\}$.  If $i_1+i_3+i_5+i_7 \neq 0\;(\mathrm{mod}\;2)$, $i_2+i_3+i_6+i_7 \neq 0\;(\mathrm{mod}\;2)$, or $i_1+i_2+i_3 \neq 0\;(\mathrm{mod}\;2)$ hold, the output quantum state is discarded and the initial-state preparation is rerun.}
    \label{fig:steane_zero_preparation}
\end{figure}

\begin{figure}[htbp]
    \centering
    \includegraphics[width=0.3\linewidth]{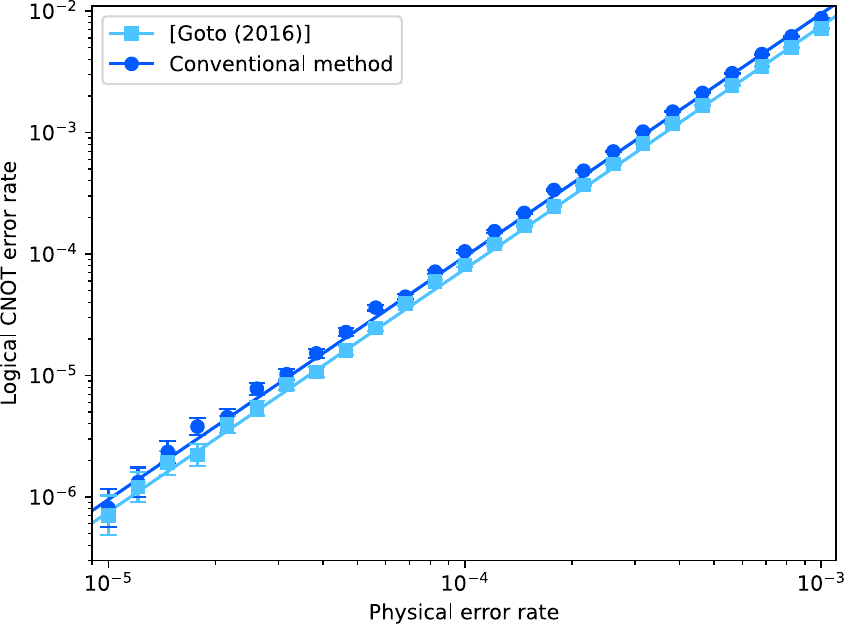}
    \caption{Comparison of the logical CNOT error rate using the conventional initial-state preparation gadget (shown in Fig.~\ref{fig:steane_zero_preparation}) and that minimizing the number of auxiliary qubits for the verification proposed by Ref.~\cite{goto2016minimizing} (shown in Fig.~\ref{fig:steane_zero_preparation_goto}). 
    Error bars in the plot represent the unbiased estimator of the standard deviation of $\log_{10} p_L$ for the logical CNOT error rates $p_L$.
    The lines in the plot are obtained from the fitting by~\eqref{eq:fitting_comparison_Steane}. 
    In our setting, the gadget in  Fig.~\ref{fig:steane_zero_preparation_goto} was slightly better than that in Fig.~\ref{fig:steane_zero_preparation}.
At the same time, we found through this numerical simulation that the conclusion as to which of the gadgets achieves better logical error rates may change highly sensitively to the details of the error model and the simulation methods since the difference between these two gadgets is too subtle; thus, it may be generally inconclusive which of the preparation gadgets to use in a practical experimental platform.}
    \label{fig:comparison_steane}
\end{figure}

The initial-state preparation gadget in  Fig.~\ref{fig:steane_zero_preparation_goto} is designed to minimize the number of auxiliary qubits for the verification, compared to the conventional method shown in Fig.~\ref{fig:steane_zero_preparation}.
To optimize the protocol, we numerically compare the performance of the two initial-state preparation gadgets by comparing the logical CNOT error rates $P_{\mathrm{Steane}}^{(1)}(p)$ for various physical error rates $p$ in our error model with fitting by
\begin{equation}
\label{eq:fitting_comparison_Steane}
P_{\mathrm{Steane}}^{(1)}(p) = a_\mathrm{Steane}^{(1)} p^{2},\\
\end{equation}
as described in Methods.
We present this numerical result in  Fig.~\ref{fig:comparison_steane}; since the method shown in Fig.~\ref{fig:steane_zero_preparation_goto} performs better than the conventional method as shown in Fig.~\ref{fig:steane_zero_preparation} in our setting, we use the former method in our simulation.
At the same time, we found through our numerical simulation that the conclusion as to which of the gadgets in Figs.~\ref{fig:steane_zero_preparation_goto} and~\ref{fig:steane_zero_preparation} achieves better logical error rates may change highly sensitively to the details of the error model and the simulation methods; thus, it may be generally inconclusive which of the preparation gadgets to use in a practical experimental platform while the gadget in  Fig.~\ref{fig:steane_zero_preparation_goto} was slightly better in the particular setting of numerical simulation in Fig.~\ref{fig:comparison_steane}.

\begin{figure}
    \centering
    \includegraphics[width=\linewidth]{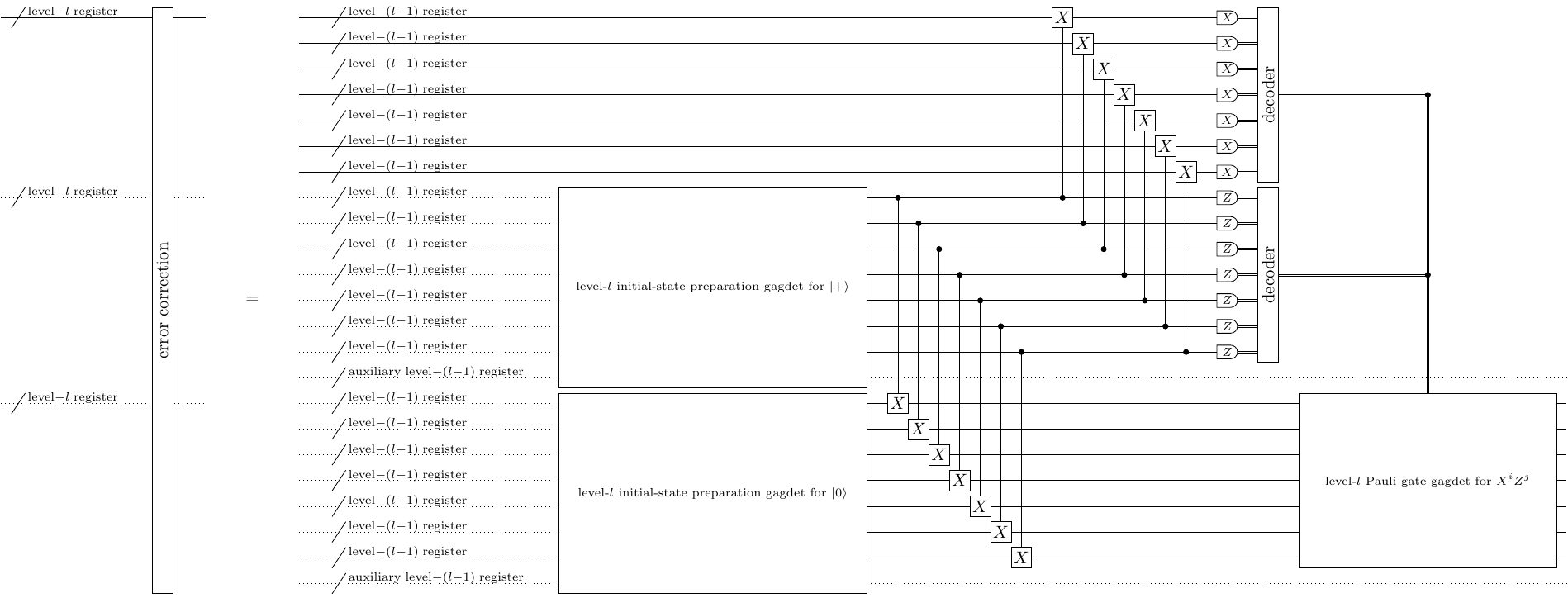}
    \caption{A level-$l$ error correction gadget for the concatenated Steane code is implemented by using the level-$(l-1)$ gadgets.}
    \label{fig:steane_ec}
\end{figure}

Also, the level-$l$ error correction gadget of the concatenated Steane code is recursively defined using the level-$(l-1)$ gadgets as shown in Fig.~\ref{fig:steane_ec}.
This gadget is called Knill's error correction gadget~\cite{knill2005quantum}.
Note that the protocol for the concatenated Steane code simulated here is different from a more optimized protocol for the concatenated Steane code simulated in Ref.~\cite{steane2003overhead}, where the syndrome extraction for quantum error correction is repeated many times to improve the threshold.
Apart from the point that we simulate the logical CNOT error rate while Ref.~\cite{steane2003overhead} the logical identity gate,
the optimization of the repetition of the syndrome extraction should also be considered to be a reason that the estimated threshold for the concatenated Steane code in Ref.~\cite{steane2003overhead} is better than that estimated in this work; however, the contribution of this work is to provide the simulation results for the simple protocol as a baseline for further comparison with more optimized protocols.

\subsubsection{$C_4$/Steane code}
\label{appendix:c4steane}
\begin{figure}
    \centering
    \includegraphics[width=\linewidth]{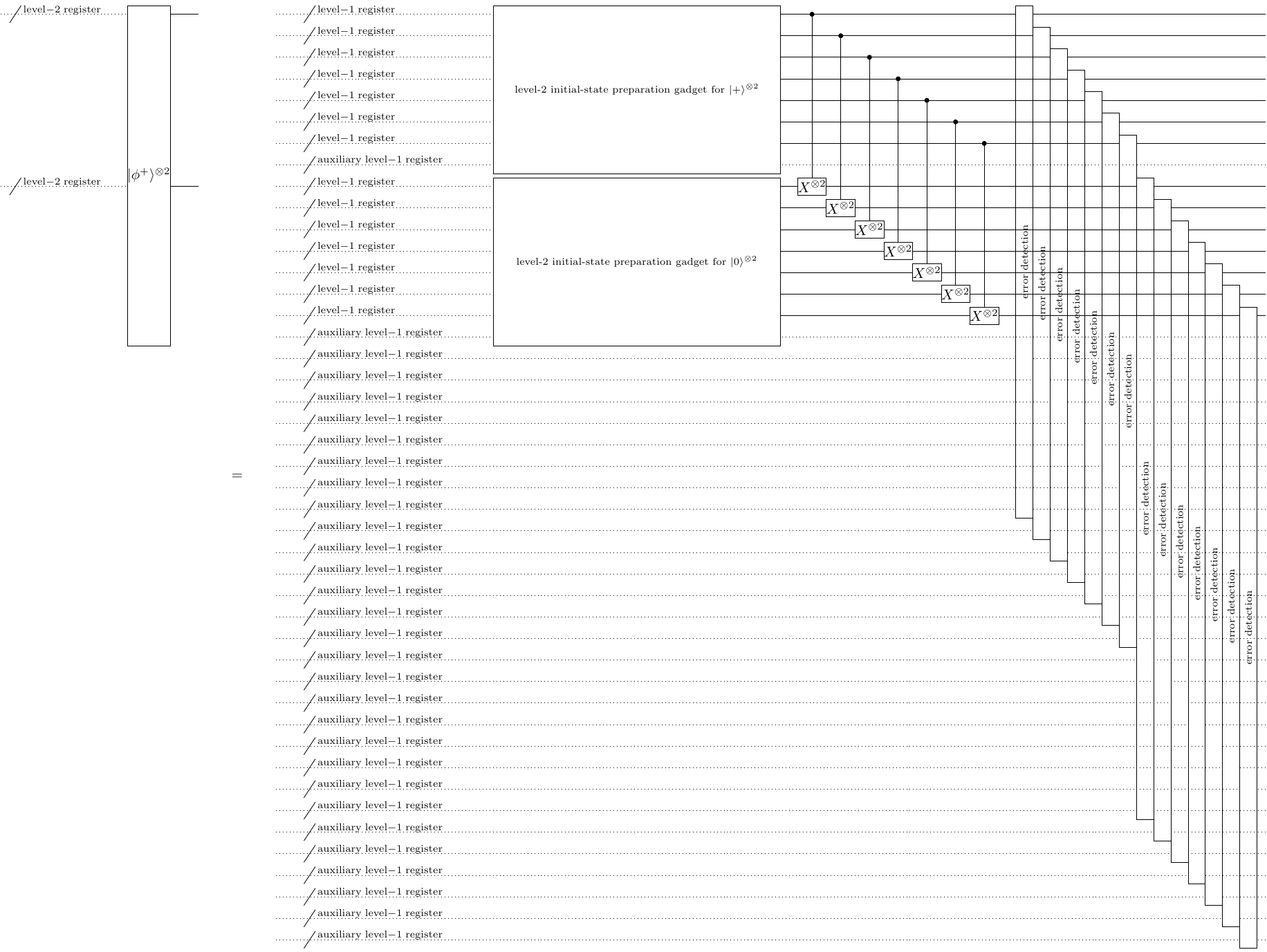}
    \caption{A level-$2$ Bell-state preparation gadget for the $C_4$/Steane code implements preparation of the logical Bell state $\ket{\phi^+} = {\ket{00}+\ket{11}\over \sqrt{2}}$.}
    \label{fig:c4steane_bell_preparation}
\end{figure}

We summarize the details of the protocol for the $C_4$/Steane code.
The protocol for the $C_4$/Steane code can be derived as a combination of the protocol for the $C_4$ code (i.e., a part of the protocol for the $C_4/C_6$ code) and the protocol for the concatenated Steane code, but for completeness, we here present the details relevant to our analysis.
A level-$1$ register is the two logical qubits of the $C_4$ code (i.e., the $[[4,2,2]]$ code).
The level-$l$ register for $l\in \{2, 3, \ldots \}$ refers to the two logical qubits of the $C_4$/Steane code.
To form a level-$l$ register, we use $7$ level-$(l-1)$ registers ($14$ qubits); in particular, similar to the concatenated quantum Hamming code in Sec.~\ref{subsec:hamming}, the first (second) qubit from each of the $7$ level-$(l-1)$ registers is picked up, and the first (second) qubit of the level-$l$ register is encoded into these picked $7$ qubits as the logical qubit of the $[[7,1,3]]$ Steane code.
The logical Pauli operators of the level-$1$ register are the same as~\eqref{eq:logical_operator_c4}.
The logical Pauli operators acting on the $i$th logical qubit of the level-$l$ register for $l\geq 2$, denoted by $P_i^{(l)}$ for $P\in\{I, X, Y, Z\}$, are given by the level-$(l-1)$ logical Pauli operators acting on the $j$th logical qubit of the $n$th level-$(l-1)$ register, denoted by $P_{n,j}^{(l-1)}$ for $P\in\{I, X, Y, Z\}$, as
\begin{align}
\begin{split}
    X_i^{(l)} &= X_{i,1}^{(l-1)} \otimes X_{i,2}^{(l-1)} \otimes X_{i,3}^{(l-1)} \otimes I_{i,4}^{(l-1)} \otimes I_{i,5}^{(l-1)} \otimes I_{i,6}^{(l-1)} \otimes I_{i,7}^{(l-1)},\\
    Z_i^{(l)} &= Z_{i,1}^{(l-1)} \otimes Z_{i,2}^{(l-1)} \otimes Z_{i,3}^{(l-1)} \otimes I_{i,4}^{(l-1)} \otimes I_{i,5}^{(l-1)} \otimes I_{i,6}^{(l-1)} \otimes I_{i,7}^{(l-1)}.
\end{split}\label{eq:logical_operator_c4steane}
\end{align}
The level-1 gadget of the $C_4$/Steane code is the same as the level-1 gadgets of the $C_4/C_6$ code, i.e., those for the $[[4,2,2]]$ code, shown in Figs.~\ref{fig:c4_zero_preparation}, \ref{fig:c6_u}, \ref{fig:c4c6_bell_preparation}, and \ref{fig:c4c6_ec}.
The level-$l$ gadget of the $C_4$/Steane code is recursively defined using the level-$(l-1)$ gadgets similarly to the concatenated Steane code shown in Figs.~\ref{fig:steane_zero_preparation}, \ref{fig:steane_ec}, except that level-$2$ error correction gadget of the $C_4$/Steane code uses the level-$2$ Bell-state preparation gadget shown in Fig.~\ref{fig:c4steane_bell_preparation}.
Since the effect of the verification failure on the logical CNOT error rate is in a sub-leading order, we omit to include this effect in the numerical simulation.

\subsection{Decoder}
\label{subsec:decoder}
We describe the decoding algorithms used in our numerical simulation for the concatenated Steane code, the $C_4/C_6$ code, the $C_4$/Steane code, and the concatenated quantum Hamming code.
Note that for the surface code, we used the minimum-weight perfect matching algorithm for decoding implemented by PyMatching package~\cite{higgott2021pymatching}.

The decoding algorithms used in our simulation for the concatenated Steane code, the $C_4/C_6$ code, the $C_4$/Steane code, and the concatenated quantum Hamming code are based on hard-decision decoders.
Note that for the concatenated Steane code, the $C_4/C_6$ code, and the $C_4$/Steane code, a soft-decision decoder is also implementable within polynomial time~\cite{PhysRevA.74.052333,goto2013fault}, which is expected to achieve higher threshold than the hard-decision decoders at the expense of computational time; in our numerical simulation, we use the hard-decision decoders to cover practical situations where the efficiency of implementing the
decoder matters.
It is unknown whether this construction of efficient soft-decision decoders for concatenated codes generalizes to the concatenated quantum Hamming code since the concatenated quantum Hamming code has a growing number of logical qubits.

For the concatenated Steane code, we use a hard-decision decoder shown in Ref.~\cite{yamasaki2024time}.
The measurement outcome of the level-$l$ measurement gadget is given by a sequence of level-$(l-1)$ logical measurement outcomes $(m_1, \ldots, m_7)$.
We check the parities $a_1 = m_1 + m_3 + m_5 + m_7 \; \mathrm{mod} \; 2$, $a_2 = m_2 + m_3 + m_6 + m_7 \; \mathrm{mod} \; 2$, and $a_3 = m_4 + m_5 + m_6 + m_7 \; \mathrm{mod} \; 2$, and if they are not all zeros, we identify the error location to be $i = a_1 + 2 a_2 + 4 a_3$.
Then, we decode the level-$l$ logical measurement outcome as
\begin{align}
\label{eq:7-qubit_decoder}
    \bar{m} =
    \begin{cases}
        m_1 + m_2 + m_3 + 1 \; \mathrm{mod} \; 2 & (i=1, 2, 3)\\
        m_1 + m_2 + m_3 \; \mathrm{mod} \; 2 & (\mathrm{otherwise})\\
    \end{cases}.
\end{align}

For the $C_4/C_6$ code, we use a hard-decision decoder shown in Ref.~\cite{goto2013fault}.
The measurement outcome of the level-$1$ measurement gadget is given by a sequence of measurement outcomes $(m^{(b)}_1, m^{(b)}_2, m^{(b)}_3, m^{(b)}_4)$, where $b\in\{X,Z\}$ represents the basis of the measurement.
The parity of the measurement outcomes is checked to detect an error, and if $m^{(b)}_1 + m^{(b)}_2 + m^{(b)}_3 + m^{(b)}_4 = 0 \;\mathrm{mod}\;2$ holds, the measurement outcome is decoded as
\begin{align}
    (\bar{m}^{(Z)}_1, \bar{m}^{(Z)}_2)
    &= (m^{(Z)}_1 + m^{(Z)}_2, m^{(Z)}_2 + m^{(Z)}_4) \;\mathrm{mod}\;2,\\
    (\bar{m}^{(X)}_1, \bar{m}^{(X)}_2)
    &= (m^{(X)}_1 + m^{(X)}_3, m^{(X)}_3 + m^{(X)}_4)\;\mathrm{mod}\;2.
\end{align}
Otherwise, we decode it as $(E,E)$, where $E$ represents that an error is detected.
The measurement outcome of the level-$l$ measurement gadget for $l\geq 2$ is given by a sequence of level-$(l-1)$ measurement outcomes $((m^{(b)}_1, m^{(b)}_2), (m^{(b)}_3, m^{(b)}_4), (m^{(b)}_5, m^{(b)}_6))$.
If errors are detected in two or three out of three code blocks, we decode it as $(E,E)$. If errors are detected in one code block, we decode it as
\begin{align}
    (\bar{m}^{(Z)}_1, \bar{m}^{(Z)}_2)
    &=
    \begin{cases}
        (m^{(Z)}_3 + m^{(Z)}_4 + m^{(Z)}_6, m^{(Z)}_4 + m^{(Z)}_5)\;\mathrm{mod}\;2 & ((m^{(Z)}_1, m^{(Z)}_2) = (E,E))\\
        (m^{(Z)}_1 + m^{(Z)}_2 + m^{(Z)}_5, m^{(Z)}_2 + m^{(Z)}_5 + m^{(Z)}_6)\;\mathrm{mod}\;2 & ((m^{(Z)}_3, m^{(Z)}_4) = (E,E))\\
        (m^{(Z)}_2 + m^{(Z)}_3, m^{(Z)}_1 + m^{(Z)}_3 + m^{(Z)}_4)\;\mathrm{mod}\;2 & ((m^{(Z)}_5, m^{(Z)}_6) = (E,E))
    \end{cases},\\
    (\bar{m}^{(X)}_1, \bar{m}^{(X)}_2)
    &=
    \begin{cases}
        (m^{(X)}_3 + m^{(X)}_4 + m^{(X)}_6, m^{(X)}_4 + m^{(X)}_5)\;\mathrm{mod}\;2 & ((m^{(X)}_1, m^{(X)}_2) = (E,E))\\
        (m^{(X)}_1 + m^{(X)}_2 + m^{(X)}_5, m^{(X)}_2 + m^{(X)}_5 + m^{(X)}_6)\;\mathrm{mod}\;2 & ((m^{(X)}_3, m^{(X)}_4) = (E,E))\\
        (m^{(X)}_2 + m^{(X)}_3, m^{(X)}_1 + m^{(X)}_3 + m^{(X)}_4)\;\mathrm{mod}\;2 & ((m^{(X)}_5, m^{(X)}_6) = (E,E))
    \end{cases}.
\end{align}
If no errors are detected, we check the parity of the measurement outcome to detect an error.
If $m^{(b)}_1+m^{(b)}_3+m^{(b)}_5 = 0 \;\mathrm{mod}\;2$ and $m^{(b)}_2+m^{(b)}_4+m^{(b)}_6 = 0 \;\mathrm{mod}\;2$ hold, we decode it as
\begin{align}
    (\bar{m}^{(Z)}_1, \bar{m}^{(Z)}_2)
    &=
        (m^{(Z)}_2 + m^{(Z)}_3, m^{(Z)}_1 + m^{(Z)}_3 + m^{(Z)}_4)\;\mathrm{mod}\;2,\\
    (\bar{m}^{(X)}_1, \bar{m}^{(X)}_2)
    &=
        (m^{(X)}_3 + m^{(X)}_4 + m^{(X)}_6, m^{(X)}_4 + m^{(X)}_5)\;\mathrm{mod}\;2.
\end{align}
Otherwise, we decode it as $(E,E)$.

For the $C_4$/Steane code, we use the same decoder as the $C_4/C_6$ code for the level-$1$ protocol and as the concatenated Steane code for the level-$l$ ($l\geq 3$) protocols.
For the level-2 measurement gadget, the measurement outcome is given as a sequence of level-1 measurement outcomes $(m_1, m_2, m_3, m_4, m_5, m_6, m_7)$.  If errors are detected in two code blocks, denoted by $i$ and $j$, we search $(m'_1, m'_2, m'_3, m'_4, m'_5, m'_6, m'_7)$ such that $m'_k = m_k$ for $k\neq i,j$ and $m'_1 + m'_3 + m'_5 + m'_7 = m'_2 + m'_3 + m'_6 + m'_7  = m'_4 + m'_5 + m'_6 + m'_7 = 0 \;\mathrm{mod}\;2$. If such a sequence is found, we decode it as
\begin{align}
    \bar{m} = m'_1 + m'_2 + m'_3 \;\mathrm{mod}\;2.
\end{align}
Otherwise, we use the same decoder as the concatenated Steane code.

For the concatenated quantum Hamming code, we use the decoder shown in Ref.~\cite{yamasaki2024time}, which is a straightforward generalization of \eqref{eq:7-qubit_decoder} and efficiently computable. See Ref.~\cite{yamasaki2024time} for details.

\section{Threshold analysis of the concatenated quantum Hamming code, the $C_4/C_6$ code, the surface code, the concatenated Steane code, and the $C_4$/Steane code}
\label{appendix_sec:threshold}

In this section, we summarize the details of our numerical results on the threshold analysis.
As described in Methods, we examine the three error models, where the ratio of the Pauli error rate on the identity gate $I$ and that on the logical gates $p$ are taken differently: (a) $\gamma=p/10$, (b) $\gamma=p/2$, and (c) $\gamma=p$.
All the simulation results in the main text are according to the error model (c).
A similar analysis in the main text for the other error models (a) and (b) is shown in Sec.~\ref{appendix_sec:other_error_models}.

\begin{figure}[]
    \centering
    \includegraphics[width=\linewidth]{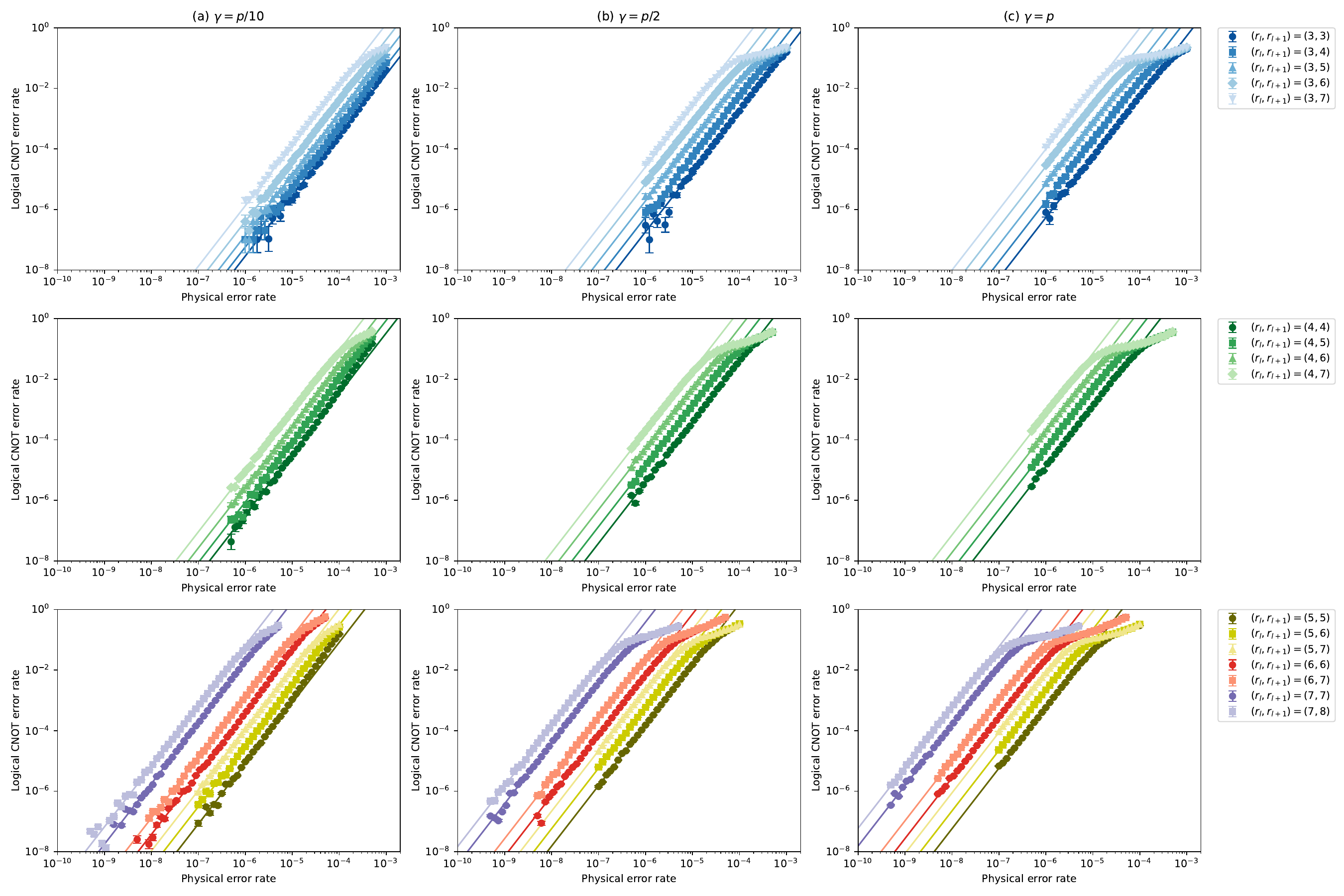}
    \caption{The logical CNOT error rates $P_{r_l}^{(r_{l+1})}$ of the quantum Hamming codes $\mcQ_{r_l}$ for $r_l = 3$ (top row), $r_l=4$ (middle row), and $r_l\in\{5,6,7\}$ (bottom row) for various physical error rates and the three error models (a) $\gamma=p/10$, (b) $\gamma=p/2$ and (c) $\gamma=p$. Each point of the logical CNOT error rate in the plot is estimated using~\eqref{eq:estimation_logical_error_rate} and~\eqref{eq:hamming_fitting}, where $P_\mathrm{CNOT}^{(0)}$ in~\eqref{eq:estimation_logical_error_rate} is given by an average over $10^6$ simulation runs, and $P_\mathrm{verification}$ and $P_\mathrm{CNOT}^{(i)}$ in~\eqref{eq:estimation_logical_error_rate} are given by averages over $10^4$ simulation runs.  The number of simulation runs counts all events including those in which the verification fails, which are discarded in the analysis.  Error bars in the plot represent the unbiased estimator of the standard deviation of $\log_{10} p_L$ for the logical CNOT error rates $p_L$. The fitting yields the parameters in Table \ref{tab:parameter_hamming}.}
    \label{fig:hamming_logical_error_rate}
\end{figure}

\begin{figure}[t]
    \centering
    \includegraphics[width=\linewidth]{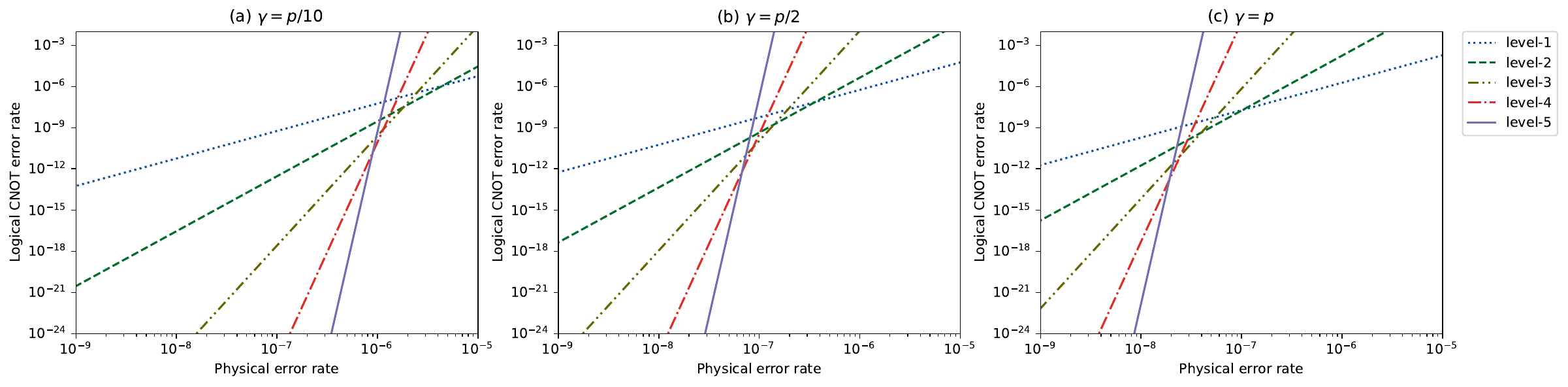}
    \caption{
    The estimation of the logical CNOT error rate of the concatenated quantum Hamming code starting from $\mcQ_3$, obtained from the fitting results in Fig.~\ref{fig:hamming_logical_error_rate} using~\eqref{eq:concatenation_formula} for the three error models (a) $\gamma=p/10$, (b) $\gamma=p/2$ and (c) $\gamma=p$.}
    \label{fig:hamming_code_threshold}
\end{figure}

We show the logical CNOT error rates of the quantum Hamming codes in Fig.~\ref{fig:hamming_logical_error_rate}, from which we obtain the threshold of the original protocol in Ref.~\cite{yamasaki2024time} based on the concatenated quantum Hamming code.
This code is obtained by concatenating the quantum Hamming code $\mcQ_{l+2}$ on the concatenation level $l\in\{1, 2, \ldots\}$.
As described in Methods, the logical error rate $P_{r_l}^{(r_{l+1})}(p)$ for the quantum Hamming code $\mcQ_{r}$ is approximated for $r_l\in\{3,4,5,6,7\}$ by the fitting curve
\begin{align}
\label{eq:hamming_fitting}
    P_{r_l}^{(r_{l+1})}(p) = a_{r_l}^{(r_{l+1})} p^2,
\end{align}
where the logical error rate of each data point is estimated using \eqref{eq:estimation_logical_error_rate}.
From our numerical results, we determine the fitting parameters as shown in Table \ref{tab:parameter_hamming}.
From these results, as described in Methods, we estimate the logical CNOT error rates for the concatenated quantum Hamming code in the original protocol starting from $\mcQ_{3}$ and that of our protocol starting from $\mcQ_{4}$ according to
\begin{align}
\label{eq:concatenation_formula}
    P_{r_{L}}^{(r_{L}+1)} \circ \cdots \circ P_{r_2}^{(r_3)} \circ P_{r_1}^{(r_2)} (p),
\end{align}
where $r_1,r_2,\ldots,r_L$ are the sequence of parameters of the quantum Hamming codes, and $p$ is the physical error rate for $\mcQ_{r_1}$.
The estimates of these logical error rates are shown in Figs.~\ref{fig:hamming_code_threshold}~(b) and (c).
The threshold values of these two concatenated quantum Hamming codes are estimated as $\sim 10^{-6}$ and $\sim 3\times 10^{-6}$, respectively.
To achieve the logical error rate $10^{-24}$ using the one for our protocol starting from $\mcQ_4$, the physical error rate for $\mcQ_4$ should be less than $P_\mathrm{target} = 2.2\times 10^{-7}$, which is the logical error rate to be achieved by the underlying quantum code in the proposed protocol.

\begin{table}[]
    \centering
    \caption{Numerical estimates of the fitting parameters for the logical error rate of quantum Hamming codes in Eq.~\eqref{eq:hamming_fitting} obtained from Fig.~\ref{fig:hamming_logical_error_rate} for the three error models (a) $\gamma=p/10$, (b) $\gamma=p/2$ and (c) $\gamma=p$.}
    \begin{ruledtabular}
    \begin{tabular}{cc|r|r|r}
         & & \multicolumn{1}{c|}{(a) $\gamma={p\over 10}$} & \multicolumn{1}{c|}{(b) $\gamma={p\over 2}$} & \multicolumn{1}{c}{(c) $\gamma=p$} \\\hline
         \multirow{5}{*}{$a_3^{(r)}$} & $r=3$ & $(2.889 \pm 0.022)\times 10^4$ & $(18.42 \pm 0.11)\times 10^4$ & $(54.21 \pm 0.03)\times 10^4$ \\
         & $r=4$ & $(5.74 \pm 0.04)\times 10^4$ & $(56.7 \pm 0.3)\times 10^4$ & $(188.8 \pm 0.8)\times 10^4$ \\
         & $r=5$ & $(13.92 \pm 0.08)\times 10^4$ & $(194.0 \pm 0.8)\times 10^4$ & $(680.6 \pm 2.3)\times 10^4$ \\
         & $r=6$ & $(40.55 \pm 0.22)\times 10^4$ & $(694.3 \pm 2.3)\times 10^4$ & $(2630 \pm 7)\times 10^4$ \\
         & $r=7$ & $(132.9 \pm 0.6)\times 10^4$ & $(2584 \pm 7)\times 10^4$ & $(9886 \pm 23)\times 10^4$ \\\hline
         \multirow{4}{*}{$a_4^{(r)}$} & $r=4$ & $(3.353 \pm 0.020)\times 10^5$ & $(38.59 \pm 0.16)\times 10^5$ & $(131.6 \pm 0.4)\times 10^5$ \\
         & $r=5$ & $(8.80 \pm 0.05)\times 10^5$ & $(136.7 \pm 0.5)\times 10^5$ & $(497.0 \pm 1.4)\times 10^5$ \\
         & $r=6$ & $(26.71 \pm 0.12)\times 10^5$ & $(499.8 \pm 1.4)\times 10^5$ & $(1882 \pm 5)\times 10^5$ \\
         & $r=7$ & $(90.9 \pm 0.3)\times 10^5$ & $(1916 \pm 4)\times 10^5$ & $(7129 \pm 13)\times 10^5$ \\\hline
         \multirow{3}{*}{$a_5^{(r)}$} & $r=5$ & $(8.27 \pm 0.04)\times 10^6$ & $(154.6 \pm 0.5)\times 10^6$ & $(574.4 \pm 1.5)\times 10^6$\\
         & $r=6$ & $(30.79 \pm 0.13)\times 10^6$ & $(582.3 \pm 1.5)\times 10^6$ & $(2224 \pm 4)\times 10^6$ \\
         & $r=7$ & $(106.1 \pm 0.4)\times 10^6$ & $(2203 \pm 5)\times 10^6$ & $(8765 \pm 15)\times 10^6$ \\\hline
         \multirow{2}{*}{$a_6^{(r)}$} & $r=6$ & $(3.719 \pm 0.014) \times 10^8$ & $(71.64 \pm 0.16)\times 10^8$ & $(273.4 \pm 0.4)\times 10^8$ \\
         & $r=7$ & $(13.00 \pm 0.04)\times 10^8$ & $(273.3 \pm 0.5)\times 10^8$ & $(1055.7 \pm 1.5)\times 10^8$ \\\hline
         \multirow{2}{*}{$a_7^{(r)}$} & $r=7$ & $(1.769 \pm  0.004)\times 10^{10}$ & $(38.89 \pm 0.06)\times 10^{10}$ & $(153.41 \pm 0.13)\times 10^{10}$ \\
         & $r=8$ & $(6.621 \pm 0.013)\times 10^{10}$ & $(150.42 \pm 0.19)\times 10^{10}$ & $(596.6 \pm 0.5)\times 10^{10}$ 
    \end{tabular}
    \end{ruledtabular}
    \label{tab:parameter_hamming}
\end{table}

\begin{figure}[]
    \centering
    \includegraphics[width=\linewidth]{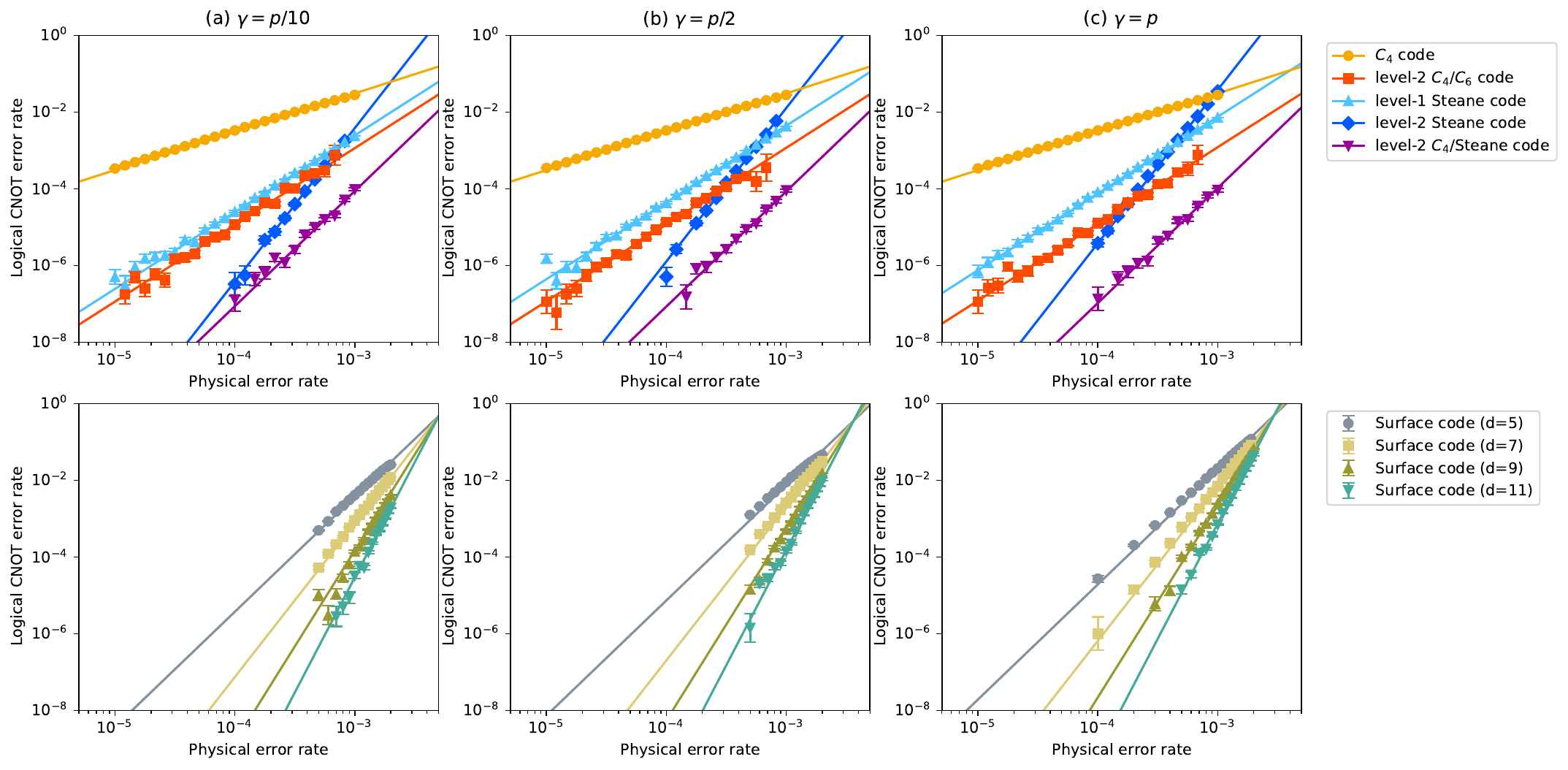}
    \caption{The logical CNOT error rates of the $C_4/C_6$ code, the surface code, the concatenated Steane code, and the $C_4$/Steane code for the three error models (a) $\gamma=p/10$, (b) $\gamma=p/2$ and (c) $\gamma=p$, which yield the parameters in Table \ref{tab:parameter_underlying}. Each point of the logical CNOT error rate in the plot is an average over $10^{6}$ simulation runs (the $C_4$ code, the level-$2$ $C_4/C_6$ code, the level-$1$ Steane code, and the surface code) or $10^{7}$ simulation runs (the level-$2$ concatenated Steane code and the level-$2$ $C_4$/Steane code). Similar to the quantum Hamming code in Fig.~\ref{fig:hamming_code_threshold},  the number of simulation runs for the $C_4/C_6$ code, the concatenated Steane code, and the $C_4$/Steane code counts all events including those in which the verification fails, which are discarded in the analysis. Error bars in the plot represent the unbiased estimator of the standard deviation of $\log_{10} p_L$ for the logical CNOT error rates $p_L$.}
    \label{fig:underlying_threshold}
\end{figure}

We also show the logical CNOT error rates of the $C_4/C_6$ code, the surface code, the Steane code, and the $C_4$/Steane code in Fig.~\ref{fig:underlying_threshold}.
Due to the limitation of the computational resources, for the numerical simulation of the level-$2$ concatenated Steane code and the level-$2$ $C_4$/Steane code, we simplified the quantum circuit shown in Fig.~3 of Methods in such a way that ten repetitions of the gate gadget of the logical $\mathrm{CNOT}^{\otimes K}$ gate followed by the error correction in Fig.~3 of Methods are replaced with one gate gadget of the logical $\mathrm{CNOT}^{\otimes K}$ gate followed by the error correction and the error-free logical $\mathrm{CNOT}^{\otimes K}$ gate.
As described in Methods, the fitting curves of the logical error rates $P_{C_4/C_6}^{(l)}$, $P_\mathrm{surface}^{(d)}$, $P_{\mathrm{Steane}}^{(l)}$, and $P_{C_4/\mathrm{Steane}}^{(l)}$ for the level-$l$ $C_4/C_6$ code, the distance-$d$ surface code, the level-$l$ concatenated Steane code, and the level-$l$ $C_4$/Steane code, respectively, are given by
\begin{align}
\label{eq:fitting_c4c6}
P_{C_4/C_6}^{(l)}(p) &= A_{C_4/C_6} (B_{C_4/C_6}p)^{F_l},\\
P_\mathrm{surface}^{(d)}(p) &= A_\mathrm{surface}(B_\mathrm{surface} p)^{d+1\over 2},\\
P_{\mathrm{Steane}}^{(l)}(p) &= a_\mathrm{Steane}^{(l)} p^{2^l},\\
P_{C_4/\mathrm{Steane}}^{(1)}(p) &= a_{C_4/\mathrm{Steane}}^{(1)} p,\\
P_{C_4/\mathrm{Steane}}^{(2)}(p) &= a_{C_4/\mathrm{Steane}}^{(2)} p^3,
\label{eq:fitting_c4steane}
\end{align}
where the notations are the same as the ones described in Methods.
From our numerical results, we determine the fitting parameters for our results as shown in Table \ref{tab:parameter_underlying}.

\begin{figure}[]
    \centering
    \includegraphics[width=\linewidth]{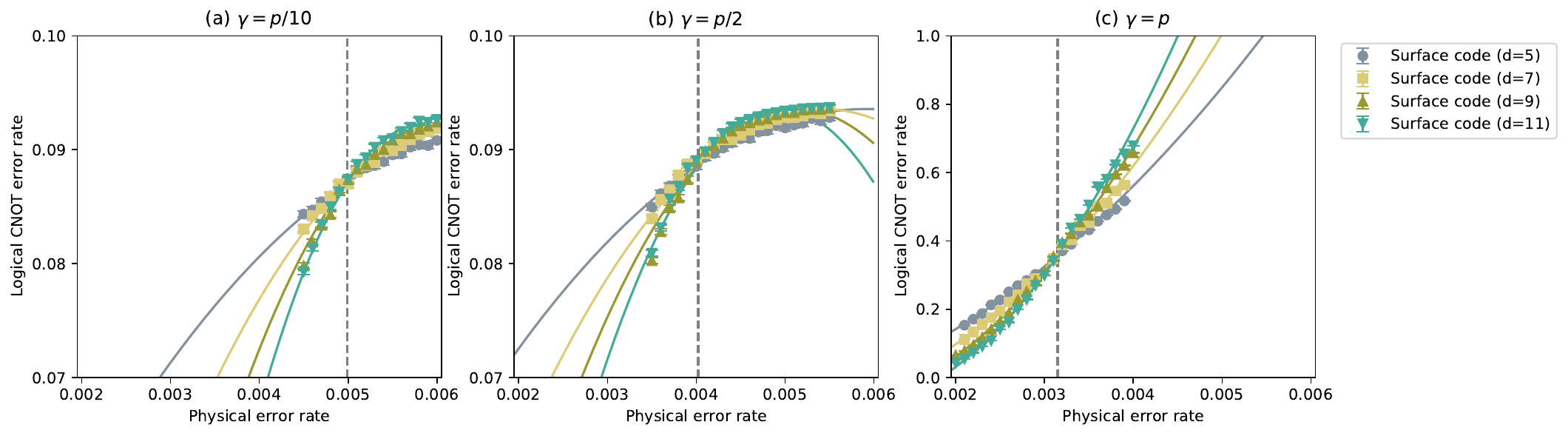}
    \caption{The fitting of the logical CNOT error rate of the surface code by~\eqref{eq:surface_code_fitting_threshold} when the physical error rate $p$ is close to the threshold for the three error models (a) $\gamma=p/10$, (b) $\gamma=p/2$ and (c) $\gamma=p$.
    These plots yield the parameters in Table \ref{tab:parameter_surface_threshold}, where the vertical dashed line represents the threshold $p^{(\mathrm{th})}_\mathrm{surface}$ in Table \ref{tab:parameter_surface_threshold}.}
    \label{fig:surface_threshold}
\end{figure}

\begin{table}[]
    \centering
    \caption{Numerical estimates of the fitting parameters for the logical error rate of the $C_4/C_6$, surface, Steane and $C_4$/Steane codes in~\eqref{eq:fitting_c4c6}--\eqref{eq:fitting_c4steane} for the three error models (a) $\gamma=p/10$, (b) $\gamma=p/2$ and (c) $\gamma=p$.}
    \begin{ruledtabular}
    \begin{tabular}{c|c|c|c}
         & (a) $\gamma={p\over 10}$ & (b) $\gamma={p\over 2}$ & (c) $\gamma=p$ \\\hline
        $(A_{C_4/C_6}, B_{C_4/C_6})$ & $(0.82 \pm 0.03, 37.3 \pm 1.3)$ & $(0.80 \pm 0.03, 38.1 \pm 1.4)$ & $(0.77 \pm 0.03, 39.6 \pm 1.4)$ \\
        $(A_\mathrm{surface}, B_\mathrm{surface})$ & $(0.477 \pm 0.005, 198.4 \pm 0.6)$ & $(0.3578 \pm 0.0022, 273.5 \pm 0.5)$ & $(0.4998 \pm 0.0018, 337.3 \pm 0.3)$\\
       $a_\mathrm{Steane}^{(1)}$ & $2447\pm 10$ & $4397\pm 14$ & $7513 \pm 18$\\
       $a_\mathrm{Steane}^{(2)}$ & $(0.379 \pm 0.011)\times 10^{10}$ & $(1.267 \pm 0.020)\times 10^{10}$ & $(3.78 \pm 0.04)\times 10^{10}$\\
       $a_{C_4/\mathrm{Steane}}^{(2)}$ & $(8.7 \pm 0.4)\times 10^4$ & $(8.5\pm 0.4)\times 10^4$ & $(10.5 \pm 0.5)\times 10^4$
    \end{tabular}
    \end{ruledtabular}
    \label{tab:parameter_underlying}
\end{table}

\begin{table}[]
    \centering
    \caption{Numerical estimates of the fitting parameters for the logical error rate of the surface code to determine the threshold in~\eqref{eq:surface_code_fitting_threshold} for the three error models (a) $\gamma=p/10$, (b) $\gamma=p/2$ and (c) $\gamma=p$.}
    \begin{ruledtabular}
    \begin{tabular}{c|c|c|c}
         & (a) $\gamma={p\over 10}$ & (b) $\gamma={p\over 2}$ & (c) $\gamma=p$ \\\hline
        $p^{(\mathrm{th})}_\mathrm{surface}$ & $(4.991 \pm 0.017)\times 10^{-3}$ & $(4.021 \pm 0.020)\times 10^{-3}$ & $(3.1480 \pm 0.0010) \times 10^{-3}$\\
        $\mu$ & $0.92\pm 0.04$ & $1.09\pm 0.06$ & $1.471 \pm 0.003$\\
        $C_\mathrm{surface}$ & $0.08713 \pm 0.00014$ & $0.08864 \pm 0.00016$ & $0.3568 \pm 0.0003$\\
        $D_\mathrm{surface}$ & $0.90\pm 0.10$ & $1.18 \pm 0.12$ & $72.7 \pm 0.2$\\
        $E_\mathrm{surface}$ & $-41\pm 10$ & $-71 \pm 15$ & $2941 \pm 16$
    \end{tabular}
    \end{ruledtabular}
    \label{tab:parameter_surface_threshold}
\end{table}

To obtain the threshold $p^{(\mathrm{th})}_\mathrm{surface}$ of the surface code, in Fig.~\ref{fig:surface_threshold}, we fit the logical error rate of the surface code when the physical error rate $p$ is close to the threshold by another fitting curve based on the critical exponent method of Ref.~\cite{Wang_2003}.
The fitting curve is given by
\begin{align}
\begin{split}
    {P^\prime}_\mathrm{surface}^{(d)}(p) &= C_\mathrm{surface} + D_\mathrm{surface} x + E_\mathrm{surface} x^2,\\
    x &= (p-p^{(\mathrm{th})}_\mathrm{surface})d^{1/\mu},
\end{split}
\label{eq:surface_code_fitting_threshold}
\end{align}
where the estimated fitting parameters are given in Table \ref{tab:parameter_surface_threshold}.
Note that it consistently holds that $p^{(\mathrm{th})}_\mathrm{surface}\approx B_\mathrm{surface}^{-1}$ for the three error models.
As discussed in the main text, since we evaluate the threshold by the logical CNOT error rate, the threshold is worse than that estimated by the quantum memory (i.e., by the logical identity gate), but our numerical simulation is motivated by the fact that the realization of the quantum memory by just implementing the logical identity gate is insufficient for universal quantum computation.
Also, Ref.~\cite{vuillot2019code} evaluated the logical CNOT error rate with the lattice surgery for the surface code and reported the threshold $0.45\%$ under the error model equivalent to our error model (c) $\gamma=p$, which is still better than the threshold evaluated here; however, as described in Sec.~\ref{appendix:surface}, our numerical simulation takes into account the logical errors attributed to the stability experiment as pointed out more recently in Ref.~\cite{Gidney2022stability} and also considers the routing space as shown in Fig.~\ref{fig:LS-circuit}~(b) unlike Ref.~\cite{vuillot2019code}, which are the reasons that the threshold evaluated here is worse than that in Ref.~\cite{vuillot2019code}.

Lastly, we remark that in general, the estimated values of the threshold largely depend on the error models and the estimation methods.
Thus, such values should be regarded as a rough guideline for a quantitative comparison between fault-tolerant protocols rather than the precise technological goal.

\section{Comparison of the error threshold and the space overhead of the proposed protocol with respect to the surface code on different error models}
\label{appendix_sec:other_error_models}

In the main text, we compare the error threshold and the space overhead of the proposed protocol with respect to the surface code under the error model described in Methods, where the error probability on the identity gate is the same as that on any one-qubit gate.
However, this may not be an appropriate error model that describes the errors occurring in the experimental setups, and simulation results can change according to different error models (see, e.g., Ref.~\cite{gidney2022benchmarking}).
In particular, the neutral atoms exhibit a much longer coherence time compared to the superconducting qubits \cite{bluvstein2023logical}, so the error on the identity gate can be much smaller than errors on logical gates.
To take this effect into account for our analysis, we consider other error models, where the error probability $\gamma$ on the identity gate is smaller than the error probability $p$ on the logical gates, which can be seen as modifications of the error models SI1000 and EM3 in Ref.~\cite{gidney2022benchmarking}.
We consider two cases where (a) $\gamma=p/10$ and (b) $\gamma=p/2$, and analyze the error threshold and the space overhead of other proposed protocols compared to the surface code.
As shown in Table~\ref{tab:threshold_different_error_models}, the $C_4/C_6/\mathrm{Hamming}$ code achieves the highest threshold and the smallest space overhead for these error models as well.

\begin{table}
    \caption{The comparison of the error threshold and the space overhead of the proposed protocol at physical error rates $p=0.01\%, 0.1\%, 1\%$ to achieve the logical error rate $10^{-24}$ with respect to those of the surface code for the two error models (a) $\gamma=p/10$ and (b) $\gamma=p/2$.
    See Table II in the main text for those values for the error model (c) $\gamma=p$.
    Bold values represent the minimum space overheads among the four quantum codes under the same physical error rates.}
    \label{tab:threshold_different_error_models}
    \begin{ruledtabular}
        \begin{tabular}{c|r|rrr|r|rrr}
             & \multicolumn{4}{c|}{(a) $\gamma=p/10$} & \multicolumn{4}{c}{(b) $\gamma=p/2$}\\\hline
            \multirow{2}{*}{} & \multirow{2}{*}{Threshold} & \multicolumn{3}{c|}{Space overhead} & \multirow{2}{*}{Threshold} & \multicolumn{3}{c}{Space overhead} \\
             & & $p=0.01\%$ & $p=0.1\%$ & $p=1\%$ & & $p=0.01\%$ & $p=0.1\%$ & $p=1\%$\\\hline
            $C_4/C_6/\mathrm{Hamming}$ & $2.7 \%$ & ${\bf 77}$ & ${\bf 2.5\times 10^2}$ & ${\bf 3.8\times 10^3}$ & $2.5 \%$ & ${\bf 93}$ & ${\bf 3.1\times 10^2}$ & ${\bf 6.1\times 10^3}$\\
            $\mathrm{Surface}/\mathrm{Hamming}$ & $0.50 \%$ & $2.3\times 10^2$ & $1.4\times 10^3$ & - & $0.40 \%$ & $3.3\times 10^2$ & $2.9\times 10^3$ & -\\
            $\mathrm{Steane}/\mathrm{Hamming}$ & $ 0.064\%$ & $6.8\times 10^2$ & - & - & $0.043 \%$ & $2.9\times 10^3$ & - & -\\
            $C_4/\mathrm{Steane}/\mathrm{Hamming}$ & $ 0.19 \%$ & $1.1\times 10^2$ & $7.3\times 10^3$ & - & $0.17 \%$ & $2.5\times 10^2$ & $2.7\times 10^4$ & -\\
            Surface & $0.50 \%$ & $7.3\times 10^2$ & $4.5\times 10^3$ & - & $0.40 \%$ & $9.6\times 10^2$ & $6.9\times 10^3$ & -
        \end{tabular}
    \end{ruledtabular}
\end{table}

\bibliography{main}

\begin{thebibliography}{80}%
\makeatletter
\providecommand \@ifxundefined [1]{%
 \@ifx{#1\undefined}
}%
\providecommand \@ifnum [1]{%
 \ifnum #1\expandafter \@firstoftwo
 \else \expandafter \@secondoftwo
 \fi
}%
\providecommand \@ifx [1]{%
 \ifx #1\expandafter \@firstoftwo
 \else \expandafter \@secondoftwo
 \fi
}%
\providecommand \natexlab [1]{#1}%
\providecommand \enquote  [1]{``#1''}%
\providecommand \bibnamefont  [1]{#1}%
\providecommand \bibfnamefont [1]{#1}%
\providecommand \citenamefont [1]{#1}%
\providecommand \href@noop [0]{\@secondoftwo}%
\providecommand \href [0]{\begingroup \@sanitize@url \@href}%
\providecommand \@href[1]{\@@startlink{#1}\@@href}%
\providecommand \@@href[1]{\endgroup#1\@@endlink}%
\providecommand \@sanitize@url [0]{\catcode `\\12\catcode `\$12\catcode `\&12\catcode `\#12\catcode `\^12\catcode `\_12\catcode `\%12\relax}%
\providecommand \@@startlink[1]{}%
\providecommand \@@endlink[0]{}%
\providecommand \url  [0]{\begingroup\@sanitize@url \@url }%
\providecommand \@url [1]{\endgroup\@href {#1}{\urlprefix }}%
\providecommand \urlprefix  [0]{URL }%
\providecommand \Eprint [0]{\href }%
\providecommand \doibase [0]{https://doi.org/}%
\providecommand \selectlanguage [0]{\@gobble}%
\providecommand \bibinfo  [0]{\@secondoftwo}%
\providecommand \bibfield  [0]{\@secondoftwo}%
\providecommand \translation [1]{[#1]}%
\providecommand \BibitemOpen [0]{}%
\providecommand \bibitemStop [0]{}%
\providecommand \bibitemNoStop [0]{.\EOS\space}%
\providecommand \EOS [0]{\spacefactor3000\relax}%
\providecommand \BibitemShut  [1]{\csname bibitem#1\endcsname}%
\let\auto@bib@innerbib\@empty
\bibitem [{\citenamefont {Kovalev}\ and\ \citenamefont {Pryadko}(2013)}]{kovalev2013fault}%
  \BibitemOpen
  \bibfield  {author} {\bibinfo {author} {\bibfnamefont {A.~A.}\ \bibnamefont {Kovalev}}\ and\ \bibinfo {author} {\bibfnamefont {L.~P.}\ \bibnamefont {Pryadko}},\ }\bibfield  {title} {\bibinfo {title} {Fault tolerance of quantum low-density parity check codes with sublinear distance scaling},\ }\href {https://doi.org/10.1103/PhysRevA.87.020304} {\bibfield  {journal} {\bibinfo  {journal} {Phys. Rev. A}\ }\textbf {\bibinfo {volume} {87}},\ \bibinfo {pages} {020304} (\bibinfo {year} {2013})}\BibitemShut {NoStop}%
\bibitem [{\citenamefont {Gottesman}(2014)}]{gottesman2013fault}%
  \BibitemOpen
  \bibfield  {author} {\bibinfo {author} {\bibfnamefont {D.}~\bibnamefont {Gottesman}},\ }\bibfield  {title} {\bibinfo {title} {Fault-tolerant quantum computation with constant overhead},\ }\href {https://doi.org/10.26421/QIC14.15-16-5} {\bibfield  {journal} {\bibinfo  {journal} {Quantum Info. Comput.}\ }\textbf {\bibinfo {volume} {14}},\ \bibinfo {pages} {1338–1372} (\bibinfo {year} {2014})}\BibitemShut {NoStop}%
\bibitem [{\citenamefont {Fawzi}\ \emph {et~al.}(2018)\citenamefont {Fawzi}, \citenamefont {Grospellier},\ and\ \citenamefont {Leverrier}}]{fawzi2018constant}%
  \BibitemOpen
  \bibfield  {author} {\bibinfo {author} {\bibfnamefont {O.}~\bibnamefont {Fawzi}}, \bibinfo {author} {\bibfnamefont {A.}~\bibnamefont {Grospellier}},\ and\ \bibinfo {author} {\bibfnamefont {A.}~\bibnamefont {Leverrier}},\ }\bibfield  {title} {\bibinfo {title} {Constant overhead quantum fault-tolerance with quantum expander codes},\ }in\ \href {https://doi.org/10.1109/FOCS.2018.00076} {\emph {\bibinfo {booktitle} {2018 IEEE 59th Annual Symposium on Foundations of Computer Science (FOCS)}}}\ (\bibinfo {year} {2018})\ pp.\ \bibinfo {pages} {743--754}\BibitemShut {NoStop}%
\bibitem [{\citenamefont {Yamasaki}\ and\ \citenamefont {Koashi}(2024)}]{yamasaki2024time}%
  \BibitemOpen
  \bibfield  {author} {\bibinfo {author} {\bibfnamefont {H.}~\bibnamefont {Yamasaki}}\ and\ \bibinfo {author} {\bibfnamefont {M.}~\bibnamefont {Koashi}},\ }\bibfield  {title} {\bibinfo {title} {Time-efficient constant-space-overhead fault-tolerant quantum computation},\ }\href {https://doi.org/10.1038/s41567-023-02325-8} {\bibfield  {journal} {\bibinfo  {journal} {Nature Physics}\ }\textbf {\bibinfo {volume} {20}},\ \bibinfo {pages} {247} (\bibinfo {year} {2024})}\BibitemShut {NoStop}%
\bibitem [{\citenamefont {Krishna}\ and\ \citenamefont {Poulin}(2021)}]{krishna2021fault}%
  \BibitemOpen
  \bibfield  {author} {\bibinfo {author} {\bibfnamefont {A.}~\bibnamefont {Krishna}}\ and\ \bibinfo {author} {\bibfnamefont {D.}~\bibnamefont {Poulin}},\ }\bibfield  {title} {\bibinfo {title} {Fault-tolerant gates on hypergraph product codes},\ }\href {https://doi.org/10.1103/PhysRevX.11.011023} {\bibfield  {journal} {\bibinfo  {journal} {Phys. Rev. X}\ }\textbf {\bibinfo {volume} {11}},\ \bibinfo {pages} {011023} (\bibinfo {year} {2021})}\BibitemShut {NoStop}%
\bibitem [{\citenamefont {Cohen}\ \emph {et~al.}(2022)\citenamefont {Cohen}, \citenamefont {Kim}, \citenamefont {Bartlett},\ and\ \citenamefont {Brown}}]{cohen2022low}%
  \BibitemOpen
  \bibfield  {author} {\bibinfo {author} {\bibfnamefont {L.~Z.}\ \bibnamefont {Cohen}}, \bibinfo {author} {\bibfnamefont {I.~H.}\ \bibnamefont {Kim}}, \bibinfo {author} {\bibfnamefont {S.~D.}\ \bibnamefont {Bartlett}},\ and\ \bibinfo {author} {\bibfnamefont {B.~J.}\ \bibnamefont {Brown}},\ }\bibfield  {title} {\bibinfo {title} {Low-overhead fault-tolerant quantum computing using long-range connectivity},\ }\href {https://doi.org/10.1126/sciadv.abn1717} {\bibfield  {journal} {\bibinfo  {journal} {Science Advances}\ }\textbf {\bibinfo {volume} {8}},\ \bibinfo {pages} {eabn1717} (\bibinfo {year} {2022})}\BibitemShut {NoStop}%
\bibitem [{\citenamefont {Tremblay}\ \emph {et~al.}(2022)\citenamefont {Tremblay}, \citenamefont {Delfosse},\ and\ \citenamefont {Beverland}}]{tremblay2022constant}%
  \BibitemOpen
  \bibfield  {author} {\bibinfo {author} {\bibfnamefont {M.~A.}\ \bibnamefont {Tremblay}}, \bibinfo {author} {\bibfnamefont {N.}~\bibnamefont {Delfosse}},\ and\ \bibinfo {author} {\bibfnamefont {M.~E.}\ \bibnamefont {Beverland}},\ }\bibfield  {title} {\bibinfo {title} {Constant-overhead quantum error correction with thin planar connectivity},\ }\href {https://doi.org/10.1103/PhysRevLett.129.050504} {\bibfield  {journal} {\bibinfo  {journal} {Phys. Rev. Lett.}\ }\textbf {\bibinfo {volume} {129}},\ \bibinfo {pages} {050504} (\bibinfo {year} {2022})}\BibitemShut {NoStop}%
\bibitem [{\citenamefont {Bravyi}\ and\ \citenamefont {Kitaev}(1998)}]{bravyi1998quantum}%
  \BibitemOpen
  \bibfield  {author} {\bibinfo {author} {\bibfnamefont {S.~B.}\ \bibnamefont {Bravyi}}\ and\ \bibinfo {author} {\bibfnamefont {A.~Y.}\ \bibnamefont {Kitaev}},\ }\bibfield  {title} {\bibinfo {title} {Quantum codes on a lattice with boundary},\ }\Eprint {https://arxiv.org/abs/quant-ph/9811052} {arXiv:quant-ph/9811052 [quant-ph]}  (\bibinfo {year} {1998})\BibitemShut {NoStop}%
\bibitem [{\citenamefont {Dennis}\ \emph {et~al.}(2002)\citenamefont {Dennis}, \citenamefont {Kitaev}, \citenamefont {Landahl},\ and\ \citenamefont {Preskill}}]{dennis2002topological}%
  \BibitemOpen
  \bibfield  {author} {\bibinfo {author} {\bibfnamefont {E.}~\bibnamefont {Dennis}}, \bibinfo {author} {\bibfnamefont {A.}~\bibnamefont {Kitaev}}, \bibinfo {author} {\bibfnamefont {A.}~\bibnamefont {Landahl}},\ and\ \bibinfo {author} {\bibfnamefont {J.}~\bibnamefont {Preskill}},\ }\bibfield  {title} {\bibinfo {title} {Topological quantum memory},\ }\href {https://doi.org/10.1063/1.1499754} {\bibfield  {journal} {\bibinfo  {journal} {Journal of Mathematical Physics}\ }\textbf {\bibinfo {volume} {43}},\ \bibinfo {pages} {4452} (\bibinfo {year} {2002})}\BibitemShut {NoStop}%
\bibitem [{\citenamefont {Steane}(1996)}]{steane1996simple}%
  \BibitemOpen
  \bibfield  {author} {\bibinfo {author} {\bibfnamefont {A.~M.}\ \bibnamefont {Steane}},\ }\bibfield  {title} {\bibinfo {title} {Simple quantum error-correcting codes},\ }\href {https://doi.org/10.1103/PhysRevA.54.4741} {\bibfield  {journal} {\bibinfo  {journal} {Phys. Rev. A}\ }\textbf {\bibinfo {volume} {54}},\ \bibinfo {pages} {4741} (\bibinfo {year} {1996})}\BibitemShut {NoStop}%
\bibitem [{\citenamefont {Hamming}(1950)}]{hamming1950error}%
  \BibitemOpen
  \bibfield  {author} {\bibinfo {author} {\bibfnamefont {R.~W.}\ \bibnamefont {Hamming}},\ }\bibfield  {title} {\bibinfo {title} {Error detecting and error correcting codes},\ }\href {https://doi.org/10.1002/j.1538-7305.1950.tb00463.x} {\bibfield  {journal} {\bibinfo  {journal} {The Bell system technical journal}\ }\textbf {\bibinfo {volume} {29}},\ \bibinfo {pages} {147} (\bibinfo {year} {1950})}\BibitemShut {NoStop}%
\bibitem [{\citenamefont {Fowler}\ \emph {et~al.}(2012)\citenamefont {Fowler}, \citenamefont {Mariantoni}, \citenamefont {Martinis},\ and\ \citenamefont {Cleland}}]{PhysRevA.86.032324}%
  \BibitemOpen
  \bibfield  {author} {\bibinfo {author} {\bibfnamefont {A.~G.}\ \bibnamefont {Fowler}}, \bibinfo {author} {\bibfnamefont {M.}~\bibnamefont {Mariantoni}}, \bibinfo {author} {\bibfnamefont {J.~M.}\ \bibnamefont {Martinis}},\ and\ \bibinfo {author} {\bibfnamefont {A.~N.}\ \bibnamefont {Cleland}},\ }\bibfield  {title} {\bibinfo {title} {Surface codes: Towards practical large-scale quantum computation},\ }\href {https://doi.org/10.1103/PhysRevA.86.032324} {\bibfield  {journal} {\bibinfo  {journal} {Phys. Rev. A}\ }\textbf {\bibinfo {volume} {86}},\ \bibinfo {pages} {032324} (\bibinfo {year} {2012})}\BibitemShut {NoStop}%
\bibitem [{\citenamefont {Knill}(2005{\natexlab{a}})}]{knill2005quantum}%
  \BibitemOpen
  \bibfield  {author} {\bibinfo {author} {\bibfnamefont {E.}~\bibnamefont {Knill}},\ }\bibfield  {title} {\bibinfo {title} {Quantum computing with realistically noisy devices},\ }\href {https://doi.org/10.1038/nature03350} {\bibfield  {journal} {\bibinfo  {journal} {Nature}\ }\textbf {\bibinfo {volume} {434}},\ \bibinfo {pages} {39} (\bibinfo {year} {2005}{\natexlab{a}})}\BibitemShut {NoStop}%
\bibitem [{\citenamefont {Paetznick}\ \emph {et~al.}(2024)\citenamefont {Paetznick}, \citenamefont {da~Silva}, \citenamefont {Ryan-Anderson}, \citenamefont {Bello-Rivas}, \citenamefont {Campora~III}, \citenamefont {Chernoguzov}, \citenamefont {Dreiling}, \citenamefont {Foltz}, \citenamefont {Frachon}, \citenamefont {Gaebler} \emph {et~al.}}]{paetznick2024demonstration}%
  \BibitemOpen
  \bibfield  {author} {\bibinfo {author} {\bibfnamefont {A.}~\bibnamefont {Paetznick}}, \bibinfo {author} {\bibfnamefont {M.}~\bibnamefont {da~Silva}}, \bibinfo {author} {\bibfnamefont {C.}~\bibnamefont {Ryan-Anderson}}, \bibinfo {author} {\bibfnamefont {J.}~\bibnamefont {Bello-Rivas}}, \bibinfo {author} {\bibfnamefont {J.}~\bibnamefont {Campora~III}}, \bibinfo {author} {\bibfnamefont {A.}~\bibnamefont {Chernoguzov}}, \bibinfo {author} {\bibfnamefont {J.}~\bibnamefont {Dreiling}}, \bibinfo {author} {\bibfnamefont {C.}~\bibnamefont {Foltz}}, \bibinfo {author} {\bibfnamefont {F.}~\bibnamefont {Frachon}}, \bibinfo {author} {\bibfnamefont {J.}~\bibnamefont {Gaebler}}, \emph {et~al.},\ }\bibfield  {title} {\bibinfo {title} {Demonstration of logical qubits and repeated error correction with better-than-physical error rates},\ }\Eprint {https://arxiv.org/abs/2404.02280} {arXiv:2404.02280}  (\bibinfo {year} {2024})\BibitemShut {NoStop}%
\bibitem [{\citenamefont {Bluvstein}\ \emph {et~al.}(2024)\citenamefont {Bluvstein}, \citenamefont {Evered}, \citenamefont {Geim}, \citenamefont {Li}, \citenamefont {Zhou}, \citenamefont {Manovitz}, \citenamefont {Ebadi}, \citenamefont {Cain}, \citenamefont {Kalinowski}, \citenamefont {Hangleiter} \emph {et~al.}}]{bluvstein2023logical}%
  \BibitemOpen
  \bibfield  {author} {\bibinfo {author} {\bibfnamefont {D.}~\bibnamefont {Bluvstein}}, \bibinfo {author} {\bibfnamefont {S.~J.}\ \bibnamefont {Evered}}, \bibinfo {author} {\bibfnamefont {A.~A.}\ \bibnamefont {Geim}}, \bibinfo {author} {\bibfnamefont {S.~H.}\ \bibnamefont {Li}}, \bibinfo {author} {\bibfnamefont {H.}~\bibnamefont {Zhou}}, \bibinfo {author} {\bibfnamefont {T.}~\bibnamefont {Manovitz}}, \bibinfo {author} {\bibfnamefont {S.}~\bibnamefont {Ebadi}}, \bibinfo {author} {\bibfnamefont {M.}~\bibnamefont {Cain}}, \bibinfo {author} {\bibfnamefont {M.}~\bibnamefont {Kalinowski}}, \bibinfo {author} {\bibfnamefont {D.}~\bibnamefont {Hangleiter}}, \emph {et~al.},\ }\bibfield  {title} {\bibinfo {title} {Logical quantum processor based on reconfigurable atom arrays},\ }\href {https://www.nature.com/articles/s41586-023-06927-3} {\bibfield  {journal} {\bibinfo  {journal} {Nature}\ }\textbf {\bibinfo {volume} {626}},\ \bibinfo {pages} {58} (\bibinfo {year} {2024})}\BibitemShut {NoStop}%
\bibitem [{\citenamefont {Ryan-Anderson}\ \emph {et~al.}(2021)\citenamefont {Ryan-Anderson}, \citenamefont {Bohnet}, \citenamefont {Lee}, \citenamefont {Gresh}, \citenamefont {Hankin}, \citenamefont {Gaebler}, \citenamefont {Francois}, \citenamefont {Chernoguzov}, \citenamefont {Lucchetti}, \citenamefont {Brown}, \citenamefont {Gatterman}, \citenamefont {Halit}, \citenamefont {Gilmore}, \citenamefont {Gerber}, \citenamefont {Neyenhuis}, \citenamefont {Hayes},\ and\ \citenamefont {Stutz}}]{PhysRevX.11.041058}%
  \BibitemOpen
  \bibfield  {author} {\bibinfo {author} {\bibfnamefont {C.}~\bibnamefont {Ryan-Anderson}}, \bibinfo {author} {\bibfnamefont {J.~G.}\ \bibnamefont {Bohnet}}, \bibinfo {author} {\bibfnamefont {K.}~\bibnamefont {Lee}}, \bibinfo {author} {\bibfnamefont {D.}~\bibnamefont {Gresh}}, \bibinfo {author} {\bibfnamefont {A.}~\bibnamefont {Hankin}}, \bibinfo {author} {\bibfnamefont {J.~P.}\ \bibnamefont {Gaebler}}, \bibinfo {author} {\bibfnamefont {D.}~\bibnamefont {Francois}}, \bibinfo {author} {\bibfnamefont {A.}~\bibnamefont {Chernoguzov}}, \bibinfo {author} {\bibfnamefont {D.}~\bibnamefont {Lucchetti}}, \bibinfo {author} {\bibfnamefont {N.~C.}\ \bibnamefont {Brown}}, \bibinfo {author} {\bibfnamefont {T.~M.}\ \bibnamefont {Gatterman}}, \bibinfo {author} {\bibfnamefont {S.~K.}\ \bibnamefont {Halit}}, \bibinfo {author} {\bibfnamefont {K.}~\bibnamefont {Gilmore}}, \bibinfo {author} {\bibfnamefont {J.~A.}\ \bibnamefont {Gerber}}, \bibinfo {author} {\bibfnamefont {B.}~\bibnamefont {Neyenhuis}}, \bibinfo {author}
  {\bibfnamefont {D.}~\bibnamefont {Hayes}},\ and\ \bibinfo {author} {\bibfnamefont {R.~P.}\ \bibnamefont {Stutz}},\ }\bibfield  {title} {\bibinfo {title} {Realization of real-time fault-tolerant quantum error correction},\ }\href {https://doi.org/10.1103/PhysRevX.11.041058} {\bibfield  {journal} {\bibinfo  {journal} {Phys. Rev. X}\ }\textbf {\bibinfo {volume} {11}},\ \bibinfo {pages} {041058} (\bibinfo {year} {2021})}\BibitemShut {NoStop}%
\bibitem [{\citenamefont {Egan}\ \emph {et~al.}(2021)\citenamefont {Egan}, \citenamefont {Debroy}, \citenamefont {Noel}, \citenamefont {Risinger}, \citenamefont {Zhu}, \citenamefont {Biswas}, \citenamefont {Newman}, \citenamefont {Li}, \citenamefont {Brown}, \citenamefont {Cetina} \emph {et~al.}}]{egan2021fault}%
  \BibitemOpen
  \bibfield  {author} {\bibinfo {author} {\bibfnamefont {L.}~\bibnamefont {Egan}}, \bibinfo {author} {\bibfnamefont {D.~M.}\ \bibnamefont {Debroy}}, \bibinfo {author} {\bibfnamefont {C.}~\bibnamefont {Noel}}, \bibinfo {author} {\bibfnamefont {A.}~\bibnamefont {Risinger}}, \bibinfo {author} {\bibfnamefont {D.}~\bibnamefont {Zhu}}, \bibinfo {author} {\bibfnamefont {D.}~\bibnamefont {Biswas}}, \bibinfo {author} {\bibfnamefont {M.}~\bibnamefont {Newman}}, \bibinfo {author} {\bibfnamefont {M.}~\bibnamefont {Li}}, \bibinfo {author} {\bibfnamefont {K.~R.}\ \bibnamefont {Brown}}, \bibinfo {author} {\bibfnamefont {M.}~\bibnamefont {Cetina}}, \emph {et~al.},\ }\bibfield  {title} {\bibinfo {title} {Fault-tolerant control of an error-corrected qubit},\ }\href {https://www.nature.com/articles/s41586-021-03928-y} {\bibfield  {journal} {\bibinfo  {journal} {Nature}\ }\textbf {\bibinfo {volume} {598}},\ \bibinfo {pages} {281} (\bibinfo {year} {2021})}\BibitemShut {NoStop}%
\bibitem [{\citenamefont {Yamasaki}\ \emph {et~al.}(2020)\citenamefont {Yamasaki}, \citenamefont {Fukui}, \citenamefont {Takeuchi}, \citenamefont {Tani},\ and\ \citenamefont {Koashi}}]{yamasaki2020polylogoverhead}%
  \BibitemOpen
  \bibfield  {author} {\bibinfo {author} {\bibfnamefont {H.}~\bibnamefont {Yamasaki}}, \bibinfo {author} {\bibfnamefont {K.}~\bibnamefont {Fukui}}, \bibinfo {author} {\bibfnamefont {Y.}~\bibnamefont {Takeuchi}}, \bibinfo {author} {\bibfnamefont {S.}~\bibnamefont {Tani}},\ and\ \bibinfo {author} {\bibfnamefont {M.}~\bibnamefont {Koashi}},\ }\bibfield  {title} {\bibinfo {title} {Polylog-overhead highly fault-tolerant measurement-based quantum computation: all-gaussian implementation with gottesman-kitaev-preskill code},\ }\Eprint {https://arxiv.org/abs/2006.05416} {arXiv:2006.05416 [quant-ph]}  (\bibinfo {year} {2020})\BibitemShut {NoStop}%
\bibitem [{\citenamefont {Bourassa}\ \emph {et~al.}(2021)\citenamefont {Bourassa}, \citenamefont {Alexander}, \citenamefont {Vasmer}, \citenamefont {Patil}, \citenamefont {Tzitrin}, \citenamefont {Matsuura}, \citenamefont {Su}, \citenamefont {Baragiola}, \citenamefont {Guha}, \citenamefont {Dauphinais}, \citenamefont {Sabapathy}, \citenamefont {Menicucci},\ and\ \citenamefont {Dhand}}]{Bourassa2021blueprintscalable}%
  \BibitemOpen
  \bibfield  {author} {\bibinfo {author} {\bibfnamefont {J.~E.}\ \bibnamefont {Bourassa}}, \bibinfo {author} {\bibfnamefont {R.~N.}\ \bibnamefont {Alexander}}, \bibinfo {author} {\bibfnamefont {M.}~\bibnamefont {Vasmer}}, \bibinfo {author} {\bibfnamefont {A.}~\bibnamefont {Patil}}, \bibinfo {author} {\bibfnamefont {I.}~\bibnamefont {Tzitrin}}, \bibinfo {author} {\bibfnamefont {T.}~\bibnamefont {Matsuura}}, \bibinfo {author} {\bibfnamefont {D.}~\bibnamefont {Su}}, \bibinfo {author} {\bibfnamefont {B.~Q.}\ \bibnamefont {Baragiola}}, \bibinfo {author} {\bibfnamefont {S.}~\bibnamefont {Guha}}, \bibinfo {author} {\bibfnamefont {G.}~\bibnamefont {Dauphinais}}, \bibinfo {author} {\bibfnamefont {K.~K.}\ \bibnamefont {Sabapathy}}, \bibinfo {author} {\bibfnamefont {N.~C.}\ \bibnamefont {Menicucci}},\ and\ \bibinfo {author} {\bibfnamefont {I.}~\bibnamefont {Dhand}},\ }\bibfield  {title} {\bibinfo {title} {Blueprint for a {S}calable {P}hotonic {F}ault-{T}olerant {Q}uantum {C}omputer},\ }\href
  {https://doi.org/10.22331/q-2021-02-04-392} {\bibfield  {journal} {\bibinfo  {journal} {{Quantum}}\ }\textbf {\bibinfo {volume} {5}},\ \bibinfo {pages} {392} (\bibinfo {year} {2021})}\BibitemShut {NoStop}%
\bibitem [{\citenamefont {Litinski}\ and\ \citenamefont {Nickerson}(2022)}]{litinski2022active}%
  \BibitemOpen
  \bibfield  {author} {\bibinfo {author} {\bibfnamefont {D.}~\bibnamefont {Litinski}}\ and\ \bibinfo {author} {\bibfnamefont {N.}~\bibnamefont {Nickerson}},\ }\bibfield  {title} {\bibinfo {title} {Active volume: An architecture for efficient fault-tolerant quantum computers with limited non-local connections},\ }\Eprint {https://arxiv.org/abs/2211.15465} {arXiv:2211.15465 [quant-ph]}  (\bibinfo {year} {2022})\BibitemShut {NoStop}%
\bibitem [{\citenamefont {Goto}\ and\ \citenamefont {Uchikawa}(2013)}]{goto2013fault}%
  \BibitemOpen
  \bibfield  {author} {\bibinfo {author} {\bibfnamefont {H.}~\bibnamefont {Goto}}\ and\ \bibinfo {author} {\bibfnamefont {H.}~\bibnamefont {Uchikawa}},\ }\bibfield  {title} {\bibinfo {title} {Fault-tolerant quantum computation with a soft-decision decoder for error correction and detection by teleportation},\ }\href {https://doi.org/10.1038/srep02044} {\bibfield  {journal} {\bibinfo  {journal} {Scientific reports}\ }\textbf {\bibinfo {volume} {3}},\ \bibinfo {pages} {2044} (\bibinfo {year} {2013})}\BibitemShut {NoStop}%
\bibitem [{\citenamefont {Goto}(2016)}]{goto2016minimizing}%
  \BibitemOpen
  \bibfield  {author} {\bibinfo {author} {\bibfnamefont {H.}~\bibnamefont {Goto}},\ }\bibfield  {title} {\bibinfo {title} {Minimizing resource overheads for fault-tolerant preparation of encoded states of the steane code},\ }\href {https://doi.org/10.1038/srep19578} {\bibfield  {journal} {\bibinfo  {journal} {Scientific Reports}\ }\textbf {\bibinfo {volume} {6}},\ \bibinfo {pages} {19578} (\bibinfo {year} {2016})}\BibitemShut {NoStop}%
\bibitem [{\citenamefont {Vuillot}\ \emph {et~al.}(2019)\citenamefont {Vuillot}, \citenamefont {Lao}, \citenamefont {Criger}, \citenamefont {Almud{\'e}ver}, \citenamefont {Bertels},\ and\ \citenamefont {Terhal}}]{vuillot2019code}%
  \BibitemOpen
  \bibfield  {author} {\bibinfo {author} {\bibfnamefont {C.}~\bibnamefont {Vuillot}}, \bibinfo {author} {\bibfnamefont {L.}~\bibnamefont {Lao}}, \bibinfo {author} {\bibfnamefont {B.}~\bibnamefont {Criger}}, \bibinfo {author} {\bibfnamefont {C.~G.}\ \bibnamefont {Almud{\'e}ver}}, \bibinfo {author} {\bibfnamefont {K.}~\bibnamefont {Bertels}},\ and\ \bibinfo {author} {\bibfnamefont {B.~M.}\ \bibnamefont {Terhal}},\ }\bibfield  {title} {\bibinfo {title} {Code deformation and lattice surgery are gauge fixing},\ }\href {https://doi.org/10.1088/1367-2630/ab0199} {\bibfield  {journal} {\bibinfo  {journal} {New Journal of Physics}\ }\textbf {\bibinfo {volume} {21}},\ \bibinfo {pages} {033028} (\bibinfo {year} {2019})}\BibitemShut {NoStop}%
\bibitem [{\citenamefont {Goto}(2014)}]{goto2014step}%
  \BibitemOpen
  \bibfield  {author} {\bibinfo {author} {\bibfnamefont {H.}~\bibnamefont {Goto}},\ }\bibfield  {title} {\bibinfo {title} {Step-by-step magic state encoding for efficient fault-tolerant quantum computation},\ }\href {https://doi.org/10.1038/srep07501} {\bibfield  {journal} {\bibinfo  {journal} {Scientific Reports}\ }\textbf {\bibinfo {volume} {4}},\ \bibinfo {pages} {7501} (\bibinfo {year} {2014})}\BibitemShut {NoStop}%
\bibitem [{\citenamefont {Schumacher}(1996)}]{schumacher1996sending}%
  \BibitemOpen
  \bibfield  {author} {\bibinfo {author} {\bibfnamefont {B.}~\bibnamefont {Schumacher}},\ }\bibfield  {title} {\bibinfo {title} {Sending entanglement through noisy quantum channels},\ }\href {https://doi.org/10.1103/PhysRevA.54.2614} {\bibfield  {journal} {\bibinfo  {journal} {Phys. Rev. A}\ }\textbf {\bibinfo {volume} {54}},\ \bibinfo {pages} {2614} (\bibinfo {year} {1996})}\BibitemShut {NoStop}%
\bibitem [{\citenamefont {Shor}(1994)}]{shor1994algorithms}%
  \BibitemOpen
  \bibfield  {author} {\bibinfo {author} {\bibfnamefont {P.}~\bibnamefont {Shor}},\ }\bibfield  {title} {\bibinfo {title} {Algorithms for quantum computation: discrete logarithms and factoring},\ }in\ \href {https://doi.org/10.1109/SFCS.1994.365700} {\emph {\bibinfo {booktitle} {Proceedings 35th Annual Symposium on Foundations of Computer Science}}}\ (\bibinfo {year} {1994})\ pp.\ \bibinfo {pages} {124--134}\BibitemShut {NoStop}%
\bibitem [{\citenamefont {Gidney}\ and\ \citenamefont {Eker{\aa}}(2021)}]{gidney2021factor}%
  \BibitemOpen
  \bibfield  {author} {\bibinfo {author} {\bibfnamefont {C.}~\bibnamefont {Gidney}}\ and\ \bibinfo {author} {\bibfnamefont {M.}~\bibnamefont {Eker{\aa}}},\ }\bibfield  {title} {\bibinfo {title} {How to factor 2048 bit rsa integers in 8 hours using 20 million noisy qubits},\ }\href {https://doi.org/10.22331/q-2021-04-15-433} {\bibfield  {journal} {\bibinfo  {journal} {Quantum}\ }\textbf {\bibinfo {volume} {5}},\ \bibinfo {pages} {433} (\bibinfo {year} {2021})}\BibitemShut {NoStop}%
\bibitem [{\citenamefont {Rivest}\ \emph {et~al.}(1978)\citenamefont {Rivest}, \citenamefont {Shamir},\ and\ \citenamefont {Adleman}}]{rivest1978method}%
  \BibitemOpen
  \bibfield  {author} {\bibinfo {author} {\bibfnamefont {R.~L.}\ \bibnamefont {Rivest}}, \bibinfo {author} {\bibfnamefont {A.}~\bibnamefont {Shamir}},\ and\ \bibinfo {author} {\bibfnamefont {L.}~\bibnamefont {Adleman}},\ }\bibfield  {title} {\bibinfo {title} {A method for obtaining digital signatures and public-key cryptosystems},\ }\href {https://doi.org/10.1145/359340.359342} {\bibfield  {journal} {\bibinfo  {journal} {Communications of the ACM}\ }\textbf {\bibinfo {volume} {21}},\ \bibinfo {pages} {120} (\bibinfo {year} {1978})}\BibitemShut {NoStop}%
\bibitem [{\citenamefont {Barker}\ and\ \citenamefont {Dang}(2016)}]{barker2016nist}%
  \BibitemOpen
  \bibfield  {author} {\bibinfo {author} {\bibfnamefont {E.}~\bibnamefont {Barker}}\ and\ \bibinfo {author} {\bibfnamefont {Q.}~\bibnamefont {Dang}},\ }\bibfield  {title} {\bibinfo {title} {Nist special publication 800-57 part 1, revision 4},\ }\href@noop {} {\bibfield  {journal} {\bibinfo  {journal} {NIST, Tech. Rep}\ }\textbf {\bibinfo {volume} {16}} (\bibinfo {year} {2016})}\BibitemShut {NoStop}%
\bibitem [{\citenamefont {Steane}(2003)}]{steane2003overhead}%
  \BibitemOpen
  \bibfield  {author} {\bibinfo {author} {\bibfnamefont {A.~M.}\ \bibnamefont {Steane}},\ }\bibfield  {title} {\bibinfo {title} {Overhead and noise threshold of fault-tolerant quantum error correction},\ }\href {https://doi.org/10.1103/PhysRevA.68.042322} {\bibfield  {journal} {\bibinfo  {journal} {Phys. Rev. A}\ }\textbf {\bibinfo {volume} {68}},\ \bibinfo {pages} {042322} (\bibinfo {year} {2003})}\BibitemShut {NoStop}%
\bibitem [{\citenamefont {Higgott}(2021)}]{higgott2021pymatching}%
  \BibitemOpen
  \bibfield  {author} {\bibinfo {author} {\bibfnamefont {O.}~\bibnamefont {Higgott}},\ }\bibfield  {title} {\bibinfo {title} {Pymatching: A python package for decoding quantum codes with minimum-weight perfect matching},\ }\Eprint {https://arxiv.org/abs/2105.13082} {arXiv:2105.13082 [quant-ph]}  (\bibinfo {year} {2021})\BibitemShut {NoStop}%
\bibitem [{\citenamefont {Higgott}\ and\ \citenamefont {Gidney}(2023)}]{higgott2023sparse}%
  \BibitemOpen
  \bibfield  {author} {\bibinfo {author} {\bibfnamefont {O.}~\bibnamefont {Higgott}}\ and\ \bibinfo {author} {\bibfnamefont {C.}~\bibnamefont {Gidney}},\ }\bibfield  {title} {\bibinfo {title} {Sparse blossom: correcting a million errors per core second with minimum-weight matching},\ }\Eprint {https://arxiv.org/abs/2303.15933} {arXiv:2303.15933 [quant-ph]}  (\bibinfo {year} {2023})\BibitemShut {NoStop}%
\bibitem [{\citenamefont {Horsman}\ \emph {et~al.}(2012)\citenamefont {Horsman}, \citenamefont {Fowler}, \citenamefont {Devitt},\ and\ \citenamefont {Van~Meter}}]{horsman2012surface}%
  \BibitemOpen
  \bibfield  {author} {\bibinfo {author} {\bibfnamefont {D.}~\bibnamefont {Horsman}}, \bibinfo {author} {\bibfnamefont {A.~G.}\ \bibnamefont {Fowler}}, \bibinfo {author} {\bibfnamefont {S.}~\bibnamefont {Devitt}},\ and\ \bibinfo {author} {\bibfnamefont {R.}~\bibnamefont {Van~Meter}},\ }\bibfield  {title} {\bibinfo {title} {Surface code quantum computing by lattice surgery},\ }\href {https://doi.org/10.1088/1367-2630/14/12/123011} {\bibfield  {journal} {\bibinfo  {journal} {New Journal of Physics}\ }\textbf {\bibinfo {volume} {14}},\ \bibinfo {pages} {123011} (\bibinfo {year} {2012})}\BibitemShut {NoStop}%
\bibitem [{\citenamefont {Gidney}(2022{\natexlab{a}})}]{Gidney2022stability}%
  \BibitemOpen
  \bibfield  {author} {\bibinfo {author} {\bibfnamefont {C.}~\bibnamefont {Gidney}},\ }\bibfield  {title} {\bibinfo {title} {Stability {E}xperiments: {T}he {O}verlooked {D}ual of {M}emory {E}xperiments},\ }\href {https://doi.org/10.22331/q-2022-08-24-786} {\bibfield  {journal} {\bibinfo  {journal} {{Quantum}}\ }\textbf {\bibinfo {volume} {6}},\ \bibinfo {pages} {786} (\bibinfo {year} {2022}{\natexlab{a}})}\BibitemShut {NoStop}%
\bibitem [{\citenamefont {Cross}\ \emph {et~al.}(2009)\citenamefont {Cross}, \citenamefont {Divincenzo},\ and\ \citenamefont {Terhal}}]{10.5555/2011814.2011815}%
  \BibitemOpen
  \bibfield  {author} {\bibinfo {author} {\bibfnamefont {A.~W.}\ \bibnamefont {Cross}}, \bibinfo {author} {\bibfnamefont {D.~P.}\ \bibnamefont {Divincenzo}},\ and\ \bibinfo {author} {\bibfnamefont {B.~M.}\ \bibnamefont {Terhal}},\ }\bibfield  {title} {\bibinfo {title} {A comparative code study for quantum fault tolerance},\ }\href {https://doi.org/10.26421/QIC9.7-8-1} {\bibfield  {journal} {\bibinfo  {journal} {Quantum Info. Comput.}\ }\textbf {\bibinfo {volume} {9}},\ \bibinfo {pages} {541–572} (\bibinfo {year} {2009})}\BibitemShut {NoStop}%
\bibitem [{\citenamefont {Chamberland}\ and\ \citenamefont {Ronagh}(2018)}]{chamberland2018deep}%
  \BibitemOpen
  \bibfield  {author} {\bibinfo {author} {\bibfnamefont {C.}~\bibnamefont {Chamberland}}\ and\ \bibinfo {author} {\bibfnamefont {P.}~\bibnamefont {Ronagh}},\ }\bibfield  {title} {\bibinfo {title} {Deep neural decoders for near term fault-tolerant experiments},\ }\href {https://iopscience.iop.org/article/10.1088/2058-9565/aad1f7} {\bibfield  {journal} {\bibinfo  {journal} {Quantum Science and Technology}\ }\textbf {\bibinfo {volume} {3}},\ \bibinfo {pages} {044002} (\bibinfo {year} {2018})}\BibitemShut {NoStop}%
\bibitem [{\citenamefont {Xu}\ \emph {et~al.}(2023)\citenamefont {Xu}, \citenamefont {Ataides}, \citenamefont {Pattison}, \citenamefont {Raveendran}, \citenamefont {Bluvstein}, \citenamefont {Wurtz}, \citenamefont {Vasic}, \citenamefont {Lukin}, \citenamefont {Jiang},\ and\ \citenamefont {Zhou}}]{xu2023constantoverhead}%
  \BibitemOpen
  \bibfield  {author} {\bibinfo {author} {\bibfnamefont {Q.}~\bibnamefont {Xu}}, \bibinfo {author} {\bibfnamefont {J.~P.~B.}\ \bibnamefont {Ataides}}, \bibinfo {author} {\bibfnamefont {C.~A.}\ \bibnamefont {Pattison}}, \bibinfo {author} {\bibfnamefont {N.}~\bibnamefont {Raveendran}}, \bibinfo {author} {\bibfnamefont {D.}~\bibnamefont {Bluvstein}}, \bibinfo {author} {\bibfnamefont {J.}~\bibnamefont {Wurtz}}, \bibinfo {author} {\bibfnamefont {B.}~\bibnamefont {Vasic}}, \bibinfo {author} {\bibfnamefont {M.~D.}\ \bibnamefont {Lukin}}, \bibinfo {author} {\bibfnamefont {L.}~\bibnamefont {Jiang}},\ and\ \bibinfo {author} {\bibfnamefont {H.}~\bibnamefont {Zhou}},\ }\bibfield  {title} {\bibinfo {title} {Constant-overhead fault-tolerant quantum computation with reconfigurable atom arrays},\ }\Eprint {https://arxiv.org/abs/2308.08648} {arXiv:2308.08648 [quant-ph]}  (\bibinfo {year} {2023})\BibitemShut {NoStop}%
\bibitem [{\citenamefont {Fellous-Asiani}\ \emph {et~al.}(2023)\citenamefont {Fellous-Asiani}, \citenamefont {Chai}, \citenamefont {Thonnart}, \citenamefont {Ng}, \citenamefont {Whitney},\ and\ \citenamefont {Auff\`eves}}]{PRXQuantum.4.040319}%
  \BibitemOpen
  \bibfield  {author} {\bibinfo {author} {\bibfnamefont {M.}~\bibnamefont {Fellous-Asiani}}, \bibinfo {author} {\bibfnamefont {J.~H.}\ \bibnamefont {Chai}}, \bibinfo {author} {\bibfnamefont {Y.}~\bibnamefont {Thonnart}}, \bibinfo {author} {\bibfnamefont {H.~K.}\ \bibnamefont {Ng}}, \bibinfo {author} {\bibfnamefont {R.~S.}\ \bibnamefont {Whitney}},\ and\ \bibinfo {author} {\bibfnamefont {A.}~\bibnamefont {Auff\`eves}},\ }\bibfield  {title} {\bibinfo {title} {Optimizing resource efficiencies for scalable full-stack quantum computers},\ }\href {https://doi.org/10.1103/PRXQuantum.4.040319} {\bibfield  {journal} {\bibinfo  {journal} {PRX Quantum}\ }\textbf {\bibinfo {volume} {4}},\ \bibinfo {pages} {040319} (\bibinfo {year} {2023})}\BibitemShut {NoStop}%
\bibitem [{\citenamefont {Pattison}\ \emph {et~al.}(2023)\citenamefont {Pattison}, \citenamefont {Krishna},\ and\ \citenamefont {Preskill}}]{pattison2023hierarchical}%
  \BibitemOpen
  \bibfield  {author} {\bibinfo {author} {\bibfnamefont {C.~A.}\ \bibnamefont {Pattison}}, \bibinfo {author} {\bibfnamefont {A.}~\bibnamefont {Krishna}},\ and\ \bibinfo {author} {\bibfnamefont {J.}~\bibnamefont {Preskill}},\ }\bibfield  {title} {\bibinfo {title} {Hierarchical memories: Simulating quantum ldpc codes with local gates},\ }\Eprint {https://arxiv.org/abs/2303.04798} {arXiv:2303.04798 [quant-ph]}  (\bibinfo {year} {2023})\BibitemShut {NoStop}%
\bibitem [{\citenamefont {Bravyi}\ \emph {et~al.}(2023)\citenamefont {Bravyi}, \citenamefont {Cross}, \citenamefont {Gambetta}, \citenamefont {Maslov}, \citenamefont {Rall},\ and\ \citenamefont {Yoder}}]{bravyi2023highthreshold}%
  \BibitemOpen
  \bibfield  {author} {\bibinfo {author} {\bibfnamefont {S.}~\bibnamefont {Bravyi}}, \bibinfo {author} {\bibfnamefont {A.~W.}\ \bibnamefont {Cross}}, \bibinfo {author} {\bibfnamefont {J.~M.}\ \bibnamefont {Gambetta}}, \bibinfo {author} {\bibfnamefont {D.}~\bibnamefont {Maslov}}, \bibinfo {author} {\bibfnamefont {P.}~\bibnamefont {Rall}},\ and\ \bibinfo {author} {\bibfnamefont {T.~J.}\ \bibnamefont {Yoder}},\ }\bibfield  {title} {\bibinfo {title} {High-threshold and low-overhead fault-tolerant quantum memory},\ }\Eprint {https://arxiv.org/abs/2308.07915} {arXiv:2308.07915 [quant-ph]}  (\bibinfo {year} {2023})\BibitemShut {NoStop}%
\bibitem [{\citenamefont {Christandl}\ and\ \citenamefont {Müller-Hermes}(2022)}]{christandl2022fault}%
  \BibitemOpen
  \bibfield  {author} {\bibinfo {author} {\bibfnamefont {M.}~\bibnamefont {Christandl}}\ and\ \bibinfo {author} {\bibfnamefont {A.}~\bibnamefont {Müller-Hermes}},\ }\bibfield  {title} {\bibinfo {title} {Fault-tolerant coding for quantum communication},\ }\href {https://doi.org/10.1109/TIT.2022.3169438} {\bibfield  {journal} {\bibinfo  {journal} {IEEE Transactions on Information Theory}\ }\textbf {\bibinfo {volume} {70}},\ \bibinfo {pages} {282} (\bibinfo {year} {2022})}\BibitemShut {NoStop}%
\bibitem [{\citenamefont {Gottesman}(2000)}]{gottesman2000fault}%
  \BibitemOpen
  \bibfield  {author} {\bibinfo {author} {\bibfnamefont {D.}~\bibnamefont {Gottesman}},\ }\bibfield  {title} {\bibinfo {title} {Fault-tolerant quantum computation with local gates},\ }\href {https://doi.org/10.1080/09500340008244046} {\bibfield  {journal} {\bibinfo  {journal} {Journal of Modern Optics}\ }\textbf {\bibinfo {volume} {47}},\ \bibinfo {pages} {333} (\bibinfo {year} {2000})}\BibitemShut {NoStop}%
\bibitem [{\citenamefont {Baspin}\ \emph {et~al.}(2023)\citenamefont {Baspin}, \citenamefont {Fawzi},\ and\ \citenamefont {Shayeghi}}]{baspin2023lower}%
  \BibitemOpen
  \bibfield  {author} {\bibinfo {author} {\bibfnamefont {N.}~\bibnamefont {Baspin}}, \bibinfo {author} {\bibfnamefont {O.}~\bibnamefont {Fawzi}},\ and\ \bibinfo {author} {\bibfnamefont {A.}~\bibnamefont {Shayeghi}},\ }\bibfield  {title} {\bibinfo {title} {A lower bound on the overhead of quantum error correction in low dimensions},\ }\Eprint {https://arxiv.org/abs/2302.04317} {arXiv:2302.04317 [quant-ph]}  (\bibinfo {year} {2023})\BibitemShut {NoStop}%
\bibitem [{\citenamefont {Svore}\ \emph {et~al.}(2007)\citenamefont {Svore}, \citenamefont {Divincenzo},\ and\ \citenamefont {Terhal}}]{svore2007noise}%
  \BibitemOpen
  \bibfield  {author} {\bibinfo {author} {\bibfnamefont {K.~M.}\ \bibnamefont {Svore}}, \bibinfo {author} {\bibfnamefont {D.~P.}\ \bibnamefont {Divincenzo}},\ and\ \bibinfo {author} {\bibfnamefont {B.~M.}\ \bibnamefont {Terhal}},\ }\bibfield  {title} {\bibinfo {title} {Noise threshold for a fault-tolerant two-dimensional lattice architecture},\ }\href@noop {} {\bibfield  {journal} {\bibinfo  {journal} {Quantum Info. Comput.}\ }\textbf {\bibinfo {volume} {7}},\ \bibinfo {pages} {297–318} (\bibinfo {year} {2007})}\BibitemShut {NoStop}%
\bibitem [{\citenamefont {Chao}\ and\ \citenamefont {Reichardt}(2018{\natexlab{a}})}]{chao2018fault}%
  \BibitemOpen
  \bibfield  {author} {\bibinfo {author} {\bibfnamefont {R.}~\bibnamefont {Chao}}\ and\ \bibinfo {author} {\bibfnamefont {B.~W.}\ \bibnamefont {Reichardt}},\ }\bibfield  {title} {\bibinfo {title} {Fault-tolerant quantum computation with few qubits},\ }\href {https://doi.org/10.1038/s41534-018-0085-z} {\bibfield  {journal} {\bibinfo  {journal} {npj Quantum Information}\ }\textbf {\bibinfo {volume} {4}},\ \bibinfo {pages} {42} (\bibinfo {year} {2018}{\natexlab{a}})}\BibitemShut {NoStop}%
\bibitem [{\citenamefont {Reichardt}(2020)}]{reichardt2020fault}%
  \BibitemOpen
  \bibfield  {author} {\bibinfo {author} {\bibfnamefont {B.~W.}\ \bibnamefont {Reichardt}},\ }\bibfield  {title} {\bibinfo {title} {{Fault-tolerant quantum error correction for Steane's seven-qubit color code with few or no extra qubits}},\ }\href {https://doi.org/10.1088/2058-9565/abc6f4} {\bibfield  {journal} {\bibinfo  {journal} {Quantum Science and Technology}\ }\textbf {\bibinfo {volume} {6}},\ \bibinfo {pages} {015007} (\bibinfo {year} {2020})}\BibitemShut {NoStop}%
\bibitem [{\citenamefont {Yoder}\ and\ \citenamefont {Kim}(2017)}]{yoder2017surface}%
  \BibitemOpen
  \bibfield  {author} {\bibinfo {author} {\bibfnamefont {T.~J.}\ \bibnamefont {Yoder}}\ and\ \bibinfo {author} {\bibfnamefont {I.~H.}\ \bibnamefont {Kim}},\ }\bibfield  {title} {\bibinfo {title} {The surface code with a twist},\ }\href {https://doi.org/10.22331/q-2017-04-25-2} {\bibfield  {journal} {\bibinfo  {journal} {Quantum}\ }\textbf {\bibinfo {volume} {1}},\ \bibinfo {pages} {2} (\bibinfo {year} {2017})}\BibitemShut {NoStop}%
\bibitem [{\citenamefont {Chao}\ and\ \citenamefont {Reichardt}(2018{\natexlab{b}})}]{chao2018quantum}%
  \BibitemOpen
  \bibfield  {author} {\bibinfo {author} {\bibfnamefont {R.}~\bibnamefont {Chao}}\ and\ \bibinfo {author} {\bibfnamefont {B.~W.}\ \bibnamefont {Reichardt}},\ }\bibfield  {title} {\bibinfo {title} {{Quantum Error Correction with Only Two Extra Qubits}},\ }\href {https://doi.org/10.1103/PhysRevLett.121.050502} {\bibfield  {journal} {\bibinfo  {journal} {Phys. Rev. Lett.}\ }\textbf {\bibinfo {volume} {121}},\ \bibinfo {pages} {050502} (\bibinfo {year} {2018}{\natexlab{b}})}\BibitemShut {NoStop}%
\bibitem [{\citenamefont {Chamberland}\ and\ \citenamefont {Beverland}(2018)}]{chamberland2018flag}%
  \BibitemOpen
  \bibfield  {author} {\bibinfo {author} {\bibfnamefont {C.}~\bibnamefont {Chamberland}}\ and\ \bibinfo {author} {\bibfnamefont {M.~E.}\ \bibnamefont {Beverland}},\ }\bibfield  {title} {\bibinfo {title} {Flag fault-tolerant error correction with arbitrary distance codes},\ }\href {https://doi.org/10.22331/q-2018-02-08-53} {\bibfield  {journal} {\bibinfo  {journal} {Quantum}\ }\textbf {\bibinfo {volume} {2}},\ \bibinfo {pages} {53} (\bibinfo {year} {2018})}\BibitemShut {NoStop}%
\bibitem [{\citenamefont {Chao}\ and\ \citenamefont {Reichardt}(2020)}]{chao2020flag}%
  \BibitemOpen
  \bibfield  {author} {\bibinfo {author} {\bibfnamefont {R.}~\bibnamefont {Chao}}\ and\ \bibinfo {author} {\bibfnamefont {B.~W.}\ \bibnamefont {Reichardt}},\ }\bibfield  {title} {\bibinfo {title} {Flag fault-tolerant error correction for any stabilizer code},\ }\href {https://doi.org/10.1103/PRXQuantum.1.010302} {\bibfield  {journal} {\bibinfo  {journal} {PRX Quantum}\ }\textbf {\bibinfo {volume} {1}},\ \bibinfo {pages} {010302} (\bibinfo {year} {2020})}\BibitemShut {NoStop}%
\bibitem [{\citenamefont {Chamberland}\ and\ \citenamefont {Cross}(2019)}]{chamberland2019fault}%
  \BibitemOpen
  \bibfield  {author} {\bibinfo {author} {\bibfnamefont {C.}~\bibnamefont {Chamberland}}\ and\ \bibinfo {author} {\bibfnamefont {A.~W.}\ \bibnamefont {Cross}},\ }\bibfield  {title} {\bibinfo {title} {Fault-tolerant magic state preparation with flag qubits},\ }\href {https://doi.org/10.22331/q-2019-05-20-143} {\bibfield  {journal} {\bibinfo  {journal} {Quantum}\ }\textbf {\bibinfo {volume} {3}},\ \bibinfo {pages} {143} (\bibinfo {year} {2019})}\BibitemShut {NoStop}%
\bibitem [{\citenamefont {Goto}(2024)}]{goto2024high}%
  \BibitemOpen
  \bibfield  {author} {\bibinfo {author} {\bibfnamefont {H.}~\bibnamefont {Goto}},\ }\bibfield  {title} {\bibinfo {title} {High-performance fault-tolerant quantum computing with many-hypercube codes},\ }\href {https://doi.org/10.1126/sciadv.adp6388} {\bibfield  {journal} {\bibinfo  {journal} {Science Advances}\ }\textbf {\bibinfo {volume} {10}},\ \bibinfo {pages} {eadp6388} (\bibinfo {year} {2024})}\BibitemShut {NoStop}%
\bibitem [{\citenamefont {Nielsen}\ and\ \citenamefont {Chuang}(2010)}]{nielsen2010quantum}%
  \BibitemOpen
  \bibfield  {author} {\bibinfo {author} {\bibfnamefont {M.~A.}\ \bibnamefont {Nielsen}}\ and\ \bibinfo {author} {\bibfnamefont {I.~L.}\ \bibnamefont {Chuang}},\ }\href@noop {} {\emph {\bibinfo {title} {Quantum computation and quantum information}}}\ (\bibinfo  {publisher} {Cambridge university press},\ \bibinfo {year} {2010})\BibitemShut {NoStop}%
\bibitem [{\citenamefont {Gottesman}(2010)}]{gottesman2010introduction}%
  \BibitemOpen
  \bibfield  {author} {\bibinfo {author} {\bibfnamefont {D.}~\bibnamefont {Gottesman}},\ }\bibfield  {title} {\bibinfo {title} {An introduction to quantum error correction and fault-tolerant quantum computation},\ }in\ \href {http://www.ams.org/books/psapm/068/} {\emph {\bibinfo {booktitle} {Quantum information science and its contributions to mathematics, Proceedings of Symposia in Applied Mathematics}}},\ Vol.~\bibinfo {volume} {68}\ (\bibinfo {year} {2010})\ pp.\ \bibinfo {pages} {13--58}\BibitemShut {NoStop}%
\bibitem [{\citenamefont {Steane}(1997)}]{PhysRevLett.78.2252}%
  \BibitemOpen
  \bibfield  {author} {\bibinfo {author} {\bibfnamefont {A.~M.}\ \bibnamefont {Steane}},\ }\bibfield  {title} {\bibinfo {title} {Active stabilization, quantum computation, and quantum state synthesis},\ }\href {https://doi.org/10.1103/PhysRevLett.78.2252} {\bibfield  {journal} {\bibinfo  {journal} {Phys. Rev. Lett.}\ }\textbf {\bibinfo {volume} {78}},\ \bibinfo {pages} {2252} (\bibinfo {year} {1997})}\BibitemShut {NoStop}%
\bibitem [{\citenamefont {Knill}(2005{\natexlab{b}})}]{PhysRevA.71.042322}%
  \BibitemOpen
  \bibfield  {author} {\bibinfo {author} {\bibfnamefont {E.}~\bibnamefont {Knill}},\ }\bibfield  {title} {\bibinfo {title} {Scalable quantum computing in the presence of large detected-error rates},\ }\href {https://doi.org/10.1103/PhysRevA.71.042322} {\bibfield  {journal} {\bibinfo  {journal} {Phys. Rev. A}\ }\textbf {\bibinfo {volume} {71}},\ \bibinfo {pages} {042322} (\bibinfo {year} {2005}{\natexlab{b}})}\BibitemShut {NoStop}%
\bibitem [{\citenamefont {Glancy}\ \emph {et~al.}(2006)\citenamefont {Glancy}, \citenamefont {Knill},\ and\ \citenamefont {Vasconcelos}}]{glancy2006entanglement}%
  \BibitemOpen
  \bibfield  {author} {\bibinfo {author} {\bibfnamefont {S.}~\bibnamefont {Glancy}}, \bibinfo {author} {\bibfnamefont {E.}~\bibnamefont {Knill}},\ and\ \bibinfo {author} {\bibfnamefont {H.~M.}\ \bibnamefont {Vasconcelos}},\ }\bibfield  {title} {\bibinfo {title} {Entanglement purification of any stabilizer state},\ }\href {https://doi.org/10.1103/PhysRevA.74.032319} {\bibfield  {journal} {\bibinfo  {journal} {Phys. Rev. A}\ }\textbf {\bibinfo {volume} {74}},\ \bibinfo {pages} {032319} (\bibinfo {year} {2006})}\BibitemShut {NoStop}%
\bibitem [{\citenamefont {Sahay}\ \emph {et~al.}(2024)\citenamefont {Sahay}, \citenamefont {Lin}, \citenamefont {Huang}, \citenamefont {Brown},\ and\ \citenamefont {Puri}}]{sahay2024error}%
  \BibitemOpen
  \bibfield  {author} {\bibinfo {author} {\bibfnamefont {K.}~\bibnamefont {Sahay}}, \bibinfo {author} {\bibfnamefont {Y.}~\bibnamefont {Lin}}, \bibinfo {author} {\bibfnamefont {S.}~\bibnamefont {Huang}}, \bibinfo {author} {\bibfnamefont {K.~R.}\ \bibnamefont {Brown}},\ and\ \bibinfo {author} {\bibfnamefont {S.}~\bibnamefont {Puri}},\ }\bibfield  {title} {\bibinfo {title} {{Error correction of transversal CNOT gates for scalable surface code computation}},\ }\Eprint {https://arxiv.org/abs/2408.01393} {arXiv:2408.01393}  (\bibinfo {year} {2024})\BibitemShut {NoStop}%
\bibitem [{\citenamefont {Lee}\ \emph {et~al.}(2021)\citenamefont {Lee}, \citenamefont {Berry}, \citenamefont {Gidney}, \citenamefont {Huggins}, \citenamefont {McClean}, \citenamefont {Wiebe},\ and\ \citenamefont {Babbush}}]{PRXQuantum.2.030305}%
  \BibitemOpen
  \bibfield  {author} {\bibinfo {author} {\bibfnamefont {J.}~\bibnamefont {Lee}}, \bibinfo {author} {\bibfnamefont {D.~W.}\ \bibnamefont {Berry}}, \bibinfo {author} {\bibfnamefont {C.}~\bibnamefont {Gidney}}, \bibinfo {author} {\bibfnamefont {W.~J.}\ \bibnamefont {Huggins}}, \bibinfo {author} {\bibfnamefont {J.~R.}\ \bibnamefont {McClean}}, \bibinfo {author} {\bibfnamefont {N.}~\bibnamefont {Wiebe}},\ and\ \bibinfo {author} {\bibfnamefont {R.}~\bibnamefont {Babbush}},\ }\bibfield  {title} {\bibinfo {title} {Even more efficient quantum computations of chemistry through tensor hypercontraction},\ }\href {https://doi.org/10.1103/PRXQuantum.2.030305} {\bibfield  {journal} {\bibinfo  {journal} {PRX Quantum}\ }\textbf {\bibinfo {volume} {2}},\ \bibinfo {pages} {030305} (\bibinfo {year} {2021})}\BibitemShut {NoStop}%
\bibitem [{\citenamefont {Yoshioka}\ \emph {et~al.}(2023)\citenamefont {Yoshioka}, \citenamefont {Okubo}, \citenamefont {Suzuki}, \citenamefont {Koizumi},\ and\ \citenamefont {Mizukami}}]{yoshioka2023hunting}%
  \BibitemOpen
  \bibfield  {author} {\bibinfo {author} {\bibfnamefont {N.}~\bibnamefont {Yoshioka}}, \bibinfo {author} {\bibfnamefont {T.}~\bibnamefont {Okubo}}, \bibinfo {author} {\bibfnamefont {Y.}~\bibnamefont {Suzuki}}, \bibinfo {author} {\bibfnamefont {Y.}~\bibnamefont {Koizumi}},\ and\ \bibinfo {author} {\bibfnamefont {W.}~\bibnamefont {Mizukami}},\ }\bibfield  {title} {\bibinfo {title} {Hunting for quantum-classical crossover in condensed matter problems},\ }\Eprint {https://arxiv.org/abs/2210.14109} {arXiv:2210.14109 [quant-ph]}  (\bibinfo {year} {2023})\BibitemShut {NoStop}%
\bibitem [{\citenamefont {Gidney}(2021)}]{gidney2021stim}%
  \BibitemOpen
  \bibfield  {author} {\bibinfo {author} {\bibfnamefont {C.}~\bibnamefont {Gidney}},\ }\bibfield  {title} {\bibinfo {title} {Stim: a fast stabilizer circuit simulator},\ }\href {https://doi.org/10.22331/q-2021-07-06-497} {\bibfield  {journal} {\bibinfo  {journal} {Quantum}\ }\textbf {\bibinfo {volume} {5}},\ \bibinfo {pages} {497} (\bibinfo {year} {2021})}\BibitemShut {NoStop}%
\bibitem [{fug(2018)}]{fugaku}%
  \BibitemOpen
  \href@noop {} {\bibinfo {title} {Specifications - supercomputer fugaku: Fujitsu global}},\ \bibinfo {howpublished} {\url{https://www.fujitsu.com/global/about/innovation/fugaku/specifications/}} (\bibinfo {year} {2018})\BibitemShut {NoStop}%
\bibitem [{\citenamefont {Wang}\ \emph {et~al.}(2003)\citenamefont {Wang}, \citenamefont {Harrington},\ and\ \citenamefont {Preskill}}]{Wang_2003}%
  \BibitemOpen
  \bibfield  {author} {\bibinfo {author} {\bibfnamefont {C.}~\bibnamefont {Wang}}, \bibinfo {author} {\bibfnamefont {J.}~\bibnamefont {Harrington}},\ and\ \bibinfo {author} {\bibfnamefont {J.}~\bibnamefont {Preskill}},\ }\bibfield  {title} {\bibinfo {title} {Confinement-higgs transition in a disordered gauge theory and the accuracy threshold for quantum memory},\ }\href {https://doi.org/10.1016/s0003-4916(02)00019-2} {\bibfield  {journal} {\bibinfo  {journal} {Annals of Physics}\ }\textbf {\bibinfo {volume} {303}},\ \bibinfo {pages} {31–58} (\bibinfo {year} {2003})}\BibitemShut {NoStop}%
\bibitem [{qpi(2016)}]{qpic}%
  \BibitemOpen
  \href@noop {} {\bibinfo {title} {qpic}},\ \bibinfo {howpublished} {\url{https://github.com/qpic/qpic}} (\bibinfo {year} {2016})\BibitemShut {NoStop}%
\bibitem [{\citenamefont {Gottesman}(1997)}]{gottesman1997stabilizer}%
  \BibitemOpen
  \bibfield  {author} {\bibinfo {author} {\bibfnamefont {D.}~\bibnamefont {Gottesman}},\ }\emph {\bibinfo {title} {Stabilizer codes and quantum error correction}},\ \href {https://doi.org/10.48550/arXiv.quant-ph/9705052} {Ph.D. thesis},\ \bibinfo  {school} {California Institute of Technology} (\bibinfo {year} {1997})\BibitemShut {NoStop}%
\bibitem [{\citenamefont {Wilde}(2009)}]{wilde2009logical}%
  \BibitemOpen
  \bibfield  {author} {\bibinfo {author} {\bibfnamefont {M.~M.}\ \bibnamefont {Wilde}},\ }\bibfield  {title} {\bibinfo {title} {Logical operators of quantum codes},\ }\href {https://doi.org/10.1103/PhysRevA.79.062322} {\bibfield  {journal} {\bibinfo  {journal} {Phys. Rev. A}\ }\textbf {\bibinfo {volume} {79}},\ \bibinfo {pages} {062322} (\bibinfo {year} {2009})}\BibitemShut {NoStop}%
\bibitem [{\citenamefont {Steane}(2002)}]{steane2002fast}%
  \BibitemOpen
  \bibfield  {author} {\bibinfo {author} {\bibfnamefont {A.~M.}\ \bibnamefont {Steane}},\ }\bibfield  {title} {\bibinfo {title} {Fast fault-tolerant filtering of quantum codewords},\ }\Eprint {https://arxiv.org/abs/quant-ph/0202036} {arXiv:quant-ph/0202036}  (\bibinfo {year} {2002})\BibitemShut {NoStop}%
\bibitem [{\citenamefont {Paetznick}\ and\ \citenamefont {Reichardt}(2011)}]{paetznick2011fault}%
  \BibitemOpen
  \bibfield  {author} {\bibinfo {author} {\bibfnamefont {A.}~\bibnamefont {Paetznick}}\ and\ \bibinfo {author} {\bibfnamefont {B.~W.}\ \bibnamefont {Reichardt}},\ }\bibfield  {title} {\bibinfo {title} {Fault-tolerant ancilla preparation and noise threshold lower bounds for the 23-qubit golay code},\ }\href {https://api.semanticscholar.org/CorpusID:22526275} {\bibfield  {journal} {\bibinfo  {journal} {Quantum Inf. Comput.}\ }\textbf {\bibinfo {volume} {12}},\ \bibinfo {pages} {1034} (\bibinfo {year} {2011})}\BibitemShut {NoStop}%
\bibitem [{\citenamefont {Reichardt}(2006)}]{4031377}%
  \BibitemOpen
  \bibfield  {author} {\bibinfo {author} {\bibfnamefont {B.~W.}\ \bibnamefont {Reichardt}},\ }\bibfield  {title} {\bibinfo {title} {Postselection threshold against biased noise},\ }in\ \href {https://doi.org/10.1109/FOCS.2006.64} {\emph {\bibinfo {booktitle} {2006 47th Annual IEEE Symposium on Foundations of Computer Science (FOCS'06)}}}\ (\bibinfo {year} {2006})\ pp.\ \bibinfo {pages} {420--428}\BibitemShut {NoStop}%
\bibitem [{\citenamefont {Aliferis}\ \emph {et~al.}(2008)\citenamefont {Aliferis}, \citenamefont {Gottesman},\ and\ \citenamefont {Preskill}}]{DBLP:journals/qic/AliferisGP08}%
  \BibitemOpen
  \bibfield  {author} {\bibinfo {author} {\bibfnamefont {P.}~\bibnamefont {Aliferis}}, \bibinfo {author} {\bibfnamefont {D.}~\bibnamefont {Gottesman}},\ and\ \bibinfo {author} {\bibfnamefont {J.}~\bibnamefont {Preskill}},\ }\bibfield  {title} {\bibinfo {title} {Accuracy threshold for postselected quantum computation},\ }\href {https://doi.org/10.26421/QIC8.3-4-1} {\bibfield  {journal} {\bibinfo  {journal} {Quantum Inf. Comput.}\ }\textbf {\bibinfo {volume} {8}},\ \bibinfo {pages} {181} (\bibinfo {year} {2008})}\BibitemShut {NoStop}%
\bibitem [{\citenamefont {Reichardt}(2009)}]{reichardt2009error}%
  \BibitemOpen
  \bibfield  {author} {\bibinfo {author} {\bibfnamefont {B.~W.}\ \bibnamefont {Reichardt}},\ }\bibfield  {title} {\bibinfo {title} {Error-detection-based quantum fault-tolerance threshold},\ }\href {https://link.springer.com/article/10.1007/s00453-007-9069-7} {\bibfield  {journal} {\bibinfo  {journal} {Algorithmica}\ }\textbf {\bibinfo {volume} {55}},\ \bibinfo {pages} {517} (\bibinfo {year} {2009})}\BibitemShut {NoStop}%
\bibitem [{\citenamefont {Freedman}\ and\ \citenamefont {Meyer}(2001)}]{freedman2001projective}%
  \BibitemOpen
  \bibfield  {author} {\bibinfo {author} {\bibfnamefont {M.~H.}\ \bibnamefont {Freedman}}\ and\ \bibinfo {author} {\bibfnamefont {D.~A.}\ \bibnamefont {Meyer}},\ }\bibfield  {title} {\bibinfo {title} {Projective plane and planar quantum codes},\ }\href {https://doi.org/10.1007/s102080010013} {\bibfield  {journal} {\bibinfo  {journal} {Foundations of Computational Mathematics}\ }\textbf {\bibinfo {volume} {1}},\ \bibinfo {pages} {325} (\bibinfo {year} {2001})}\BibitemShut {NoStop}%
\bibitem [{\citenamefont {Kitaev}(1997)}]{Kitaev1997QuantumEC}%
  \BibitemOpen
  \bibfield  {author} {\bibinfo {author} {\bibfnamefont {A.~Y.}\ \bibnamefont {Kitaev}},\ }\bibinfo {title} {Quantum error correction with imperfect gates},\ in\ \href {https://doi.org/10.1007/978-1-4615-5923-8_19} {\emph {\bibinfo {booktitle} {Quantum Communication, Computing, and Measurement}}},\ \bibinfo {editor} {edited by\ \bibinfo {editor} {\bibfnamefont {O.}~\bibnamefont {Hirota}}, \bibinfo {editor} {\bibfnamefont {A.~S.}\ \bibnamefont {Holevo}},\ and\ \bibinfo {editor} {\bibfnamefont {C.~M.}\ \bibnamefont {Caves}}}\ (\bibinfo  {publisher} {Springer US},\ \bibinfo {address} {Boston, MA},\ \bibinfo {year} {1997})\ pp.\ \bibinfo {pages} {181--188}\BibitemShut {NoStop}%
\bibitem [{\citenamefont {Kitaev}(2003)}]{kitaev2003fault}%
  \BibitemOpen
  \bibfield  {author} {\bibinfo {author} {\bibfnamefont {A.~Y.}\ \bibnamefont {Kitaev}},\ }\bibfield  {title} {\bibinfo {title} {Fault-tolerant quantum computation by anyons},\ }\href {https://doi.org/10.1016/S0003-4916(02)00018-0} {\bibfield  {journal} {\bibinfo  {journal} {Annals of physics}\ }\textbf {\bibinfo {volume} {303}},\ \bibinfo {pages} {2} (\bibinfo {year} {2003})}\BibitemShut {NoStop}%
\bibitem [{\citenamefont {Bombin}\ and\ \citenamefont {Martin-Delgado}(2007)}]{Bombin_2007}%
  \BibitemOpen
  \bibfield  {author} {\bibinfo {author} {\bibfnamefont {H.}~\bibnamefont {Bombin}}\ and\ \bibinfo {author} {\bibfnamefont {M.~A.}\ \bibnamefont {Martin-Delgado}},\ }\bibfield  {title} {\bibinfo {title} {Optimal resources for topological two-dimensional stabilizer codes: Comparative study},\ }\href {http://dx.doi.org/10.1103/PhysRevA.76.012305} {\bibfield  {journal} {\bibinfo  {journal} {Physical Review A}\ }\textbf {\bibinfo {volume} {76}} (\bibinfo {year} {2007})}\BibitemShut {NoStop}%
\bibitem [{\citenamefont {Chamberland}\ and\ \citenamefont {Campbell}(2022)}]{Chamberland_2022}%
  \BibitemOpen
  \bibfield  {author} {\bibinfo {author} {\bibfnamefont {C.}~\bibnamefont {Chamberland}}\ and\ \bibinfo {author} {\bibfnamefont {E.~T.}\ \bibnamefont {Campbell}},\ }\bibfield  {title} {\bibinfo {title} {Universal quantum computing with twist-free and temporally encoded lattice surgery},\ }\href {https://doi.org/10.1103/prxquantum.3.010331} {\bibfield  {journal} {\bibinfo  {journal} {PRX Quantum}\ }\textbf {\bibinfo {volume} {3}},\ \bibinfo {pages} {010331} (\bibinfo {year} {2022})}\BibitemShut {NoStop}%
\bibitem [{\citenamefont {Fowler}\ and\ \citenamefont {Gidney}(2019)}]{fowler2019low}%
  \BibitemOpen
  \bibfield  {author} {\bibinfo {author} {\bibfnamefont {A.~G.}\ \bibnamefont {Fowler}}\ and\ \bibinfo {author} {\bibfnamefont {C.}~\bibnamefont {Gidney}},\ }\bibfield  {title} {\bibinfo {title} {Low overhead quantum computation using lattice surgery},\ }\Eprint {https://arxiv.org/abs/1808.06709} {arXiv:1808.06709 [quant-ph]}  (\bibinfo {year} {2019})\BibitemShut {NoStop}%
\bibitem [{\citenamefont {Gidney}(2022{\natexlab{b}})}]{Gidney_2022}%
  \BibitemOpen
  \bibfield  {author} {\bibinfo {author} {\bibfnamefont {C.}~\bibnamefont {Gidney}},\ }\bibfield  {title} {\bibinfo {title} {Stability experiments: The overlooked dual of memory experiments},\ }\href {https://doi.org/10.22331/q-2022-08-24-786} {\bibfield  {journal} {\bibinfo  {journal} {Quantum}\ }\textbf {\bibinfo {volume} {6}},\ \bibinfo {pages} {786} (\bibinfo {year} {2022}{\natexlab{b}})}\BibitemShut {NoStop}%
\bibitem [{\citenamefont {Poulin}(2006)}]{PhysRevA.74.052333}%
  \BibitemOpen
  \bibfield  {author} {\bibinfo {author} {\bibfnamefont {D.}~\bibnamefont {Poulin}},\ }\bibfield  {title} {\bibinfo {title} {Optimal and efficient decoding of concatenated quantum block codes},\ }\href {https://doi.org/10.1103/PhysRevA.74.052333} {\bibfield  {journal} {\bibinfo  {journal} {Phys. Rev. A}\ }\textbf {\bibinfo {volume} {74}},\ \bibinfo {pages} {052333} (\bibinfo {year} {2006})}\BibitemShut {NoStop}%
\bibitem [{\citenamefont {Gidney}\ \emph {et~al.}(2022)\citenamefont {Gidney}, \citenamefont {Newman},\ and\ \citenamefont {McEwen}}]{gidney2022benchmarking}%
  \BibitemOpen
  \bibfield  {author} {\bibinfo {author} {\bibfnamefont {C.}~\bibnamefont {Gidney}}, \bibinfo {author} {\bibfnamefont {M.}~\bibnamefont {Newman}},\ and\ \bibinfo {author} {\bibfnamefont {M.}~\bibnamefont {McEwen}},\ }\bibfield  {title} {\bibinfo {title} {Benchmarking the planar honeycomb code},\ }\href {https://doi.org/10.22331/q-2022-09-21-813} {\bibfield  {journal} {\bibinfo  {journal} {Quantum}\ }\textbf {\bibinfo {volume} {6}},\ \bibinfo {pages} {813} (\bibinfo {year} {2022})}\BibitemShut {NoStop}%
\end{thebibliography}%

\end{document}